\documentclass[aps,pra,twocolumn,,superscriptaddress,floatfix]{revtex4-1}
\usepackage[svgnames]{xcolor}
\usepackage[colorlinks=true,urlcolor = purple,linkcolor=teal,citecolor=magenta,bookmarks=false]{hyperref}
\usepackage{amsfonts,amsmath,amssymb}
\usepackage{graphicx}
\usepackage{dsfont}

\usepackage{tikz}
\usetikzlibrary{arrows,topaths,shapes.geometric,decorations.markings,shadows,positioning,
plotmarks}

\usepackage{pgfplots}

\def\ii{{\rm i}}
\newcommand{\dd}{{\rm d}}
\def\bra#1{\mathinner{\langle{#1}|}}
\def\ket#1{\mathinner{|{#1}\rangle}}

\def\expect#1{\langle#1\rangle}

\def\ol#1{\bar{#1}}

\newcommand{\fs}{\mathfrak{s}}
\newcommand{\mass}{\mathfrak{m}}

\begin{document}

\title{Anomalous spin diffusion in one-dimensional antiferromagnets}

\author{Jacopo De Nardis}
\affiliation{Department of Physics and Astronomy, University of Ghent, 
Krijgslaan 281, 9000 Gent, Belgium}

\author{Marko Medenjak}
\affiliation
{Institut de Physique Th\'eorique Philippe Meyer, \'Ecole Normale Sup\'erieure, \\ PSL University, Sorbonne Universit\'es, CNRS, 75005 Paris, France}

\author{Christoph Karrasch}
\affiliation{Technische Universit\"at Braunschweig, Institut f\"ur Mathematische Physik, Mendelssohnstra\ss e 3, 38106 Braunschweig, Germany}

\author{Enej Ilievski}
\affiliation{Institute for Theoretical Physics Amsterdam and Delta Institute for Theoretical Physics,
University of Amsterdam, Science Park 904, 1098 XH Amsterdam, The Netherlands}

\date{\today}

\begin{abstract}
The problem of characterizing low-temperature spin dynamics in antiferromagnetic spin chains has so far remained elusive.
Here we reinvestigate it by focusing on isotropic antiferromagnetic chains whose low-energy effective field theory is governed by
the quantum non-linear sigma model. Employing an exact non-perturbative theoretical approach, we analyze the low-temperature behaviour 
in the vicinity of non-magnetized states and obtain exact expressions for the spin diffusion 
constant and the NMR relaxation rate, which we compare with previous theoretical results in the literature. Surprisingly, 
in $SU(2)$-invariant spin chains in the vicinity of half-filling we find a crossover from the semi-classical regime to a 
strongly interacting quantum regime characterized by zero spin Drude weight and diverging spin conductivity, indicating
super-diffusive spin dynamics. The dynamical exponent of spin fluctuations is argued to belong to the Kardar-Parisi-Zhang universality 
class. Furthermore, by employing numerical tDMRG simulations, we find robust evidence that the anomalous spin transport 
persists also at high temperatures, irrespectively of the spectral gap and integrability of the model.
\end{abstract}

\pacs{02.30.Ik,05.70.Ln,75.10.Jm}

\maketitle

One-dimensional isotropic antiferromagnets reveal several remarkable aspects, which made them a subject of very intense
experimental and theoretical investigations in the past. One of the most profound features is a fundamental distinction
between spin systems with odd and integer spin. In one dimension, the latter exhibit dynamically generated gapped spectrum while
the former is characterised by gapless excitations with fractional statistics \cite{Haldane1983,haldane1983continuum,Shankar1990}.

In the context of non-equilibrium physics, the main focus has been to explain the peculiar properties of the spin relaxation dynamics of the Haldane-gapped spin chain compounds. In spite of various theoretical approaches,
ranging from field-theoretical techniques such as the form-factor expansions \cite{Konik2003,Altshuler2006},
to the semi-classical approximations \cite{Sachdev1997,Damle1998,Sachdev2000,Cuccoli2000,Damle2005},
the status of the topic remained controversial, with a number of conflicting statements concerning the spin Drude 
weight, spin diffusion constant, and the nuclear magnetic resonance (NMR) rate.

{
Recent years have brought many theoretical advancements in the domain of non-equilibrium phenomena in exactly solvable 
interacting systems. One of the key achievements amongst is the formalism of the generalized hydrodynamics \cite{PhysRevX.6.041065,PhysRevLett.117.207201}, see also
\cite{Bulchandani2018,Bulchandani2017,PhysRevLett.120.045301,Schemmer2019,PhysRevLett.119.195301,
Alba2019,SciPostPhys.2.2.014,PhysRevB.96.115124,Mestyan2019,Mazza2018,Doyon2018,PhysRevB.96.020403,Bastianello2019},
which offers an efficient and universal language to tackle various
non-equilibrium problems. Among others, it enables us to obtain closed-form analytic expressions for transport coefficients, such
as Drude weights \cite{PhysRevLett.82.1764,IN_Drude,SciPostPhys.3.6.039,IN_Hubbard} (see also \cite{KlumperDrude}) and, more recently,
diffusion constants in interacting quantum systems \cite{DeNardis2018,Gopalakrishnan2018,1812.00767,1812.02701}.
This powerful toolbox puts us in a position to address a number of perennial issues which fall outside of the scope of
previous approaches.}


{
In this work, we revisit and resolve the problem of spin transport in antiferromagnetic spin chains at low
temperatures in the half-filled sector, investigated previously in \cite{Fujimoto1999,Konik2003,essler2009finite}.
Here we focus our attention to two physically relevant quantities, the spin diffusion constant and the nuclear spin relaxation rate.
We concentrate entirely to locally-interacting quantum spin-$S$ chains with $SU(2)$-symmetric Hamiltonians where our findings markedly 
differ from previous predictions. We demonstrate that in the experimentally relevant regime $h/T \ll 1$, where $T$ is the 
temperature and $h$ the external magnetic field, the spin dynamics is dominated by \textit{collective magnonic bound-state excitations} as described by the full many-body scattering matrix of the underlying effective field theory.}
This has several far-reaching physical consequences, most prominently the \textit{divergent} spin (charge) diffusion constant and 
spin conductivity at any finite temperature, which signals {\textit{super-diffusive} spin transport, with
time-dependent DC conductivity growing as $t^{1/3}$ at large times}.
{
This anomalous feature was initially observed numerically
in an \emph{integrable} isotropic Heisenberg model \cite{znidarivc2011spin,Znidaric2011},
and established rigorously in \cite{ilievski2018superdiffusion}. A recent numerical study in the same model \cite{Ljubotina2019} gives 
a strong evidence that the spin relaxation dynamics falls into the Kardar-Parisi-Zhang (KPZ) universality class,
otherwise better known from the physics of growing interfaces \cite{PhysRevLett.56.889,CORWIN2012,Takeuchi2018}.

By performing exact non-perturbative calculations, we argue that this type of anomalous spin transport is
a distinguished feature of spin/charge transport at low temperatures even in generic one-dimensional \textit{non-integrable} isotropic 
antiferromagnetic compounds and regardless of whether the low-lying theory is gapped or gapless.
Moreover, our numerical tDMRG simulations give evidence that the anomalous spin relaxation also persists at higher temperatures. This 
indicates that non-Abelian global symmetry of spin interaction can have a profound consequence
on the nature of spin transport on sub-ballistic time scales irrespectively of integrability.}

\paragraph*{\bf Spin diffusion constant from integrability.}
Let $\hat{H}$ be a spin-chain Hamiltonian with the conserved total magnetization $\hat{S}^{z} = \sum_i \hat{s}^z_i$.
The linear-response spin diffusion constant $\mathfrak{D}$ is computed as the spatio-temporal integrated spin current autocorrelation function \cite{Kubo57,Znidari2019},
\begin{equation}\label{eq:diffusionconst}
\mathfrak{D}(T,h) =   \frac{1}{T \chi_{{h}} (T,h)}
\int_{0}^{\infty}\!\!\dd t \left(  \big\langle \hat{J}(t)\hat{j}_{0}(0) \big\rangle_{T,h}  - \mathcal{D}  \right),
\end{equation}
where $\hat{J} =\sum_i \hat{j}_{i}$ is the total spin current with density $\hat{j}_{i}$ at site $i$,
$\langle \bullet \rangle_{T,h}$ corresponds to the equilibrium average
with respect to the grand-canonical Gibbs ensemble $\hat{\varrho}_{\rm GC}(T,h)\simeq \exp{(-(\hat{H}-h\hat{S}^{z})/T)}$, while
$\chi_{h}(T,h)=-\partial^{2} f(T,h)/\partial h^{2}$
is the static spin susceptibility, where
$f(T,h) = -T\log\,{\rm Tr}(\hat{\varrho}_{\rm GC}(T,h))$, and $\mathcal{D}(T,h)$ is the spin Drude weight which has been
subtracted in order to ensure that $\mathfrak{D}(T,h)$ is well-defined.
The spin Drude weight is defined as the large-time limit of the spatially-integrated current-current correlator
in Eq.~\eqref{eq:diffusionconst}, and is generically finite in integrable systems.
However, in a non-magnetized sector (i.e. at half-filling $h=0$) which is of our interest here,
$\mathcal{D}(T,0)=0$ essentially due to particle-hole symmetry of local conservation laws \cite{ilievski2016quasilocal,IN_Drude,SM}. 
This is in agreement with the prediction of the semi-classical theory \cite{Sachdev2000}.

The task of computing the exact diffusion constants in integrable models remains, on the other hand, a challenging open question.
Just very recently, exact explicit expression for the diffusion matrix in a general equilibrium state has been derived
in \cite{1812.00767} using the thermal form factor expansion and in \cite{1812.02701} within the kinetic theory approach.
In this work, we employ the general formula for the exact spin diffusion constant obtained in \cite{1812.00767,1812.02701}.
Here we specifically examine the vicinity of the \emph{half-filled} equilibrium states where, remarkably, we found that the formula
further simplifies and in fact exactly coincides with the curvature of
the zero-frequency noise (or Drude self-weight) \cite{Blanter2000,SciPostPhys.3.6.039},
\begin{equation}\label{eq:spin_Drude}
\mathcal{D}^{\rm self}(T,h) = 2\int_0^\infty \dd t \big\langle \hat{j}_{0}(t) \hat{j}_{0}(0) \big\rangle_{T,h},
\end{equation}
with respect to the magnetization $\nu(T,h) \equiv 4{T} \expect{\hat{S}^{z}}_{T,h}$,
\begin{equation}\label{eq:mainResDiffusion}
\mathfrak{D} \equiv \mathfrak{D}(T,0) =  \frac{\partial^{2} \mathcal{D}^{\rm self}(T,\nu)}{\partial \nu^{2}} \Big|_{\nu=0}.
\end{equation} 
The obtained expression can alternatively by viewed as the optimized diffusion-lower derived in \cite{MKP17}.
We note that Eq.~\eqref{eq:mainResDiffusion} remains valid also for small $h$, up to corrections of the order $\mathcal{O}(h^{2})$.
The spin diffusion constant can accordingly be expressed in terms of equilibrium state functions via the hydrodynamic mode resolution
\begin{equation}\label{eq:D_sum}
\mathfrak{D} = \sum_{s} \mathfrak{D}_{s},
\end{equation}
with $\mathfrak{D}_{s}\! =\! \int \tfrac{\dd p_s(\theta)}{2\pi} n_{s}(\theta)
[1-n_s(\theta)]\times |v^{\rm eff}_s(\theta)| \partial^2_{\nu}(m^{\rm dr}_s)^{2}|_{\nu=0}$.
Here the integer label $s$ runs over all distinct quasi-particle species \cite{1812.00767,SciPostPhys.2.2.014},
$n_{s}(\theta)$ correspond to their (thermal) Fermi occupation functions, $p_s(\theta)$ are their effective (i.e. dressed) momenta
parametrized by rapidity variable $\theta$, $v^{\rm eff}_{s}(\theta)=\partial \varepsilon_{s}(\theta)/\partial p_{s}(\theta)$
are the effective (group) velocities and finally $m^{\rm dr}_{s}$ the dressed magnetization (spin) with respect to a thermal 
background, see \cite{SM}. We will now apply this formula to models with different particle contents and in the low temperature 
regime.

\begin{figure}[t!]
\center
\includegraphics[width=0.49\textwidth]{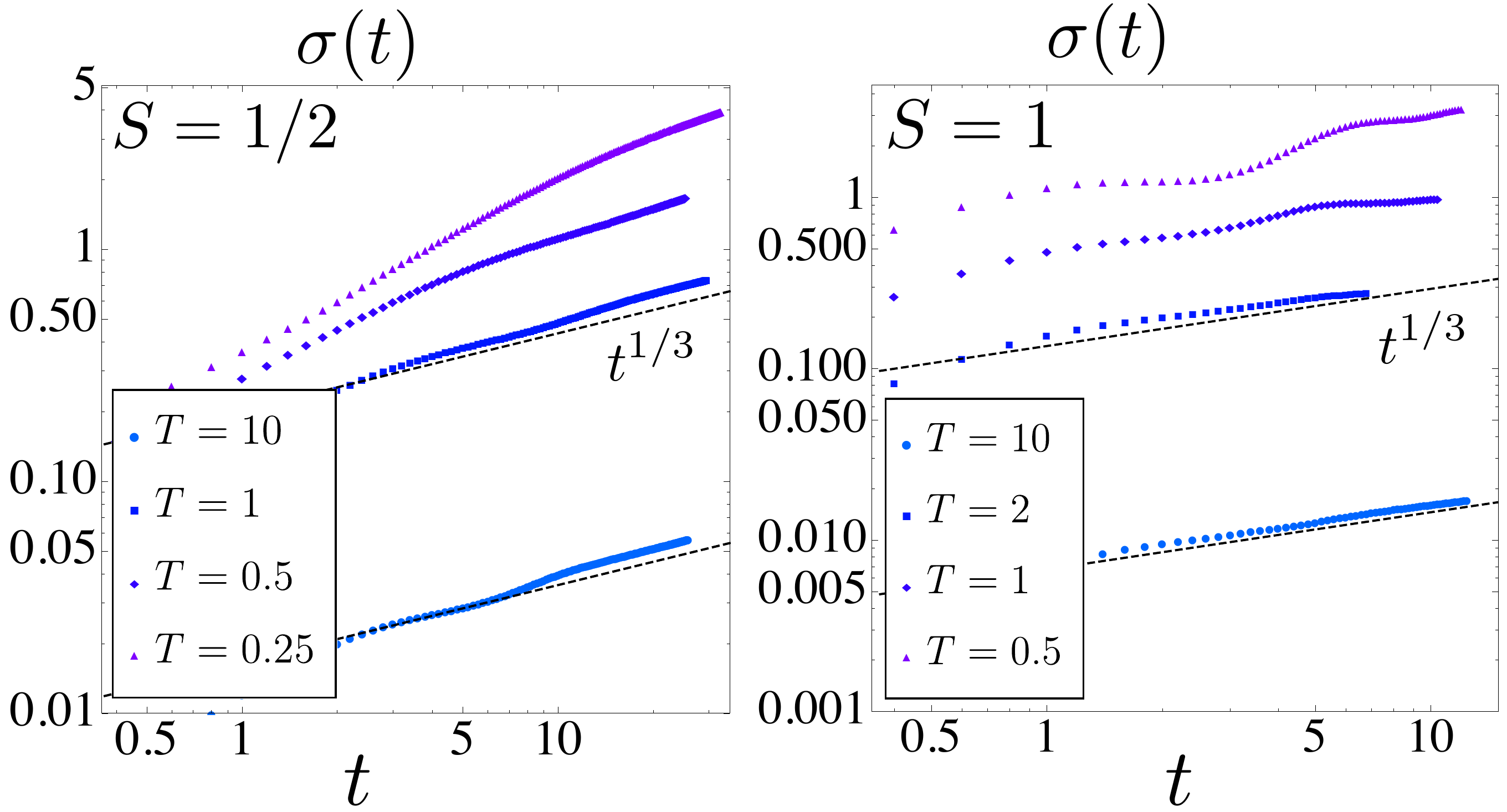}
\caption{Time-dependent spin conductivity (in units of exchange coupling $J$) for the isotropic \textit{gapless} Heisenberg spin 
$S=1/2$ (left) and the spin $S=1$ (right) \textit{(non-integrable) gapped} chain at half-filling $h=0$,
displayed for several different temperatures (increasing from top to bottom) computed using tDMRG simulations.
Both cases exhibit an algebraic law $\sigma(t) \sim t^{1/3}$, indicating that the spin super-diffusion is unrelated to
the spectral gap and integrability of the model.}
\label{Fig:spin}
\end{figure}

\paragraph*{\bf Non-integrable isotropic antiferromagnetic chains.}
We now consider the low-temperature spin dynamics in \emph{generic} antiferromagnetic spin chains with isotropic
spin interactions. For definiteness, we focus on the $SU(2)$-symmetric Heisenberg spin-$S$ chains
$
\hat{H}_{S} = J\sum_{i}\hat{\bf s}_{i}\cdot \hat{\bf s}_{i+1},
$
with $\hat{\bf s}\cdot \hat{\bf s}=S(S+1)$.
In the large-$S$ limit, the effective low-energy action which describes the evolution of the
staggered and ferromagnetic fluctuations $\hat{\bf s}_{i} \approx S(-1)^{i}\hat{\bf n} + \hat{\bf m}$
yields a non-abelian quantum field theory known as
the $O(3)$ non-linear sigma model (NLSM) \cite{Haldane1983,haldane1983continuum,Affleck1987}.
In dimensionless units $v=2JS\to 1$ and coupling parameter $g=2/S$, the Hamiltonian reads,
\begin{equation}
\hat{H}^{(\Theta)}_{\Sigma} = \frac{v}{2}\int \dd x \left[g\Big(\hat{\bf m} +
\frac{\Theta}{4\pi}\partial_{x}\hat{\bf n}\Big)^{2} + \frac{1}{g}(\partial_{x}\hat{\bf n})^{2}\right],
\end{equation}
where ferromagnetic magnetization $\hat{\bf m} = \hat{\bf n}\times \hat{\bf p}$ generates spatial rotations of the unit vector
field $\hat{\bf n}=(\hat{n}^{x},\hat{n}^{y},\hat{n}^{z})$, with the canonically-conjugate momentum
$\hat{\bf p}=(1/g)\partial_{t}\hat{\bf n} + (\Theta/4\pi)\hat{\bf n}\times \partial_{x}\hat{\bf n}$ and
$\Theta=2\pi S$ is the topological angle. For $\Theta\in\{0,\pi\}$ the $O(3)$ NLSM model is an \emph{integrable} QFT
with a completely factorizable scattering matrix~\cite{ZZ1979,ZZ1992}.
Specifically, at $\Theta = 0$ the model yields the effective low-energy theory for the staggered ($k\approx \pi$) and the 
ferromagnetic ($k\approx 0$) fluctuations in the Haldane--gapped integer spin chains.
The $k\to 0$ component of the spin-lattice magnetization corresponds to the conserved Noether charge $\hat{\bf m}$,
obeying continuity equation $\partial_{t}\hat{\bf m}+\partial_{x}(\hat{\bf n}\times (1/g)\partial_{x}\hat{\bf n})=0$.
The elementary excitations are a \emph{massive} triplet of bosons with a relativistic dispersion 
$e(k)=\sqrt{k^{2}+\mass^{2}}$, with $\mass$ being a dynamically-generated mass $\mass\sim \Lambda\,e^{-\pi\,S}$ whose magnitude
is determined by the underlying spin-$S$ lattice model at momentum scale $\Lambda$.
While the NLSM has no physical bound states in the spectrum, the scattering is non-diagonal and governed by
a non-trivial exchange of spin degrees of freedom.
At $\Theta = \pi$, the $O(3)$ NLSM describes the low-energy continuum theory of the half-integer
spin chains with \emph{massless} elementary excitations \cite{Affleck1987,Shankar1990}.

\paragraph*{Low-temperature spin transport.}
Hydrodynamic description of transport is based on the notion of quasi-particles. The physical excitations of the $O(3)$ NLSM
are spin-full boson which interacts via a non-trivial spin exchange. This is conventionally understood in terms of
interacting spin waves (magnons) which are regarded as additional auxiliary quasi-particles and are characterized by internal quantum 
numbers $s>0$ corresponding to a quantized amount of bare spin they carry. The elementary bosonic excitation is ascribed $s=0$.

In the low-temperature limit and small $h$, with ratio $h/T \gg 1$ large, the contributions of spin-carrying auxiliary quasi-particles
become suppressed, and a dilute gas of spin-full bosons serves as a good approximation. In this regime we accordingly recover
the prediction of the semi-classical theory (cf. \cite{SM}) 
\begin{equation}\label{eq:sachdevus}
\mathfrak{D}_{\Sigma}  \simeq \mathfrak{D}_0 =\mathfrak{D}_{\rm cl}(T,h) , \quad  \quad h/T \gg 1,
\end{equation}
where $\mathfrak{D}_{\rm cl}(T,h)=(e^{\mass/T}/\mass)/[1+2\cosh{(h/T)}]$, see  \cite{Sachdev1997}.
In contrast, the behaviour of the spin diffusion constant in the regime $h/T \ll 1 $ is fundamentally different
and the sub-leading corrections attributed to internal magnonic excitations can no longer be neglected. Even worse,
their net contribution to the diffusion constant diverges at small field as $\sim 1/|h|$.
The correct expression for the spin diffusion constant is then given by Eq.~\eqref{eq:mainResDiffusion},
\begin{equation}\label{eq:SachdevResulto3}
\mathfrak{D}_{\Sigma} = \sum_{s \geq 0} \mathfrak{D}_s \sim  \frac{e^{\mass/T}}{3 \mathfrak{m}\ |h|} +\mathcal{O}(h^{0}), \quad  \quad h/T \ll 1.
\end{equation}
In particular the spin DC conductivity \cite{SM} reads
$\sigma(T,h) = \mathfrak{D} (T,h)\chi_{{h}}(T,h)= \kappa(T)|h|^{-1} + \mathcal{O}(h^{0})$,
with $\kappa(T) \sim T^{-1/2}$ at small $T$. Then one can check that $\kappa(T)>0$ for any $T$, see \cite{SM}, implying that spin transport in the NLSM at half-filling $h=0$
and $T>0$ is \textit{super-diffusive}.
For half-integer gapless spin chains we can repeat the same logic for the NLSM with the topological angle $\Theta=\pi$, and
once again find a diverging spin conductivity. This leads us to conclude that the presence or absence of the spectral
gap plays no essential role for this observed super-diffusive spin dynamics in isotropic antiferromagnetic chains.

\paragraph*{Spin transport at intermediate and high temperatures.}{
Characterizing spin dynamics at intermediate and high temperatures in physical spin chains goes beyond a simple effective QFT description and thus
poses a more challenging task. Here we rely on tDMRG simulations. In Fig.~\ref{Fig:spin} we display the time-dependent
spin DC conductivity
$
\sigma(t) =\frac{1}{T}\int_{0}^{t}\!\!\dd t'    \big\langle \hat{J}(t')\hat{j}_{0}(0) \big\rangle_{T,h=0}
$
at half-filling and for various temperatures. The latter can be deduced from the growth rate of the spin current following
a quench from an initial bi-partitioned state with a tiny magnetization imbalance $\delta s^z$, namely
$\sigma(t) = \lim_{\delta s^z \to 0}  \langle \sum_{x} \hat{j}_{x}(t) \rangle_{T,\delta s^z}/\delta s^z$, which is
simpler from the numerics standpoint. While very low temperatures cannot be reached by this numerical technique, at higher
temperatures we find a clear signature of superdiffusion, characterized by time-dependent
conductivity $\sigma(t)\sim t^{1/3}$ at large times, see Fig.~\ref{Fig:spin} as well as \cite{SM}, both for the gapless spin-$1/2$
and the gapped (non-integrable) spin-$1$ XXX chain.
In our simulations, we have employed the finite-temperature time-dependent density matrix renormalization group
algorithm \cite{PhysRevB.90.155104,Karrasch2013}, using a fixed discarded weight and the maximum bond dimension of 4000 for spin $1/2$ 
and 2000 for spin $1$, with system size large compare to the causality light cone at the largest simulation time.
}

\paragraph*{\bf Comparison with previous results.}

To further elaborate on the physical implications of our findings, we now discuss our theoretical predictions in a broader context
and clarify the pitfalls of the previous approaches.

\paragraph*{Semi-classical approach.}

It is instructive to first shortly summarize the semi-classical approach to the low-$T$ quantum transport developed in 
refs.~\cite{Sachdev1997,Damle1998} (see also \cite{Sachdev19972,Rieger2011,1712.09466}). Using that in the regime $T,h\ll \mass$ the 
mean collision time (i.e. the inverse density) becomes exponentially large ($\sim T^{-1}\,e^{\mass/T}$), it 
has been argued that on large spatio-temporal scales (compared to inverse temperature $t\gg T^{-1}$ and
the thermal de Broglie wavelength $x\gg \lambda_{\rm T}$) the spin dynamics essentially becomes `universal' and can
be accurately described in terms of classical trajectories. By accordingly keeping only the zero-momentum part of the full quantum 
scattering matrix in the gapped $O(3)$ NLSM ($\Theta = 0$), \cite{Sachdev1997} predicts a large but \emph{finite} spin diffusion 
constant
$
\mathfrak{D}_{\rm cl} \sim e^{\mass/T}/3\mass,
$
valid in the regime $h\ll T\ll \mass$ which corresponds to the contribution of massive physical excitations corresponding
to $s=0$, see Eg.~\eqref{eq:sachdevus}. It is important to keep in mind however that the semi-classical scattering theory effectively 
interchanges the non-commuting $T\to 0$ and $t\to \infty$ limits and, as a consequence, it is blind to the coherent contributions
of the internal magnonic degrees of freedom (terms with $s>0$ in Eq.~\eqref{eq:D_sum}). It turns out that there are crucial to 
correctly determine the nature of spin transport in the regime $h/T\ll 1$.

Normal spin diffusion at finite temperatures is on the other hand restored upon adding interaction \textit{anisotropy}.
To clarify this aspect, we briefly consider the XXZ spin-$1/2$ chain with anisotropy $\Delta$, assuming $\Delta > 1$ 
where the quasi-particles pertain to compounds of $s$ bound magnons~\cite{SM}. In the low-temperature limit and small $h$, with 
$h/T\gg 1$ large, the bound-state contributions ($s>1$) are \textit{suppressed} and from Eq.~\eqref{eq:mainResDiffusion} we find (cf. \cite{SM})
$
\mathfrak{D}_{\rm XXZ} \simeq \mathfrak{A}\,{e^{\mathfrak{m}/T}}
$
where $\mathfrak{A}=\mathfrak{c}^{2}/(\mathfrak{n}\,\mass)$, $\mathfrak{n}=2$ is the number of low-energy degrees of freedom 
with the low-momentum dispersion law $\varepsilon_{1}(k) \approx \mass + (\mathfrak{c}\,k)^{2}/2\mass$, 
where $\mass$ denotes the spectral gap, with
$\mass = \tfrac{1}{2}\sinh{(\eta)}\times \sum_{k \in \mathbb{Z}} (-1)^{k}/\cosh{(k \eta)}$, 
$\eta =\cosh^{-1} \Delta$. The obtained result agrees with the semi-classical result of ref.~\cite{Damle1998}
and it provides the first direct confirmation of the semi-classical approximation in an \textit{anisotropic} chain.

\paragraph*{Dressed versus bare form factors.}

Form-factor expansions established themselves as a powerful theoretical tool for studying integrable QFTs \cite{PhysRevB.78.100403,9789810202446,0305-4470-34-13-102,CastroAlvaredo2002,0305-4470-34-36-319}. In the form-factor formalism
one traditionally operates with the trivial (bare) Fock vacuum as the reference state. In contrast, a more general expansion with 
respect to e.g. a thermal background is a more delicate and technical subject which has not been fully
developed yet \cite{Leclair1996,Saleur2000,Doyon2007}.
In context of low-temperature transport, many previous works \cite{Konik2003,Essler2005,Altshuler2006,essler2009finite}
thus employed a series expansion with respect to the bare vacuum, with the reasoning that the spectral gap
renders a summation over multi-particle excitations quickly convergent. Based on this, it has been further advocated
that the ground-state dynamical structure factor experiences a small thermal broadening at finite $T$,
which for $T\ll \mass$ matches the diffusive (Lorentzian) peak predicted by the semi-classical approach.
Strictly speaking, however, such a \emph{dilute} gas picture only adequately describes physics at zero temperature. The computation of equilibrium
correlation functions instead necessitates an expansion based on \textit{dressed} (instead of bare) form factors
of local densities, and these are given by matrix elements of particle-hole excitations on top of a finite-density thermal
background \cite{Doyon2005,Doyon_correlations,PhysRevLett.120.217206,1812.00767,Cubero2019}. Considering the longitudinal magnetization component $\hat{s}^{z}$, the matrix element between a thermal state 
$\ket{\hat{\varrho}_{T,h}}$ and an excited state with a single particle-hole excitation of `type $s$',
with momenta $\Delta k_{s} = k_{s}(\theta^{+}_{s}) -  k_{s}(\theta^{-}_{s})$, reads 
\begin{equation}\label{eq:dressedFF}
\bra{\hat{\varrho}_{T,h}}\hat{s}_x^{z} \ket{\hat{\varrho}_{T,h};\theta^{+}_{s},\theta^{-}_{s}}
= e^{i x \Delta k_s} m_{s}^{\rm dr}  + \mathcal{O}( \Delta k_{s}) .
\end{equation}
Here the quantity $m_{s}^{\rm dr}$ denotes the renormalised (dressed) value of magnetization of a quasi-particle of type $s$ immersed
in a finite-density thermal background, can be radically different from the bare value $m^{\rm bare}_{s}=s$.
This effect is particularly pronounced in the vicinity of half-filled thermal equilibria,
where the effective magnetization exhibits a crossover from paramagnetic $m^{\rm dr}_{s}\sim s^{2}\,h$ ($s\ll |h|^{-1}$) to bare 
$m^{\rm dr}_{s}\sim s$ ($s\gg |h|^{-1}$) regime. We note that the vanishing of the spin Drude weight as $h\to 0$ can be
seen as a consequence of the paramagnetic behaviour of the dressed form factors \eqref{eq:dressedFF}, which 
are key building blocks in the approach of \cite{1812.00767}.

Furthermore, we wish to point out that non-perturbative effects attributed to the quasi-particle dressing also
have a profound influence on the NMR spin relaxation rate $1/T_1$ \cite{Jolicur1994,Sagi1996,Dupont2016,Coira2016,Dupont2018}.
Motivated by the preceding studies, see e.g.~\cite{Sagi1996,Konik2003}, we here specialize to the experimentally
relevant regime $h\ll T\ll \mathfrak{m}$, disregarding for simplicity possible effects of the single-ion anisotropy or
inter-chain couplings. The \textit{zero-momentum }contribution to the low-temperature dependence of the intra-band relaxation rate $T^{-1}_{1}$ of the longitudinal spin 
component is expressible in therms of the dressed form factors \eqref{eq:dressedFF} as
$T^{-1}_{1} =  2|A^{\rm xz}|^2  \sum_s \int   \dd p_{s}(\theta) [1- n_{s}(\theta)] n_{s}(\theta) r_{s}(\theta)$,
where $A^{\rm xz}$ denotes the hyperfine couplings and
$r_{s} (\theta) = (m_s^{\rm dr})^{2}/(\sqrt{\varepsilon_{s}^{\prime \prime}(0)} \sqrt{ \varepsilon_{s}^{\prime \prime}(0) \theta^{2} + \omega_{N}})$ with the NMR frequency $\omega_{\rm N}=h$ (in units $\mu_{\rm N}=1$).
By taking the $h\to 0$ limit \emph{after} first
performing the summation over the entire quasi-particle spectrum $s>0$, we find
\begin{equation}\label{eq:NMR_scaling}
\frac{1}{T_{1}} \sim e^{-(3/2)\mass/T}|h|^{-1/2}.
\end{equation}
This scaling plays nicely with the experimental study on the $S=1$ compound \cite{Takigawa1996} and, somewhat surprisingly,
is in qualitative agreement with the semi-classical results $T^{-1}_{1}\sim T\chi_{h}|\mathfrak{D}_{\rm cl} h|^{-1/2}$ found
in \cite{Sachdev1997,Damle1998}.
The key difference however is that within our method the activation rate $(3/2)\mass/T$ comes from the contributions of
the internal magnonic degrees of freedom. In contrast, the previous calculation from \cite{Sagi1996} based on the free spinfull bosons 
and the bare form-factor expansion carried out in \cite{Konik2003} yields the incorrect behaviour $T^{-1}_{1}\sim e^{-\mass/T}\log{h}$.

\paragraph*{\bf KPZ universality.}

The unexpected divergent spin conductivity, observed in both the $SU(2)$ symmetric spin chains and the $O(3)$ NLSM,
is rooted in anomalous properties of thermally dressed quasi-particles which carry large bare spin $s$ (see also \cite{1812.02701}).
Recalling that the spin diffusion constant \eqref{eq:mainResDiffusion} is an infinite sum over individual
quasi-particle contributions, one can readily notice that for the isotropic inter-spin interactions the summand saturates
at large $s$, $\lim_{s \to \infty} \mathfrak{D}_{s} = \mathfrak{D}_\infty>0$, thus rendering the spin diffusion constant infinite.
Furthermore, thermal fluctuations of the local spin $\delta \langle \hat{s}_x^z\rangle=  \langle \hat{s}_x^z\rangle -\langle \hat{s}^z\rangle_{T,h}$ can be directly linked to fluctuations of `giant quasi-particles' via \cite{SM}
$\delta \langle \hat{s}^z\rangle = T\chi_{{h}}(T,h) \lim_{s \to \infty}[\delta n_s/(s\,n_s(n_s - 1))]$,
with $\delta n_{s}$ denoting local fluctuations of the Fermi occupation functions.
Saturation at a finite asymptotic value $\mathfrak{D}_\infty$ may be correspondingly be interpreted as a self-interacting term
in the dynamics of fluctuations $\delta \langle \hat{s}^{z}\rangle$, in analogy to the Burger's equation
$\partial_t \delta \langle \hat{s}_x^z(t) \rangle =\partial_x[  \mathfrak{D}_{\rm reg}  \partial_x  \delta   \langle \hat{s}_x^z(t) \rangle + \lambda (\delta \langle   \hat{s}_x^z(t)  \rangle)^2 
+ \ldots]$; here $\mathfrak{D}_{\rm reg}<\infty$ is the `regularised' diffusion constant which accounts for the
finite contributions of `light' quasi-particles and $\lambda=\lambda(\mathfrak{D}_{\infty})$ is the nonlinearity
(self-interaction) coefficient such that $\lim_{\mathfrak{D}_\infty \to 0}\lambda(\mathfrak{D}_{\infty})=0$.
This provides a phenomenological model which underlies the KPZ universality class with dynamical
exponent $z=3/2$ \cite{Spohn2014,1812.02701}, i.e. $\langle \hat{s}^z_x (t)\hat{s}^z_0 \rangle_{T,h=0} \sim t^{-1/z} $,  consistently with the observed divergent time-dependent conductivity
$\sigma(t) \sim t^{1/3}$, see Fig. \ref{Fig:spin} and in agreement to what observed in the integrable spin-$1/2$ Heisenberg chain \cite{Ljubotina2019}.

\paragraph*{\bf Conclusions.}

We have outlined a theoretical framework for studying low-temperature spin dynamics in gapped and gapless
one-dimensional isotropic antiferromagnets based on the effective low-energy quantum field theory.
In the vicinity of half-filling, we found a crossover from the semi-classical regime $h/T\gg 1$ to the strongly-correlated
regime $h/T\ll 1$. In the $h\to 0$ limit, we analytically established a divergent spin diffusion constant and conjectured
a super-diffusive behaviour with fluctuations in the KPZ universality class. The phenomenon is seen in both half-integer
and integer spin chain, which rules out the importance of the spectral gap.
Instead, the anomalous behaviour can be attributed to the effective self-interaction of thermally-dressed
interacting magnonic waves. Presently, we exclude the conventional interpretation based on mode-coupling theory within
the phenomenological framework of the classical non-linear fluctuating hydrodynamics \cite{Spohn2014,Popkov2015,das2018nonlinear}
due to the vanishing diagonal terms of the Hessian in the current derivative expansion.

Our findings have direct applications in inelastic neutron scattering spectroscopy and quantum transport experiments \cite{Takigawa1996,PhysRevB.50.9174,PhysRevB.62.8921,Mourigal2013,Hirobe2016}, while they also open new venues for further
theoretical research on the microscopic mechanisms which underlie the observed anomalous spin transport
in the isotropic antiferromagnetic chains. Perhaps the most striking observation is that the phenomenon remains present even
at high temperatures. While this could be a footprint of the low-lying sigma model physics, it may as well be due to
an emergent classical hydrodynamical description. For instance, the isotropic classical Landau-Lifshitz field theory
is also known to exhibit super-diffusive spin transport both in equilibrium \cite{Bojan} and far from equilibrium \cite{Gamayun2019}. 
We leave these exciting questions to future studies.

\paragraph*{\bf Acknowledgements.} We thank D. Bernard, B. Doyon, R. Konik, M. Kormos, T. Prosen and S. Sachdev for comments on the manuscript and useful related discussions.  J.D.N. is supported by the Research Foundation Flanders (FWO).
E.I. is supported by VENI grant number 680-47-454 by the Netherlands Organisation for Scientific Research (NWO).  C.K. acknowledges support by the Deutsche Forschungsgemeinschaft through the Emmy Noether program (KA 3360/2-1). 

\bibliography{LowT}

\begin{thebibliography}{91}%
\makeatletter
\providecommand \@ifxundefined [1]{%
 \@ifx{#1\undefined}
}%
\providecommand \@ifnum [1]{%
 \ifnum #1\expandafter \@firstoftwo
 \else \expandafter \@secondoftwo
 \fi
}%
\providecommand \@ifx [1]{%
 \ifx #1\expandafter \@firstoftwo
 \else \expandafter \@secondoftwo
 \fi
}%
\providecommand \natexlab [1]{#1}%
\providecommand \enquote  [1]{``#1''}%
\providecommand \bibnamefont  [1]{#1}%
\providecommand \bibfnamefont [1]{#1}%
\providecommand \citenamefont [1]{#1}%
\providecommand \href@noop [0]{\@secondoftwo}%
\providecommand \href [0]{\begingroup \@sanitize@url \@href}%
\providecommand \@href[1]{\@@startlink{#1}\@@href}%
\providecommand \@@href[1]{\endgroup#1\@@endlink}%
\providecommand \@sanitize@url [0]{\catcode `\\12\catcode `\$12\catcode
  `\&12\catcode `\#12\catcode `\^12\catcode `\_12\catcode `\%12\relax}%
\providecommand \@@startlink[1]{}%
\providecommand \@@endlink[0]{}%
\providecommand \url  [0]{\begingroup\@sanitize@url \@url }%
\providecommand \@url [1]{\endgroup\@href {#1}{\urlprefix }}%
\providecommand \urlprefix  [0]{URL }%
\providecommand \Eprint [0]{\href }%
\providecommand \doibase [0]{http://dx.doi.org/}%
\providecommand \selectlanguage [0]{\@gobble}%
\providecommand \bibinfo  [0]{\@secondoftwo}%
\providecommand \bibfield  [0]{\@secondoftwo}%
\providecommand \translation [1]{[#1]}%
\providecommand \BibitemOpen [0]{}%
\providecommand \bibitemStop [0]{}%
\providecommand \bibitemNoStop [0]{.\EOS\space}%
\providecommand \EOS [0]{\spacefactor3000\relax}%
\providecommand \BibitemShut  [1]{\csname bibitem#1\endcsname}%
\let\auto@bib@innerbib\@empty
\bibitem [{\citenamefont {Haldane}(1983{\natexlab{a}})}]{Haldane1983}%
  \BibitemOpen
  \bibfield  {author} {\bibinfo {author} {\bibfnamefont {F.~D.~M.}\
  \bibnamefont {Haldane}},\ }\href {\doibase 10.1103/physrevlett.50.1153}
  {\bibfield  {journal} {\bibinfo  {journal} {Physical Review Letters}\
  }\textbf {\bibinfo {volume} {50}},\ \bibinfo {pages} {1153} (\bibinfo {year}
  {1983}{\natexlab{a}})}\BibitemShut {NoStop}%
\bibitem [{\citenamefont {Haldane}(1983{\natexlab{b}})}]{haldane1983continuum}%
  \BibitemOpen
  \bibfield  {author} {\bibinfo {author} {\bibfnamefont {F.}~\bibnamefont
  {Haldane}},\ }\href {\doibase 10.1016/0375-9601(83)90631-x} {\bibfield
  {journal} {\bibinfo  {journal} {Physics Letters A}\ }\textbf {\bibinfo
  {volume} {93}},\ \bibinfo {pages} {464} (\bibinfo {year}
  {1983}{\natexlab{b}})}\BibitemShut {NoStop}%
\bibitem [{\citenamefont {Shankar}\ and\ \citenamefont
  {Read}(1990)}]{Shankar1990}%
  \BibitemOpen
  \bibfield  {author} {\bibinfo {author} {\bibfnamefont {R.}~\bibnamefont
  {Shankar}}\ and\ \bibinfo {author} {\bibfnamefont {N.}~\bibnamefont {Read}},\
  }\href {\doibase 10.1016/0550-3213(90)90437-i} {\bibfield  {journal}
  {\bibinfo  {journal} {Nuclear Physics B}\ }\textbf {\bibinfo {volume}
  {336}},\ \bibinfo {pages} {457} (\bibinfo {year} {1990})}\BibitemShut
  {NoStop}%
\bibitem [{\citenamefont {Konik}(2003)}]{Konik2003}%
  \BibitemOpen
  \bibfield  {author} {\bibinfo {author} {\bibfnamefont {R.~M.}\ \bibnamefont
  {Konik}},\ }\href {\doibase 10.1103/physrevb.68.104435} {\bibfield  {journal}
  {\bibinfo  {journal} {Physical Review B}\ }\textbf {\bibinfo {volume} {68}}
  (\bibinfo {year} {2003}),\ 10.1103/physrevb.68.104435}\BibitemShut {NoStop}%
\bibitem [{\citenamefont {Altshuler}\ \emph {et~al.}(2006)\citenamefont
  {Altshuler}, \citenamefont {Konik},\ and\ \citenamefont
  {Tsvelik}}]{Altshuler2006}%
  \BibitemOpen
  \bibfield  {author} {\bibinfo {author} {\bibfnamefont {B.}~\bibnamefont
  {Altshuler}}, \bibinfo {author} {\bibfnamefont {R.}~\bibnamefont {Konik}}, \
  and\ \bibinfo {author} {\bibfnamefont {A.}~\bibnamefont {Tsvelik}},\ }\href
  {\doibase 10.1016/j.nuclphysb.2006.01.022} {\bibfield  {journal} {\bibinfo
  {journal} {Nuclear Physics B}\ }\textbf {\bibinfo {volume} {739}},\ \bibinfo
  {pages} {311} (\bibinfo {year} {2006})}\BibitemShut {NoStop}%
\bibitem [{\citenamefont {Sachdev}\ and\ \citenamefont
  {Damle}(1997)}]{Sachdev1997}%
  \BibitemOpen
  \bibfield  {author} {\bibinfo {author} {\bibfnamefont {S.}~\bibnamefont
  {Sachdev}}\ and\ \bibinfo {author} {\bibfnamefont {K.}~\bibnamefont
  {Damle}},\ }\href {\doibase 10.1103/physrevlett.78.943} {\bibfield  {journal}
  {\bibinfo  {journal} {Physical Review Letters}\ }\textbf {\bibinfo {volume}
  {78}},\ \bibinfo {pages} {943} (\bibinfo {year} {1997})}\BibitemShut
  {NoStop}%
\bibitem [{\citenamefont {Damle}\ and\ \citenamefont
  {Sachdev}(1998)}]{Damle1998}%
  \BibitemOpen
  \bibfield  {author} {\bibinfo {author} {\bibfnamefont {K.}~\bibnamefont
  {Damle}}\ and\ \bibinfo {author} {\bibfnamefont {S.}~\bibnamefont
  {Sachdev}},\ }\href {\doibase 10.1103/physrevb.57.8307} {\bibfield  {journal}
  {\bibinfo  {journal} {Physical Review B}\ }\textbf {\bibinfo {volume} {57}},\
  \bibinfo {pages} {8307} (\bibinfo {year} {1998})}\BibitemShut {NoStop}%
\bibitem [{\citenamefont {Sachdev}\ and\ \citenamefont
  {Damle}(2000)}]{Sachdev2000}%
  \BibitemOpen
  \bibfield  {author} {\bibinfo {author} {\bibfnamefont {S.}~\bibnamefont
  {Sachdev}}\ and\ \bibinfo {author} {\bibfnamefont {K.}~\bibnamefont
  {Damle}},\ }\href {\doibase 10.1143/jpsj.69.2712} {\bibfield  {journal}
  {\bibinfo  {journal} {Journal of the Physical Society of Japan}\ }\textbf
  {\bibinfo {volume} {69}},\ \bibinfo {pages} {2712} (\bibinfo {year}
  {2000})}\BibitemShut {NoStop}%
\bibitem [{\citenamefont {Cuccoli}\ \emph {et~al.}(2000)\citenamefont
  {Cuccoli}, \citenamefont {Tognetti}, \citenamefont {Verrucchi},\ and\
  \citenamefont {Vaia}}]{Cuccoli2000}%
  \BibitemOpen
  \bibfield  {author} {\bibinfo {author} {\bibfnamefont {A.}~\bibnamefont
  {Cuccoli}}, \bibinfo {author} {\bibfnamefont {V.}~\bibnamefont {Tognetti}},
  \bibinfo {author} {\bibfnamefont {P.}~\bibnamefont {Verrucchi}}, \ and\
  \bibinfo {author} {\bibfnamefont {R.}~\bibnamefont {Vaia}},\ }\href {\doibase
  10.1103/physrevb.62.57} {\bibfield  {journal} {\bibinfo  {journal} {Physical
  Review B}\ }\textbf {\bibinfo {volume} {62}},\ \bibinfo {pages} {57}
  (\bibinfo {year} {2000})}\BibitemShut {NoStop}%
\bibitem [{\citenamefont {Damle}\ and\ \citenamefont
  {Sachdev}(2005)}]{Damle2005}%
  \BibitemOpen
  \bibfield  {author} {\bibinfo {author} {\bibfnamefont {K.}~\bibnamefont
  {Damle}}\ and\ \bibinfo {author} {\bibfnamefont {S.}~\bibnamefont
  {Sachdev}},\ }\href {\doibase 10.1103/physrevlett.95.187201} {\bibfield
  {journal} {\bibinfo  {journal} {Physical Review Letters}\ }\textbf {\bibinfo
  {volume} {95}} (\bibinfo {year} {2005}),\
  10.1103/physrevlett.95.187201}\BibitemShut {NoStop}%
\bibitem [{\citenamefont {Castro-Alvaredo}\ \emph {et~al.}(2016)\citenamefont
  {Castro-Alvaredo}, \citenamefont {Doyon},\ and\ \citenamefont
  {Yoshimura}}]{PhysRevX.6.041065}%
  \BibitemOpen
  \bibfield  {author} {\bibinfo {author} {\bibfnamefont {O.~A.}\ \bibnamefont
  {Castro-Alvaredo}}, \bibinfo {author} {\bibfnamefont {B.}~\bibnamefont
  {Doyon}}, \ and\ \bibinfo {author} {\bibfnamefont {T.}~\bibnamefont
  {Yoshimura}},\ }\href {\doibase 10.1103/PhysRevX.6.041065} {\bibfield
  {journal} {\bibinfo  {journal} {Phys. Rev. X}\ }\textbf {\bibinfo {volume}
  {6}},\ \bibinfo {pages} {041065} (\bibinfo {year} {2016})}\BibitemShut
  {NoStop}%
\bibitem [{\citenamefont {Bertini}\ \emph {et~al.}(2016)\citenamefont
  {Bertini}, \citenamefont {Collura}, \citenamefont {{De Nardis}},\ and\
  \citenamefont {Fagotti}}]{PhysRevLett.117.207201}%
  \BibitemOpen
  \bibfield  {author} {\bibinfo {author} {\bibfnamefont {B.}~\bibnamefont
  {Bertini}}, \bibinfo {author} {\bibfnamefont {M.}~\bibnamefont {Collura}},
  \bibinfo {author} {\bibfnamefont {J.}~\bibnamefont {{De Nardis}}}, \ and\
  \bibinfo {author} {\bibfnamefont {M.}~\bibnamefont {Fagotti}},\ }\href
  {\doibase 10.1103/PhysRevLett.117.207201} {\bibfield  {journal} {\bibinfo
  {journal} {Phys. Rev. Lett.}\ }\textbf {\bibinfo {volume} {117}},\ \bibinfo
  {pages} {207201} (\bibinfo {year} {2016})}\BibitemShut {NoStop}%
\bibitem [{\citenamefont {Bulchandani}\ \emph {et~al.}(2018)\citenamefont
  {Bulchandani}, \citenamefont {Vasseur}, \citenamefont {Karrasch},\ and\
  \citenamefont {Moore}}]{Bulchandani2018}%
  \BibitemOpen
  \bibfield  {author} {\bibinfo {author} {\bibfnamefont {V.~B.}\ \bibnamefont
  {Bulchandani}}, \bibinfo {author} {\bibfnamefont {R.}~\bibnamefont
  {Vasseur}}, \bibinfo {author} {\bibfnamefont {C.}~\bibnamefont {Karrasch}}, \
  and\ \bibinfo {author} {\bibfnamefont {J.~E.}\ \bibnamefont {Moore}},\ }\href
  {\doibase 10.1103/physrevb.97.045407} {\bibfield  {journal} {\bibinfo
  {journal} {Physical Review B}\ }\textbf {\bibinfo {volume} {97}} (\bibinfo
  {year} {2018}),\ 10.1103/physrevb.97.045407}\BibitemShut {NoStop}%
\bibitem [{\citenamefont {Bulchandani}\ \emph {et~al.}(2017)\citenamefont
  {Bulchandani}, \citenamefont {Vasseur}, \citenamefont {Karrasch},\ and\
  \citenamefont {Moore}}]{Bulchandani2017}%
  \BibitemOpen
  \bibfield  {author} {\bibinfo {author} {\bibfnamefont {V.~B.}\ \bibnamefont
  {Bulchandani}}, \bibinfo {author} {\bibfnamefont {R.}~\bibnamefont
  {Vasseur}}, \bibinfo {author} {\bibfnamefont {C.}~\bibnamefont {Karrasch}}, \
  and\ \bibinfo {author} {\bibfnamefont {J.~E.}\ \bibnamefont {Moore}},\ }\href
  {\doibase 10.1103/physrevlett.119.220604} {\bibfield  {journal} {\bibinfo
  {journal} {Physical Review Letters}\ }\textbf {\bibinfo {volume} {119}}
  (\bibinfo {year} {2017}),\ 10.1103/physrevlett.119.220604}\BibitemShut
  {NoStop}%
\bibitem [{\citenamefont {Doyon}\ \emph
  {et~al.}(2018{\natexlab{a}})\citenamefont {Doyon}, \citenamefont
  {Yoshimura},\ and\ \citenamefont {Caux}}]{PhysRevLett.120.045301}%
  \BibitemOpen
  \bibfield  {author} {\bibinfo {author} {\bibfnamefont {B.}~\bibnamefont
  {Doyon}}, \bibinfo {author} {\bibfnamefont {T.}~\bibnamefont {Yoshimura}}, \
  and\ \bibinfo {author} {\bibfnamefont {J.-S.}\ \bibnamefont {Caux}},\ }\href
  {\doibase 10.1103/PhysRevLett.120.045301} {\bibfield  {journal} {\bibinfo
  {journal} {Phys. Rev. Lett.}\ }\textbf {\bibinfo {volume} {120}},\ \bibinfo
  {pages} {045301} (\bibinfo {year} {2018}{\natexlab{a}})}\BibitemShut
  {NoStop}%
\bibitem [{\citenamefont {Schemmer}\ \emph {et~al.}(2019)\citenamefont
  {Schemmer}, \citenamefont {Bouchoule}, \citenamefont {Doyon},\ and\
  \citenamefont {Dubail}}]{Schemmer2019}%
  \BibitemOpen
  \bibfield  {author} {\bibinfo {author} {\bibfnamefont {M.}~\bibnamefont
  {Schemmer}}, \bibinfo {author} {\bibfnamefont {I.}~\bibnamefont {Bouchoule}},
  \bibinfo {author} {\bibfnamefont {B.}~\bibnamefont {Doyon}}, \ and\ \bibinfo
  {author} {\bibfnamefont {J.}~\bibnamefont {Dubail}},\ }\href {\doibase
  10.1103/physrevlett.122.090601} {\bibfield  {journal} {\bibinfo  {journal}
  {Physical Review Letters}\ }\textbf {\bibinfo {volume} {122}} (\bibinfo
  {year} {2019}),\ 10.1103/physrevlett.122.090601}\BibitemShut {NoStop}%
\bibitem [{\citenamefont {Doyon}\ \emph {et~al.}(2017)\citenamefont {Doyon},
  \citenamefont {Dubail}, \citenamefont {Konik},\ and\ \citenamefont
  {Yoshimura}}]{PhysRevLett.119.195301}%
  \BibitemOpen
  \bibfield  {author} {\bibinfo {author} {\bibfnamefont {B.}~\bibnamefont
  {Doyon}}, \bibinfo {author} {\bibfnamefont {J.}~\bibnamefont {Dubail}},
  \bibinfo {author} {\bibfnamefont {R.}~\bibnamefont {Konik}}, \ and\ \bibinfo
  {author} {\bibfnamefont {T.}~\bibnamefont {Yoshimura}},\ }\href {\doibase
  10.1103/PhysRevLett.119.195301} {\bibfield  {journal} {\bibinfo  {journal}
  {Phys. Rev. Lett.}\ }\textbf {\bibinfo {volume} {119}},\ \bibinfo {pages}
  {195301} (\bibinfo {year} {2017})}\BibitemShut {NoStop}%
\bibitem [{\citenamefont {Alba}(2019)}]{Alba2019}%
  \BibitemOpen
  \bibfield  {author} {\bibinfo {author} {\bibfnamefont {V.}~\bibnamefont
  {Alba}},\ }\href {\doibase 10.1103/physrevb.99.045150} {\bibfield  {journal}
  {\bibinfo  {journal} {Physical Review B}\ }\textbf {\bibinfo {volume} {99}}
  (\bibinfo {year} {2019}),\ 10.1103/physrevb.99.045150}\BibitemShut {NoStop}%
\bibitem [{\citenamefont {Doyon}\ and\ \citenamefont
  {Yoshimura}(2017)}]{SciPostPhys.2.2.014}%
  \BibitemOpen
  \bibfield  {author} {\bibinfo {author} {\bibfnamefont {B.}~\bibnamefont
  {Doyon}}\ and\ \bibinfo {author} {\bibfnamefont {T.}~\bibnamefont
  {Yoshimura}},\ }\href {\doibase 10.21468/SciPostPhys.2.2.014} {\bibfield
  {journal} {\bibinfo  {journal} {SciPost Phys.}\ }\textbf {\bibinfo {volume}
  {2}},\ \bibinfo {pages} {014} (\bibinfo {year} {2017})}\BibitemShut {NoStop}%
\bibitem [{\citenamefont {Piroli}\ \emph {et~al.}(2017)\citenamefont {Piroli},
  \citenamefont {{De Nardis}}, \citenamefont {Collura}, \citenamefont
  {Bertini},\ and\ \citenamefont {Fagotti}}]{PhysRevB.96.115124}%
  \BibitemOpen
  \bibfield  {author} {\bibinfo {author} {\bibfnamefont {L.}~\bibnamefont
  {Piroli}}, \bibinfo {author} {\bibfnamefont {J.}~\bibnamefont {{De Nardis}}},
  \bibinfo {author} {\bibfnamefont {M.}~\bibnamefont {Collura}}, \bibinfo
  {author} {\bibfnamefont {B.}~\bibnamefont {Bertini}}, \ and\ \bibinfo
  {author} {\bibfnamefont {M.}~\bibnamefont {Fagotti}},\ }\href {\doibase
  10.1103/PhysRevB.96.115124} {\bibfield  {journal} {\bibinfo  {journal} {Phys.
  Rev. B}\ }\textbf {\bibinfo {volume} {96}},\ \bibinfo {pages} {115124}
  (\bibinfo {year} {2017})}\BibitemShut {NoStop}%
\bibitem [{\citenamefont {Mesty{\'{a}}n}\ \emph {et~al.}(2019)\citenamefont
  {Mesty{\'{a}}n}, \citenamefont {Bertini}, \citenamefont {Piroli},\ and\
  \citenamefont {Calabrese}}]{Mestyan2019}%
  \BibitemOpen
  \bibfield  {author} {\bibinfo {author} {\bibfnamefont {M.}~\bibnamefont
  {Mesty{\'{a}}n}}, \bibinfo {author} {\bibfnamefont {B.}~\bibnamefont
  {Bertini}}, \bibinfo {author} {\bibfnamefont {L.}~\bibnamefont {Piroli}}, \
  and\ \bibinfo {author} {\bibfnamefont {P.}~\bibnamefont {Calabrese}},\ }\href
  {\doibase 10.1103/physrevb.99.014305} {\bibfield  {journal} {\bibinfo
  {journal} {Physical Review B}\ }\textbf {\bibinfo {volume} {99}} (\bibinfo
  {year} {2019}),\ 10.1103/physrevb.99.014305}\BibitemShut {NoStop}%
\bibitem [{\citenamefont {Mazza}\ \emph {et~al.}(2018)\citenamefont {Mazza},
  \citenamefont {Viti}, \citenamefont {Carrega}, \citenamefont {Rossini},\ and\
  \citenamefont {Luca}}]{Mazza2018}%
  \BibitemOpen
  \bibfield  {author} {\bibinfo {author} {\bibfnamefont {L.}~\bibnamefont
  {Mazza}}, \bibinfo {author} {\bibfnamefont {J.}~\bibnamefont {Viti}},
  \bibinfo {author} {\bibfnamefont {M.}~\bibnamefont {Carrega}}, \bibinfo
  {author} {\bibfnamefont {D.}~\bibnamefont {Rossini}}, \ and\ \bibinfo
  {author} {\bibfnamefont {A.~D.}\ \bibnamefont {Luca}},\ }\href {\doibase
  10.1103/physrevb.98.075421} {\bibfield  {journal} {\bibinfo  {journal}
  {Physical Review B}\ }\textbf {\bibinfo {volume} {98}} (\bibinfo {year}
  {2018}),\ 10.1103/physrevb.98.075421}\BibitemShut {NoStop}%
\bibitem [{\citenamefont {Doyon}\ \emph
  {et~al.}(2018{\natexlab{b}})\citenamefont {Doyon}, \citenamefont {Spohn},\
  and\ \citenamefont {Yoshimura}}]{Doyon2018}%
  \BibitemOpen
  \bibfield  {author} {\bibinfo {author} {\bibfnamefont {B.}~\bibnamefont
  {Doyon}}, \bibinfo {author} {\bibfnamefont {H.}~\bibnamefont {Spohn}}, \ and\
  \bibinfo {author} {\bibfnamefont {T.}~\bibnamefont {Yoshimura}},\ }\href
  {\doibase 10.1016/j.nuclphysb.2017.12.002} {\bibfield  {journal} {\bibinfo
  {journal} {Nucl. Phys. B}\ }\textbf {\bibinfo {volume} {926}},\ \bibinfo
  {pages} {570} (\bibinfo {year} {2018}{\natexlab{b}})}\BibitemShut {NoStop}%
\bibitem [{\citenamefont {De~Luca}\ \emph {et~al.}(2017)\citenamefont
  {De~Luca}, \citenamefont {Collura},\ and\ \citenamefont {{De
  Nardis}}}]{PhysRevB.96.020403}%
  \BibitemOpen
  \bibfield  {author} {\bibinfo {author} {\bibfnamefont {A.}~\bibnamefont
  {De~Luca}}, \bibinfo {author} {\bibfnamefont {M.}~\bibnamefont {Collura}}, \
  and\ \bibinfo {author} {\bibfnamefont {J.}~\bibnamefont {{De Nardis}}},\
  }\href {\doibase 10.1103/PhysRevB.96.020403} {\bibfield  {journal} {\bibinfo
  {journal} {Phys. Rev. B}\ }\textbf {\bibinfo {volume} {96}},\ \bibinfo
  {pages} {020403} (\bibinfo {year} {2017})}\BibitemShut {NoStop}%
\bibitem [{\citenamefont {Bastianello}\ and\ \citenamefont
  {Luca}(2019)}]{Bastianello2019}%
  \BibitemOpen
  \bibfield  {author} {\bibinfo {author} {\bibfnamefont {A.}~\bibnamefont
  {Bastianello}}\ and\ \bibinfo {author} {\bibfnamefont {A.~D.}\ \bibnamefont
  {Luca}},\ }\href {\doibase 10.1103/physrevlett.122.240606} {\bibfield
  {journal} {\bibinfo  {journal} {Physical Review Letters}\ }\textbf {\bibinfo
  {volume} {122}} (\bibinfo {year} {2019}),\
  10.1103/physrevlett.122.240606}\BibitemShut {NoStop}%
\bibitem [{\citenamefont {Zotos}(1999)}]{PhysRevLett.82.1764}%
  \BibitemOpen
  \bibfield  {author} {\bibinfo {author} {\bibfnamefont {X.}~\bibnamefont
  {Zotos}},\ }\href {\doibase 10.1103/PhysRevLett.82.1764} {\bibfield
  {journal} {\bibinfo  {journal} {Phys. Rev. Lett.}\ }\textbf {\bibinfo
  {volume} {82}},\ \bibinfo {pages} {1764} (\bibinfo {year}
  {1999})}\BibitemShut {NoStop}%
\bibitem [{\citenamefont {Ilievski}\ and\ \citenamefont {{De
  Nardis}}(2017{\natexlab{a}})}]{IN_Drude}%
  \BibitemOpen
  \bibfield  {author} {\bibinfo {author} {\bibfnamefont {E.}~\bibnamefont
  {Ilievski}}\ and\ \bibinfo {author} {\bibfnamefont {J.}~\bibnamefont {{De
  Nardis}}},\ }\href {\doibase 10.1103/physrevlett.119.020602} {\bibfield
  {journal} {\bibinfo  {journal} {Physical Review Letters}\ }\textbf {\bibinfo
  {volume} {119}} (\bibinfo {year} {2017}{\natexlab{a}}),\
  10.1103/physrevlett.119.020602}\BibitemShut {NoStop}%
\bibitem [{\citenamefont {Doyon}\ and\ \citenamefont
  {Spohn}(2017)}]{SciPostPhys.3.6.039}%
  \BibitemOpen
  \bibfield  {author} {\bibinfo {author} {\bibfnamefont {B.}~\bibnamefont
  {Doyon}}\ and\ \bibinfo {author} {\bibfnamefont {H.}~\bibnamefont {Spohn}},\
  }\href {\doibase 10.21468/SciPostPhys.3.6.039} {\bibfield  {journal}
  {\bibinfo  {journal} {SciPost Phys.}\ }\textbf {\bibinfo {volume} {3}},\
  \bibinfo {pages} {039} (\bibinfo {year} {2017})}\BibitemShut {NoStop}%
\bibitem [{\citenamefont {Ilievski}\ and\ \citenamefont {{De
  Nardis}}(2017{\natexlab{b}})}]{IN_Hubbard}%
  \BibitemOpen
  \bibfield  {author} {\bibinfo {author} {\bibfnamefont {E.}~\bibnamefont
  {Ilievski}}\ and\ \bibinfo {author} {\bibfnamefont {J.}~\bibnamefont {{De
  Nardis}}},\ }\href {\doibase 10.1103/physrevb.96.081118} {\bibfield
  {journal} {\bibinfo  {journal} {Physical Review B}\ }\textbf {\bibinfo
  {volume} {96}} (\bibinfo {year} {2017}{\natexlab{b}}),\
  10.1103/physrevb.96.081118}\BibitemShut {NoStop}%
\bibitem [{\citenamefont {Urichuk}\ \emph {et~al.}(2019)\citenamefont
  {Urichuk}, \citenamefont {\"{O}z}, \citenamefont {Kl\"{u}mper},\ and\
  \citenamefont {Sirker}}]{KlumperDrude}%
  \BibitemOpen
  \bibfield  {author} {\bibinfo {author} {\bibfnamefont {A.}~\bibnamefont
  {Urichuk}}, \bibinfo {author} {\bibfnamefont {Y.}~\bibnamefont {\"{O}z}},
  \bibinfo {author} {\bibfnamefont {A.}~\bibnamefont {Kl\"{u}mper}}, \ and\
  \bibinfo {author} {\bibfnamefont {J.}~\bibnamefont {Sirker}},\ }\href
  {\doibase 10.21468/SciPostPhys.6.1.005} {\bibfield  {journal} {\bibinfo
  {journal} {SciPost Phys.}\ }\textbf {\bibinfo {volume} {6}},\ \bibinfo
  {pages} {5} (\bibinfo {year} {2019})}\BibitemShut {NoStop}%
\bibitem [{\citenamefont {{De Nardis}}\ \emph {et~al.}(2018)\citenamefont {{De
  Nardis}}, \citenamefont {Bernard},\ and\ \citenamefont
  {Doyon}}]{DeNardis2018}%
  \BibitemOpen
  \bibfield  {author} {\bibinfo {author} {\bibfnamefont {J.}~\bibnamefont {{De
  Nardis}}}, \bibinfo {author} {\bibfnamefont {D.}~\bibnamefont {Bernard}}, \
  and\ \bibinfo {author} {\bibfnamefont {B.}~\bibnamefont {Doyon}},\ }\href
  {\doibase 10.1103/physrevlett.121.160603} {\bibfield  {journal} {\bibinfo
  {journal} {Physical Review Letters}\ }\textbf {\bibinfo {volume} {121}}
  (\bibinfo {year} {2018}),\ 10.1103/physrevlett.121.160603}\BibitemShut
  {NoStop}%
\bibitem [{\citenamefont {Gopalakrishnan}\ \emph {et~al.}(2018)\citenamefont
  {Gopalakrishnan}, \citenamefont {Huse}, \citenamefont {Khemani},\ and\
  \citenamefont {Vasseur}}]{Gopalakrishnan2018}%
  \BibitemOpen
  \bibfield  {author} {\bibinfo {author} {\bibfnamefont {S.}~\bibnamefont
  {Gopalakrishnan}}, \bibinfo {author} {\bibfnamefont {D.~A.}\ \bibnamefont
  {Huse}}, \bibinfo {author} {\bibfnamefont {V.}~\bibnamefont {Khemani}}, \
  and\ \bibinfo {author} {\bibfnamefont {R.}~\bibnamefont {Vasseur}},\ }\href
  {\doibase 10.1103/physrevb.98.220303} {\bibfield  {journal} {\bibinfo
  {journal} {Physical Review B}\ }\textbf {\bibinfo {volume} {98}} (\bibinfo
  {year} {2018}),\ 10.1103/physrevb.98.220303}\BibitemShut {NoStop}%
\bibitem [{\citenamefont {Nardis}\ \emph {et~al.}(2019)\citenamefont {Nardis},
  \citenamefont {Bernard},\ and\ \citenamefont {Doyon}}]{1812.00767}%
  \BibitemOpen
  \bibfield  {author} {\bibinfo {author} {\bibfnamefont {J.~D.}\ \bibnamefont
  {Nardis}}, \bibinfo {author} {\bibfnamefont {D.}~\bibnamefont {Bernard}}, \
  and\ \bibinfo {author} {\bibfnamefont {B.}~\bibnamefont {Doyon}},\ }\href
  {\doibase 10.21468/SciPostPhys.6.4.049} {\bibfield  {journal} {\bibinfo
  {journal} {SciPost Phys.}\ }\textbf {\bibinfo {volume} {6}},\ \bibinfo
  {pages} {49} (\bibinfo {year} {2019})}\BibitemShut {NoStop}%
\bibitem [{\citenamefont {Gopalakrishnan}\ and\ \citenamefont
  {Vasseur}(2019)}]{1812.02701}%
  \BibitemOpen
  \bibfield  {author} {\bibinfo {author} {\bibfnamefont {S.}~\bibnamefont
  {Gopalakrishnan}}\ and\ \bibinfo {author} {\bibfnamefont {R.}~\bibnamefont
  {Vasseur}},\ }\href {\doibase 10.1103/PhysRevLett.122.127202} {\bibfield
  {journal} {\bibinfo  {journal} {Phys. Rev. Lett.}\ }\textbf {\bibinfo
  {volume} {122}},\ \bibinfo {pages} {127202} (\bibinfo {year}
  {2019})}\BibitemShut {NoStop}%
\bibitem [{\citenamefont {Fujimoto}(1999)}]{Fujimoto1999}%
  \BibitemOpen
  \bibfield  {author} {\bibinfo {author} {\bibfnamefont {S.}~\bibnamefont
  {Fujimoto}},\ }\href {\doibase 10.1143/jpsj.68.2810} {\bibfield  {journal}
  {\bibinfo  {journal} {Journal of the Physical Society of Japan}\ }\textbf
  {\bibinfo {volume} {68}},\ \bibinfo {pages} {2810} (\bibinfo {year}
  {1999})}\BibitemShut {NoStop}%
\bibitem [{\citenamefont {Essler}\ and\ \citenamefont
  {Konik}(2009)}]{essler2009finite}%
  \BibitemOpen
  \bibfield  {author} {\bibinfo {author} {\bibfnamefont {F.~H.~L.}\
  \bibnamefont {Essler}}\ and\ \bibinfo {author} {\bibfnamefont {R.~M.}\
  \bibnamefont {Konik}},\ }\href {\doibase 10.1088/1742-5468/2009/09/p09018}
  {\bibfield  {journal} {\bibinfo  {journal} {Journal of Statistical Mechanics:
  Theory and Experiment}\ }\textbf {\bibinfo {volume} {2009}},\ \bibinfo
  {pages} {P09018} (\bibinfo {year} {2009})}\BibitemShut {NoStop}%
\bibitem [{\citenamefont
  {{\v{Z}}nidari{\v{c}}}(2011{\natexlab{a}})}]{znidarivc2011spin}%
  \BibitemOpen
  \bibfield  {author} {\bibinfo {author} {\bibfnamefont {M.}~\bibnamefont
  {{\v{Z}}nidari{\v{c}}}},\ }\href {\doibase 10.1103/physrevlett.106.220601}
  {\bibfield  {journal} {\bibinfo  {journal} {Physical Review Letters}\
  }\textbf {\bibinfo {volume} {106}} (\bibinfo {year} {2011}{\natexlab{a}}),\
  10.1103/physrevlett.106.220601}\BibitemShut {NoStop}%
\bibitem [{\citenamefont
  {{\v{Z}}nidari{\v{c}}}(2011{\natexlab{b}})}]{Znidaric2011}%
  \BibitemOpen
  \bibfield  {author} {\bibinfo {author} {\bibfnamefont {M.}~\bibnamefont
  {{\v{Z}}nidari{\v{c}}}},\ }\href {\doibase 10.1088/1742-5468/2011/12/p12008}
  {\bibfield  {journal} {\bibinfo  {journal} {Journal of Statistical Mechanics:
  Theory and Experiment}\ }\textbf {\bibinfo {volume} {2011}},\ \bibinfo
  {pages} {P12008} (\bibinfo {year} {2011}{\natexlab{b}})}\BibitemShut
  {NoStop}%
\bibitem [{\citenamefont {Ilievski}\ \emph {et~al.}(2018)\citenamefont
  {Ilievski}, \citenamefont {Nardis}, \citenamefont {Medenjak},\ and\
  \citenamefont {Prosen}}]{ilievski2018superdiffusion}%
  \BibitemOpen
  \bibfield  {author} {\bibinfo {author} {\bibfnamefont {E.}~\bibnamefont
  {Ilievski}}, \bibinfo {author} {\bibfnamefont {J.~D.}\ \bibnamefont
  {Nardis}}, \bibinfo {author} {\bibfnamefont {M.}~\bibnamefont {Medenjak}}, \
  and\ \bibinfo {author} {\bibfnamefont {T.}~\bibnamefont {Prosen}},\ }\href
  {\doibase 10.1103/physrevlett.121.230602} {\bibfield  {journal} {\bibinfo
  {journal} {Physical Review Letters}\ }\textbf {\bibinfo {volume} {121}}
  (\bibinfo {year} {2018}),\ 10.1103/physrevlett.121.230602}\BibitemShut
  {NoStop}%
\bibitem [{\citenamefont {Ljubotina}\ \emph {et~al.}(2019)\citenamefont
  {Ljubotina}, \citenamefont {{\v{Z}}nidari{\v{c}}},\ and\ \citenamefont
  {Prosen}}]{Ljubotina2019}%
  \BibitemOpen
  \bibfield  {author} {\bibinfo {author} {\bibfnamefont {M.}~\bibnamefont
  {Ljubotina}}, \bibinfo {author} {\bibfnamefont {M.}~\bibnamefont
  {{\v{Z}}nidari{\v{c}}}}, \ and\ \bibinfo {author} {\bibfnamefont
  {T.}~\bibnamefont {Prosen}},\ }\href {\doibase
  10.1103/physrevlett.122.210602} {\bibfield  {journal} {\bibinfo  {journal}
  {Physical Review Letters}\ }\textbf {\bibinfo {volume} {122}} (\bibinfo
  {year} {2019}),\ 10.1103/physrevlett.122.210602}\BibitemShut {NoStop}%
\bibitem [{\citenamefont {Kardar}\ \emph {et~al.}(1986)\citenamefont {Kardar},
  \citenamefont {Parisi},\ and\ \citenamefont {Zhang}}]{PhysRevLett.56.889}%
  \BibitemOpen
  \bibfield  {author} {\bibinfo {author} {\bibfnamefont {M.}~\bibnamefont
  {Kardar}}, \bibinfo {author} {\bibfnamefont {G.}~\bibnamefont {Parisi}}, \
  and\ \bibinfo {author} {\bibfnamefont {Y.-C.}\ \bibnamefont {Zhang}},\ }\href
  {\doibase 10.1103/PhysRevLett.56.889} {\bibfield  {journal} {\bibinfo
  {journal} {Phys. Rev. Lett.}\ }\textbf {\bibinfo {volume} {56}},\ \bibinfo
  {pages} {889} (\bibinfo {year} {1986})}\BibitemShut {NoStop}%
\bibitem [{\citenamefont {Corwin}(2012)}]{CORWIN2012}%
  \BibitemOpen
  \bibfield  {author} {\bibinfo {author} {\bibfnamefont {I.}~\bibnamefont
  {Corwin}},\ }\href {\doibase 10.1142/s2010326311300014} {\bibfield  {journal}
  {\bibinfo  {journal} {Random Matrices: Theory and Applications}\ }\textbf
  {\bibinfo {volume} {01}},\ \bibinfo {pages} {1130001} (\bibinfo {year}
  {2012})}\BibitemShut {NoStop}%
\bibitem [{\citenamefont {Takeuchi}(2018)}]{Takeuchi2018}%
  \BibitemOpen
  \bibfield  {author} {\bibinfo {author} {\bibfnamefont {K.~A.}\ \bibnamefont
  {Takeuchi}},\ }\href {\doibase 10.1016/j.physa.2018.03.009} {\bibfield
  {journal} {\bibinfo  {journal} {Physica A: Statistical Mechanics and its
  Applications}\ }\textbf {\bibinfo {volume} {504}},\ \bibinfo {pages} {77}
  (\bibinfo {year} {2018})}\BibitemShut {NoStop}%
\bibitem [{\citenamefont {Kubo}(1957)}]{Kubo57}%
  \BibitemOpen
  \bibfield  {author} {\bibinfo {author} {\bibfnamefont {R.}~\bibnamefont
  {Kubo}},\ }\href {\doibase 10.1143/jpsj.12.570} {\bibfield  {journal}
  {\bibinfo  {journal} {Journal of the Physical Society of Japan}\ }\textbf
  {\bibinfo {volume} {12}},\ \bibinfo {pages} {570} (\bibinfo {year}
  {1957})}\BibitemShut {NoStop}%
\bibitem [{\citenamefont {{\v{Z}}nidari{\v{c}}}(2019)}]{Znidari2019}%
  \BibitemOpen
  \bibfield  {author} {\bibinfo {author} {\bibfnamefont {M.}~\bibnamefont
  {{\v{Z}}nidari{\v{c}}}},\ }\href {\doibase 10.1103/physrevb.99.035143}
  {\bibfield  {journal} {\bibinfo  {journal} {Physical Review B}\ }\textbf
  {\bibinfo {volume} {99}} (\bibinfo {year} {2019}),\
  10.1103/physrevb.99.035143}\BibitemShut {NoStop}%
\bibitem [{\citenamefont {Ilievski}\ \emph {et~al.}(2016)\citenamefont
  {Ilievski}, \citenamefont {Medenjak}, \citenamefont {Prosen},\ and\
  \citenamefont {Zadnik}}]{ilievski2016quasilocal}%
  \BibitemOpen
  \bibfield  {author} {\bibinfo {author} {\bibfnamefont {E.}~\bibnamefont
  {Ilievski}}, \bibinfo {author} {\bibfnamefont {M.}~\bibnamefont {Medenjak}},
  \bibinfo {author} {\bibfnamefont {T.}~\bibnamefont {Prosen}}, \ and\ \bibinfo
  {author} {\bibfnamefont {L.}~\bibnamefont {Zadnik}},\ }\href {\doibase
  10.1088/1742-5468/2016/06/064008} {\bibfield  {journal} {\bibinfo  {journal}
  {Journal of Statistical Mechanics: Theory and Experiment}\ }\textbf {\bibinfo
  {volume} {2016}},\ \bibinfo {pages} {064008} (\bibinfo {year}
  {2016})}\BibitemShut {NoStop}%
\bibitem [{SM()}]{SM}%
  \BibitemOpen
  \href@noop {} {}\bibinfo {note} {Supplemental Material associated with this
  manuscript.}\BibitemShut {Stop}%
\bibitem [{\citenamefont {Blanter}\ and\ \citenamefont
  {B\"{u}ttiker}(2000)}]{Blanter2000}%
  \BibitemOpen
  \bibfield  {author} {\bibinfo {author} {\bibfnamefont {Y.}~\bibnamefont
  {Blanter}}\ and\ \bibinfo {author} {\bibfnamefont {M.}~\bibnamefont
  {B\"{u}ttiker}},\ }\href {\doibase 10.1016/s0370-1573(99)00123-4} {\bibfield
  {journal} {\bibinfo  {journal} {Physics Reports}\ }\textbf {\bibinfo {volume}
  {336}},\ \bibinfo {pages} {1} (\bibinfo {year} {2000})}\BibitemShut {NoStop}%
\bibitem [{\citenamefont {Medenjak}\ \emph {et~al.}(2017)\citenamefont
  {Medenjak}, \citenamefont {Karrasch},\ and\ \citenamefont {Prosen}}]{MKP17}%
  \BibitemOpen
  \bibfield  {author} {\bibinfo {author} {\bibfnamefont {M.}~\bibnamefont
  {Medenjak}}, \bibinfo {author} {\bibfnamefont {C.}~\bibnamefont {Karrasch}},
  \ and\ \bibinfo {author} {\bibfnamefont {T.}~\bibnamefont {Prosen}},\ }\href
  {\doibase 10.1103/physrevlett.119.080602} {\bibfield  {journal} {\bibinfo
  {journal} {Physical Review Letters}\ }\textbf {\bibinfo {volume} {119}}
  (\bibinfo {year} {2017}),\ 10.1103/physrevlett.119.080602}\BibitemShut
  {NoStop}%
\bibitem [{\citenamefont {Affleck}\ and\ \citenamefont
  {Haldane}(1987)}]{Affleck1987}%
  \BibitemOpen
  \bibfield  {author} {\bibinfo {author} {\bibfnamefont {I.}~\bibnamefont
  {Affleck}}\ and\ \bibinfo {author} {\bibfnamefont {F.~D.~M.}\ \bibnamefont
  {Haldane}},\ }\href {\doibase 10.1103/physrevb.36.5291} {\bibfield  {journal}
  {\bibinfo  {journal} {Physical Review B}\ }\textbf {\bibinfo {volume} {36}},\
  \bibinfo {pages} {5291} (\bibinfo {year} {1987})}\BibitemShut {NoStop}%
\bibitem [{\citenamefont {Zamolodchikov}\ and\ \citenamefont
  {Zamolodchikov}(1979)}]{ZZ1979}%
  \BibitemOpen
  \bibfield  {author} {\bibinfo {author} {\bibfnamefont {A.~B.}\ \bibnamefont
  {Zamolodchikov}}\ and\ \bibinfo {author} {\bibfnamefont {A.~B.}\ \bibnamefont
  {Zamolodchikov}},\ }\href {\doibase 10.1016/0003-4916(79)90391-9} {\bibfield
  {journal} {\bibinfo  {journal} {Annals of Physics}\ }\textbf {\bibinfo
  {volume} {120}},\ \bibinfo {pages} {253} (\bibinfo {year}
  {1979})}\BibitemShut {NoStop}%
\bibitem [{\citenamefont {Zamolodchikov}\ and\ \citenamefont
  {Zamolodchikov}(1992)}]{ZZ1992}%
  \BibitemOpen
  \bibfield  {author} {\bibinfo {author} {\bibfnamefont {A.}~\bibnamefont
  {Zamolodchikov}}\ and\ \bibinfo {author} {\bibfnamefont {A.}~\bibnamefont
  {Zamolodchikov}},\ }\href {\doibase 10.1016/0550-3213(92)90136-y} {\bibfield
  {journal} {\bibinfo  {journal} {Nuclear Physics B}\ }\textbf {\bibinfo
  {volume} {379}},\ \bibinfo {pages} {602} (\bibinfo {year}
  {1992})}\BibitemShut {NoStop}%
\bibitem [{\citenamefont {Karrasch}\ \emph {et~al.}(2014)\citenamefont
  {Karrasch}, \citenamefont {Kennes},\ and\ \citenamefont
  {Moore}}]{PhysRevB.90.155104}%
  \BibitemOpen
  \bibfield  {author} {\bibinfo {author} {\bibfnamefont {C.}~\bibnamefont
  {Karrasch}}, \bibinfo {author} {\bibfnamefont {D.~M.}\ \bibnamefont
  {Kennes}}, \ and\ \bibinfo {author} {\bibfnamefont {J.~E.}\ \bibnamefont
  {Moore}},\ }\href {\doibase 10.1103/PhysRevB.90.155104} {\bibfield  {journal}
  {\bibinfo  {journal} {Phys. Rev. B}\ }\textbf {\bibinfo {volume} {90}},\
  \bibinfo {pages} {155104} (\bibinfo {year} {2014})}\BibitemShut {NoStop}%
\bibitem [{\citenamefont {Karrasch}\ \emph {et~al.}(2013)\citenamefont
  {Karrasch}, \citenamefont {Bardarson},\ and\ \citenamefont
  {Moore}}]{Karrasch2013}%
  \BibitemOpen
  \bibfield  {author} {\bibinfo {author} {\bibfnamefont {C.}~\bibnamefont
  {Karrasch}}, \bibinfo {author} {\bibfnamefont {J.~H.}\ \bibnamefont
  {Bardarson}}, \ and\ \bibinfo {author} {\bibfnamefont {J.~E.}\ \bibnamefont
  {Moore}},\ }\href {\doibase 10.1088/1367-2630/15/8/083031} {\bibfield
  {journal} {\bibinfo  {journal} {New Journal of Physics}\ }\textbf {\bibinfo
  {volume} {15}},\ \bibinfo {pages} {083031} (\bibinfo {year}
  {2013})}\BibitemShut {NoStop}%
\bibitem [{\citenamefont {Sachdev}\ and\ \citenamefont
  {Young}(1997)}]{Sachdev19972}%
  \BibitemOpen
  \bibfield  {author} {\bibinfo {author} {\bibfnamefont {S.}~\bibnamefont
  {Sachdev}}\ and\ \bibinfo {author} {\bibfnamefont {A.~P.}\ \bibnamefont
  {Young}},\ }\href {\doibase 10.1103/physrevlett.78.2220} {\bibfield
  {journal} {\bibinfo  {journal} {Physical Review Letters}\ }\textbf {\bibinfo
  {volume} {78}},\ \bibinfo {pages} {2220} (\bibinfo {year}
  {1997})}\BibitemShut {NoStop}%
\bibitem [{\citenamefont {Rieger}\ and\ \citenamefont
  {Igl{\'{o}}i}(2011)}]{Rieger2011}%
  \BibitemOpen
  \bibfield  {author} {\bibinfo {author} {\bibfnamefont {H.}~\bibnamefont
  {Rieger}}\ and\ \bibinfo {author} {\bibfnamefont {F.}~\bibnamefont
  {Igl{\'{o}}i}},\ }\href {\doibase 10.1103/physrevb.84.165117} {\bibfield
  {journal} {\bibinfo  {journal} {Physical Review B}\ }\textbf {\bibinfo
  {volume} {84}} (\bibinfo {year} {2011}),\
  10.1103/physrevb.84.165117}\BibitemShut {NoStop}%
\bibitem [{\citenamefont {Moca}\ \emph {et~al.}(2017)\citenamefont {Moca},
  \citenamefont {Kormos},\ and\ \citenamefont {Zar{\'{a}}nd}}]{1712.09466}%
  \BibitemOpen
  \bibfield  {author} {\bibinfo {author} {\bibfnamefont {C.~P.}\ \bibnamefont
  {Moca}}, \bibinfo {author} {\bibfnamefont {M.}~\bibnamefont {Kormos}}, \ and\
  \bibinfo {author} {\bibfnamefont {G.}~\bibnamefont {Zar{\'{a}}nd}},\ }\href
  {\doibase 10.1103/physrevlett.119.100603} {\bibfield  {journal} {\bibinfo
  {journal} {Phys. Rev. Lett.}\ }\textbf {\bibinfo {volume} {119}} (\bibinfo
  {year} {2017}),\ 10.1103/physrevlett.119.100603}\BibitemShut {NoStop}%
\bibitem [{\citenamefont {Essler}\ and\ \citenamefont
  {Konik}(2008)}]{PhysRevB.78.100403}%
  \BibitemOpen
  \bibfield  {author} {\bibinfo {author} {\bibfnamefont {F.~H.~L.}\
  \bibnamefont {Essler}}\ and\ \bibinfo {author} {\bibfnamefont {R.~M.}\
  \bibnamefont {Konik}},\ }\href {\doibase 10.1103/PhysRevB.78.100403}
  {\bibfield  {journal} {\bibinfo  {journal} {Phys. Rev. B}\ }\textbf {\bibinfo
  {volume} {78}},\ \bibinfo {pages} {100403} (\bibinfo {year}
  {2008})}\BibitemShut {NoStop}%
\bibitem [{\citenamefont {Smirnov}(1992)}]{9789810202446}%
  \BibitemOpen
  \bibfield  {author} {\bibinfo {author} {\bibfnamefont {F.~A.}\ \bibnamefont
  {Smirnov}},\ }\href
  {https://www.amazon.com/Factors-Completely-Integrable-Advanced-Mathematical/dp/981020244X?SubscriptionId=0JYN1NVW651KCA56C102&tag=techkie-20&linkCode=xm2&camp=2025&creative=165953&creativeASIN=981020244X}
  {\emph {\bibinfo {title} {Form Factors in Completely Integrable Models of
  Quantum Field Theory (Advanced Series in Mathematical Physics)}}}\ (\bibinfo
  {publisher} {World Scientific Pub Co Inc},\ \bibinfo {year}
  {1992})\BibitemShut {NoStop}%
\bibitem [{\citenamefont {Delfino}(2001)}]{0305-4470-34-13-102}%
  \BibitemOpen
  \bibfield  {author} {\bibinfo {author} {\bibfnamefont {G.}~\bibnamefont
  {Delfino}},\ }\href {http://stacks.iop.org/0305-4470/34/i=13/a=102}
  {\bibfield  {journal} {\bibinfo  {journal} {J. Phys. A}\ }\textbf {\bibinfo
  {volume} {34}},\ \bibinfo {pages} {L161} (\bibinfo {year}
  {2001})}\BibitemShut {NoStop}%
\bibitem [{\citenamefont {Castro-Alvaredo}\ and\ \citenamefont
  {Fring}(2002)}]{CastroAlvaredo2002}%
  \BibitemOpen
  \bibfield  {author} {\bibinfo {author} {\bibfnamefont {O.}~\bibnamefont
  {Castro-Alvaredo}}\ and\ \bibinfo {author} {\bibfnamefont {A.}~\bibnamefont
  {Fring}},\ }\href {\doibase 10.1016/s0550-3213(02)00409-1} {\bibfield
  {journal} {\bibinfo  {journal} {Nuclear Physics B}\ }\textbf {\bibinfo
  {volume} {636}},\ \bibinfo {pages} {611} (\bibinfo {year}
  {2002})}\BibitemShut {NoStop}%
\bibitem [{\citenamefont {Mussardo}(2001)}]{0305-4470-34-36-319}%
  \BibitemOpen
  \bibfield  {author} {\bibinfo {author} {\bibfnamefont {G.}~\bibnamefont
  {Mussardo}},\ }\href {http://stacks.iop.org/0305-4470/34/i=36/a=319}
  {\bibfield  {journal} {\bibinfo  {journal} {J. Phys. A}\ }\textbf {\bibinfo
  {volume} {34}},\ \bibinfo {pages} {7399} (\bibinfo {year}
  {2001})}\BibitemShut {NoStop}%
\bibitem [{\citenamefont {Leclair}\ \emph {et~al.}(1996)\citenamefont
  {Leclair}, \citenamefont {Lesage}, \citenamefont {Sachdev},\ and\
  \citenamefont {Saleur}}]{Leclair1996}%
  \BibitemOpen
  \bibfield  {author} {\bibinfo {author} {\bibfnamefont {A.}~\bibnamefont
  {Leclair}}, \bibinfo {author} {\bibfnamefont {F.}~\bibnamefont {Lesage}},
  \bibinfo {author} {\bibfnamefont {S.}~\bibnamefont {Sachdev}}, \ and\
  \bibinfo {author} {\bibfnamefont {H.}~\bibnamefont {Saleur}},\ }\href
  {\doibase 10.1016/s0550-3213(96)00456-7} {\bibfield  {journal} {\bibinfo
  {journal} {Nuclear Physics B}\ }\textbf {\bibinfo {volume} {482}},\ \bibinfo
  {pages} {579} (\bibinfo {year} {1996})}\BibitemShut {NoStop}%
\bibitem [{\citenamefont {Saleur}(2000)}]{Saleur2000}%
  \BibitemOpen
  \bibfield  {author} {\bibinfo {author} {\bibfnamefont {H.}~\bibnamefont
  {Saleur}},\ }\href {\doibase 10.1016/s0550-3213(99)00665-3} {\bibfield
  {journal} {\bibinfo  {journal} {Nuclear Physics B}\ }\textbf {\bibinfo
  {volume} {567}},\ \bibinfo {pages} {602} (\bibinfo {year}
  {2000})}\BibitemShut {NoStop}%
\bibitem [{\citenamefont {Doyon}(2007)}]{Doyon2007}%
  \BibitemOpen
  \bibfield  {author} {\bibinfo {author} {\bibfnamefont {B.}~\bibnamefont
  {Doyon}},\ }\href {\doibase 10.3842/sigma.2007.011} {\bibfield  {journal}
  {\bibinfo  {journal} {SIGMA}\ } (\bibinfo {year} {2007}),\
  10.3842/sigma.2007.011}\BibitemShut {NoStop}%
\bibitem [{\citenamefont {{Fabian H. L. Essler and Robert M.
  Konik}}(2005)}]{Essler2005}%
  \BibitemOpen
  \bibfield  {author} {\bibinfo {author} {\bibnamefont {{Fabian H. L. Essler
  and Robert M. Konik}}},\ }in\ \href {\doibase 10.1142/9789812775344_0020}
  {\emph {\bibinfo {booktitle} {From Fields to Strings: Circumnavigating
  Theoretical Physics}}}\ (\bibinfo  {publisher} {{WORLD} {SCIENTIFIC}},\
  \bibinfo {year} {2005})\ pp.\ \bibinfo {pages} {684--830}\BibitemShut
  {NoStop}%
\bibitem [{\citenamefont {Doyon}(2005)}]{Doyon2005}%
  \BibitemOpen
  \bibfield  {author} {\bibinfo {author} {\bibfnamefont {B.}~\bibnamefont
  {Doyon}},\ }\href {\doibase 10.1088/1742-5468/2005/11/p11006} {\bibfield
  {journal} {\bibinfo  {journal} {J. Stat. Mech. Theory Exp.}\ }\textbf
  {\bibinfo {volume} {2005}},\ \bibinfo {pages} {P11006} (\bibinfo {year}
  {2005})}\BibitemShut {NoStop}%
\bibitem [{\citenamefont {Doyon}(2018)}]{Doyon_correlations}%
  \BibitemOpen
  \bibfield  {author} {\bibinfo {author} {\bibfnamefont {B.}~\bibnamefont
  {Doyon}},\ }\href {\doibase 10.21468/scipostphys.5.5.054} {\bibfield
  {journal} {\bibinfo  {journal} {{SciPost} Physics}\ }\textbf {\bibinfo
  {volume} {5}} (\bibinfo {year} {2018}),\
  10.21468/scipostphys.5.5.054}\BibitemShut {NoStop}%
\bibitem [{\citenamefont {De~Nardis}\ and\ \citenamefont
  {Panfil}(2018)}]{PhysRevLett.120.217206}%
  \BibitemOpen
  \bibfield  {author} {\bibinfo {author} {\bibfnamefont {J.}~\bibnamefont
  {De~Nardis}}\ and\ \bibinfo {author} {\bibfnamefont {M.}~\bibnamefont
  {Panfil}},\ }\href {\doibase 10.1103/PhysRevLett.120.217206} {\bibfield
  {journal} {\bibinfo  {journal} {Phys. Rev. Lett.}\ }\textbf {\bibinfo
  {volume} {120}},\ \bibinfo {pages} {217206} (\bibinfo {year}
  {2018})}\BibitemShut {NoStop}%
\bibitem [{\citenamefont {Cubero}\ and\ \citenamefont
  {Panfil}(2019)}]{Cubero2019}%
  \BibitemOpen
  \bibfield  {author} {\bibinfo {author} {\bibfnamefont {A.~C.}\ \bibnamefont
  {Cubero}}\ and\ \bibinfo {author} {\bibfnamefont {M.}~\bibnamefont
  {Panfil}},\ }\href {\doibase 10.1007/jhep01(2019)104} {\bibfield  {journal}
  {\bibinfo  {journal} {Journal of High Energy Physics}\ }\textbf {\bibinfo
  {volume} {2019}} (\bibinfo {year} {2019}),\
  10.1007/jhep01(2019)104}\BibitemShut {NoStop}%
\bibitem [{\citenamefont {Jolicur}\ and\ \citenamefont
  {Golinelli}(1994)}]{Jolicur1994}%
  \BibitemOpen
  \bibfield  {author} {\bibinfo {author} {\bibfnamefont {T.}~\bibnamefont
  {Jolicur}}\ and\ \bibinfo {author} {\bibfnamefont {O.}~\bibnamefont
  {Golinelli}},\ }\href {\doibase 10.1103/physrevb.50.9265} {\bibfield
  {journal} {\bibinfo  {journal} {Physical Review B}\ }\textbf {\bibinfo
  {volume} {50}},\ \bibinfo {pages} {9265} (\bibinfo {year}
  {1994})}\BibitemShut {NoStop}%
\bibitem [{\citenamefont {Sagi}\ and\ \citenamefont
  {Affleck}(1996)}]{Sagi1996}%
  \BibitemOpen
  \bibfield  {author} {\bibinfo {author} {\bibfnamefont {J.}~\bibnamefont
  {Sagi}}\ and\ \bibinfo {author} {\bibfnamefont {I.}~\bibnamefont {Affleck}},\
  }\href {\doibase 10.1103/physrevb.53.9188} {\bibfield  {journal} {\bibinfo
  {journal} {Physical Review B}\ }\textbf {\bibinfo {volume} {53}},\ \bibinfo
  {pages} {9188} (\bibinfo {year} {1996})}\BibitemShut {NoStop}%
\bibitem [{\citenamefont {Dupont}\ \emph {et~al.}(2016)\citenamefont {Dupont},
  \citenamefont {Capponi},\ and\ \citenamefont {Laflorencie}}]{Dupont2016}%
  \BibitemOpen
  \bibfield  {author} {\bibinfo {author} {\bibfnamefont {M.}~\bibnamefont
  {Dupont}}, \bibinfo {author} {\bibfnamefont {S.}~\bibnamefont {Capponi}}, \
  and\ \bibinfo {author} {\bibfnamefont {N.}~\bibnamefont {Laflorencie}},\
  }\href {\doibase 10.1103/PhysRevB.94.144409} {\bibfield  {journal} {\bibinfo
  {journal} {Phys. Rev. B}\ }\textbf {\bibinfo {volume} {94}},\ \bibinfo
  {pages} {144409} (\bibinfo {year} {2016})}\BibitemShut {NoStop}%
\bibitem [{\citenamefont {Coira}\ \emph {et~al.}(2016)\citenamefont {Coira},
  \citenamefont {Barmettler}, \citenamefont {Giamarchi},\ and\ \citenamefont
  {Kollath}}]{Coira2016}%
  \BibitemOpen
  \bibfield  {author} {\bibinfo {author} {\bibfnamefont {E.}~\bibnamefont
  {Coira}}, \bibinfo {author} {\bibfnamefont {P.}~\bibnamefont {Barmettler}},
  \bibinfo {author} {\bibfnamefont {T.}~\bibnamefont {Giamarchi}}, \ and\
  \bibinfo {author} {\bibfnamefont {C.}~\bibnamefont {Kollath}},\ }\href
  {\doibase 10.1103/PhysRevB.94.144408} {\bibfield  {journal} {\bibinfo
  {journal} {Phys. Rev. B}\ }\textbf {\bibinfo {volume} {94}},\ \bibinfo
  {pages} {144408} (\bibinfo {year} {2016})}\BibitemShut {NoStop}%
\bibitem [{\citenamefont {Dupont}\ \emph {et~al.}(2018)\citenamefont {Dupont},
  \citenamefont {Capponi}, \citenamefont {Laflorencie},\ and\ \citenamefont
  {Orignac}}]{Dupont2018}%
  \BibitemOpen
  \bibfield  {author} {\bibinfo {author} {\bibfnamefont {M.}~\bibnamefont
  {Dupont}}, \bibinfo {author} {\bibfnamefont {S.}~\bibnamefont {Capponi}},
  \bibinfo {author} {\bibfnamefont {N.}~\bibnamefont {Laflorencie}}, \ and\
  \bibinfo {author} {\bibfnamefont {E.}~\bibnamefont {Orignac}},\ }\href
  {\doibase 10.1103/physrevb.98.094403} {\bibfield  {journal} {\bibinfo
  {journal} {Physical Review B}\ }\textbf {\bibinfo {volume} {98}} (\bibinfo
  {year} {2018}),\ 10.1103/physrevb.98.094403}\BibitemShut {NoStop}%
\bibitem [{\citenamefont {Takigawa}\ \emph {et~al.}(1996)\citenamefont
  {Takigawa}, \citenamefont {Asano}, \citenamefont {Ajiro}, \citenamefont
  {Mekata},\ and\ \citenamefont {Uemura}}]{Takigawa1996}%
  \BibitemOpen
  \bibfield  {author} {\bibinfo {author} {\bibfnamefont {M.}~\bibnamefont
  {Takigawa}}, \bibinfo {author} {\bibfnamefont {T.}~\bibnamefont {Asano}},
  \bibinfo {author} {\bibfnamefont {Y.}~\bibnamefont {Ajiro}}, \bibinfo
  {author} {\bibfnamefont {M.}~\bibnamefont {Mekata}}, \ and\ \bibinfo {author}
  {\bibfnamefont {Y.~J.}\ \bibnamefont {Uemura}},\ }\href {\doibase
  10.1103/physrevlett.76.2173} {\bibfield  {journal} {\bibinfo  {journal}
  {Physical Review Letters}\ }\textbf {\bibinfo {volume} {76}},\ \bibinfo
  {pages} {2173} (\bibinfo {year} {1996})}\BibitemShut {NoStop}%
\bibitem [{\citenamefont {Spohn}(2014)}]{Spohn2014}%
  \BibitemOpen
  \bibfield  {author} {\bibinfo {author} {\bibfnamefont {H.}~\bibnamefont
  {Spohn}},\ }\href {\doibase 10.1007/s10955-014-0933-y} {\bibfield  {journal}
  {\bibinfo  {journal} {Journal of Statistical Physics}\ }\textbf {\bibinfo
  {volume} {154}},\ \bibinfo {pages} {1191} (\bibinfo {year}
  {2014})}\BibitemShut {NoStop}%
\bibitem [{\citenamefont {Popkov}\ \emph {et~al.}(2015)\citenamefont {Popkov},
  \citenamefont {Schadschneider}, \citenamefont {Schmidt},\ and\ \citenamefont
  {Sch\"{u}tz}}]{Popkov2015}%
  \BibitemOpen
  \bibfield  {author} {\bibinfo {author} {\bibfnamefont {V.}~\bibnamefont
  {Popkov}}, \bibinfo {author} {\bibfnamefont {A.}~\bibnamefont
  {Schadschneider}}, \bibinfo {author} {\bibfnamefont {J.}~\bibnamefont
  {Schmidt}}, \ and\ \bibinfo {author} {\bibfnamefont {G.~M.}\ \bibnamefont
  {Sch\"{u}tz}},\ }\href {\doibase 10.1073/pnas.1512261112} {\bibfield
  {journal} {\bibinfo  {journal} {Proceedings of the National Academy of
  Sciences}\ }\textbf {\bibinfo {volume} {112}},\ \bibinfo {pages} {12645}
  (\bibinfo {year} {2015})}\BibitemShut {NoStop}%
\bibitem [{\citenamefont {Das}\ \emph {et~al.}(2018)\citenamefont {Das},
  \citenamefont {Damle}, \citenamefont {Dhar}, \citenamefont {Huse},
  \citenamefont {Kulkarni}, \citenamefont {Mendl},\ and\ \citenamefont
  {Spohn}}]{das2018nonlinear}%
  \BibitemOpen
  \bibfield  {author} {\bibinfo {author} {\bibfnamefont {A.}~\bibnamefont
  {Das}}, \bibinfo {author} {\bibfnamefont {K.}~\bibnamefont {Damle}}, \bibinfo
  {author} {\bibfnamefont {A.}~\bibnamefont {Dhar}}, \bibinfo {author}
  {\bibfnamefont {D.~A.}\ \bibnamefont {Huse}}, \bibinfo {author}
  {\bibfnamefont {M.}~\bibnamefont {Kulkarni}}, \bibinfo {author}
  {\bibfnamefont {C.~B.}\ \bibnamefont {Mendl}}, \ and\ \bibinfo {author}
  {\bibfnamefont {H.}~\bibnamefont {Spohn}},\ }\href@noop {} {\  (\bibinfo
  {year} {2018})},\ \Eprint {http://arxiv.org/abs/arXiv:1903.01329}
  {arXiv:1903.01329} \BibitemShut {NoStop}%
\bibitem [{\citenamefont {Regnault}\ \emph {et~al.}(1994)\citenamefont
  {Regnault}, \citenamefont {Zaliznyak}, \citenamefont {Renard},\ and\
  \citenamefont {Vettier}}]{PhysRevB.50.9174}%
  \BibitemOpen
  \bibfield  {author} {\bibinfo {author} {\bibfnamefont {L.~P.}\ \bibnamefont
  {Regnault}}, \bibinfo {author} {\bibfnamefont {I.}~\bibnamefont {Zaliznyak}},
  \bibinfo {author} {\bibfnamefont {J.~P.}\ \bibnamefont {Renard}}, \ and\
  \bibinfo {author} {\bibfnamefont {C.}~\bibnamefont {Vettier}},\ }\href
  {\doibase 10.1103/PhysRevB.50.9174} {\bibfield  {journal} {\bibinfo
  {journal} {Phys. Rev. B}\ }\textbf {\bibinfo {volume} {50}},\ \bibinfo
  {pages} {9174} (\bibinfo {year} {1994})}\BibitemShut {NoStop}%
\bibitem [{\citenamefont {Zheludev}\ \emph {et~al.}(2000)\citenamefont
  {Zheludev}, \citenamefont {Masuda}, \citenamefont {Tsukada}, \citenamefont
  {Uchiyama}, \citenamefont {Uchinokura}, \citenamefont {B\"oni},\ and\
  \citenamefont {Lee}}]{PhysRevB.62.8921}%
  \BibitemOpen
  \bibfield  {author} {\bibinfo {author} {\bibfnamefont {A.}~\bibnamefont
  {Zheludev}}, \bibinfo {author} {\bibfnamefont {T.}~\bibnamefont {Masuda}},
  \bibinfo {author} {\bibfnamefont {I.}~\bibnamefont {Tsukada}}, \bibinfo
  {author} {\bibfnamefont {Y.}~\bibnamefont {Uchiyama}}, \bibinfo {author}
  {\bibfnamefont {K.}~\bibnamefont {Uchinokura}}, \bibinfo {author}
  {\bibfnamefont {P.}~\bibnamefont {B\"oni}}, \ and\ \bibinfo {author}
  {\bibfnamefont {S.-H.}\ \bibnamefont {Lee}},\ }\href {\doibase
  10.1103/PhysRevB.62.8921} {\bibfield  {journal} {\bibinfo  {journal} {Phys.
  Rev. B}\ }\textbf {\bibinfo {volume} {62}},\ \bibinfo {pages} {8921}
  (\bibinfo {year} {2000})}\BibitemShut {NoStop}%
\bibitem [{\citenamefont {Mourigal}\ \emph {et~al.}(2013)\citenamefont
  {Mourigal}, \citenamefont {Enderle}, \citenamefont {Kl\"{o}pperpieper},
  \citenamefont {Caux}, \citenamefont {Stunault},\ and\ \citenamefont
  {R{\o}nnow}}]{Mourigal2013}%
  \BibitemOpen
  \bibfield  {author} {\bibinfo {author} {\bibfnamefont {M.}~\bibnamefont
  {Mourigal}}, \bibinfo {author} {\bibfnamefont {M.}~\bibnamefont {Enderle}},
  \bibinfo {author} {\bibfnamefont {A.}~\bibnamefont {Kl\"{o}pperpieper}},
  \bibinfo {author} {\bibfnamefont {J.-S.}\ \bibnamefont {Caux}}, \bibinfo
  {author} {\bibfnamefont {A.}~\bibnamefont {Stunault}}, \ and\ \bibinfo
  {author} {\bibfnamefont {H.~M.}\ \bibnamefont {R{\o}nnow}},\ }\href {\doibase
  10.1038/nphys2652} {\bibfield  {journal} {\bibinfo  {journal} {Nature
  Physics}\ }\textbf {\bibinfo {volume} {9}},\ \bibinfo {pages} {435} (\bibinfo
  {year} {2013})}\BibitemShut {NoStop}%
\bibitem [{\citenamefont {Hirobe}\ \emph {et~al.}(2016)\citenamefont {Hirobe},
  \citenamefont {Sato}, \citenamefont {Kawamata}, \citenamefont {Shiomi},
  \citenamefont {ichi Uchida}, \citenamefont {Iguchi}, \citenamefont {Koike},
  \citenamefont {Maekawa},\ and\ \citenamefont {Saitoh}}]{Hirobe2016}%
  \BibitemOpen
  \bibfield  {author} {\bibinfo {author} {\bibfnamefont {D.}~\bibnamefont
  {Hirobe}}, \bibinfo {author} {\bibfnamefont {M.}~\bibnamefont {Sato}},
  \bibinfo {author} {\bibfnamefont {T.}~\bibnamefont {Kawamata}}, \bibinfo
  {author} {\bibfnamefont {Y.}~\bibnamefont {Shiomi}}, \bibinfo {author}
  {\bibfnamefont {K.}~\bibnamefont {ichi Uchida}}, \bibinfo {author}
  {\bibfnamefont {R.}~\bibnamefont {Iguchi}}, \bibinfo {author} {\bibfnamefont
  {Y.}~\bibnamefont {Koike}}, \bibinfo {author} {\bibfnamefont
  {S.}~\bibnamefont {Maekawa}}, \ and\ \bibinfo {author} {\bibfnamefont
  {E.}~\bibnamefont {Saitoh}},\ }\href {\doibase 10.1038/nphys3895} {\bibfield
  {journal} {\bibinfo  {journal} {Nat. Phys.}\ }\textbf {\bibinfo {volume}
  {13}},\ \bibinfo {pages} {30} (\bibinfo {year} {2016})}\BibitemShut {NoStop}%
\bibitem [{\citenamefont {Prosen}\ and\ \citenamefont
  {{\v{Z}}unkovi{\v{c}}}(2013)}]{Bojan}%
  \BibitemOpen
  \bibfield  {author} {\bibinfo {author} {\bibfnamefont {T.}~\bibnamefont
  {Prosen}}\ and\ \bibinfo {author} {\bibfnamefont {B.}~\bibnamefont
  {{\v{Z}}unkovi{\v{c}}}},\ }\href {\doibase 10.1103/physrevlett.111.040602}
  {\bibfield  {journal} {\bibinfo  {journal} {Physical Review Letters}\
  }\textbf {\bibinfo {volume} {111}} (\bibinfo {year} {2013}),\
  10.1103/physrevlett.111.040602}\BibitemShut {NoStop}%
\bibitem [{\citenamefont {Gamayun}\ \emph {et~al.}(2019)\citenamefont
  {Gamayun}, \citenamefont {Miao},\ and\ \citenamefont
  {Ilievski}}]{Gamayun2019}%
  \BibitemOpen
  \bibfield  {author} {\bibinfo {author} {\bibfnamefont {O.}~\bibnamefont
  {Gamayun}}, \bibinfo {author} {\bibfnamefont {Y.}~\bibnamefont {Miao}}, \
  and\ \bibinfo {author} {\bibfnamefont {E.}~\bibnamefont {Ilievski}},\ }\href
  {\doibase 10.1103/physrevb.99.140301} {\bibfield  {journal} {\bibinfo
  {journal} {Physical Review B}\ }\textbf {\bibinfo {volume} {99}} (\bibinfo
  {year} {2019}),\ 10.1103/physrevb.99.140301}\BibitemShut {NoStop}%
\bibitem [{\citenamefont {Ilievski}\ and\ \citenamefont {{De
  Nardis}}(2017{\natexlab{c}})}]{PhysRevB.96.081118}%
  \BibitemOpen
  \bibfield  {author} {\bibinfo {author} {\bibfnamefont {E.}~\bibnamefont
  {Ilievski}}\ and\ \bibinfo {author} {\bibfnamefont {J.}~\bibnamefont {{De
  Nardis}}},\ }\href {\doibase 10.1103/PhysRevB.96.081118} {\bibfield
  {journal} {\bibinfo  {journal} {Phys. Rev. B}\ }\textbf {\bibinfo {volume}
  {96}},\ \bibinfo {pages} {081118} (\bibinfo {year}
  {2017}{\natexlab{c}})}\BibitemShut {NoStop}%
\bibitem [{Note1()}]{Note1}%
  \BibitemOpen
  \bibinfo {note} {Notice that in ref. \cite {DeNardis2018,1812.00767} and
  \cite {SciPostPhys.3.6.039} the notation used is $T = - K$ since the
  scattering matrix is taken to be the inverse of the one used here. Moreover
  in \cite {DeNardis2018,1812.00767} and \cite {SciPostPhys.3.6.039} the
  notation $\rho _{{\protect \rm p}, a}$, $\rho _{{\protect \rm s},a }$ is used
  for $\rho _a$, $\rho ^{\protect \rm tot}_a$ here.}\BibitemShut {Stop}%
\bibitem [{\citenamefont {Wiegmann}(1985)}]{Wiegmann1985}%
  \BibitemOpen
  \bibfield  {author} {\bibinfo {author} {\bibfnamefont {P.}~\bibnamefont
  {Wiegmann}},\ }\href {\doibase 10.1016/0370-2693(85)91171-2} {\bibfield
  {journal} {\bibinfo  {journal} {Physics Letters B}\ }\textbf {\bibinfo
  {volume} {152}},\ \bibinfo {pages} {209} (\bibinfo {year}
  {1985})}\BibitemShut {NoStop}%
\bibitem [{\citenamefont {Johnson}\ and\ \citenamefont
  {McCoy}(1972)}]{Johnson1972}%
  \BibitemOpen
  \bibfield  {author} {\bibinfo {author} {\bibfnamefont {J.~D.}\ \bibnamefont
  {Johnson}}\ and\ \bibinfo {author} {\bibfnamefont {B.~M.}\ \bibnamefont
  {McCoy}},\ }\href {\doibase 10.1103/physreva.6.1613} {\bibfield  {journal}
  {\bibinfo  {journal} {Physical Review A}\ }\textbf {\bibinfo {volume} {6}},\
  \bibinfo {pages} {1613} (\bibinfo {year} {1972})}\BibitemShut {NoStop}%
\bibitem [{\citenamefont {Bertini}\ and\ \citenamefont
  {Piroli}(2018)}]{Bertini2018}%
  \BibitemOpen
  \bibfield  {author} {\bibinfo {author} {\bibfnamefont {B.}~\bibnamefont
  {Bertini}}\ and\ \bibinfo {author} {\bibfnamefont {L.}~\bibnamefont
  {Piroli}},\ }\href {\doibase 10.1088/1742-5468/aab04b} {\bibfield  {journal}
  {\bibinfo  {journal} {Journal of Statistical Mechanics: Theory and
  Experiment}\ }\textbf {\bibinfo {volume} {2018}},\ \bibinfo {pages} {033104}
  (\bibinfo {year} {2018})}\BibitemShut {NoStop}%
\bibitem [{\citenamefont {L\"{a}uchli}\ \emph {et~al.}(2006)\citenamefont
  {L\"{a}uchli}, \citenamefont {Schmid},\ and\ \citenamefont
  {Trebst}}]{Luchli2006}%
  \BibitemOpen
  \bibfield  {author} {\bibinfo {author} {\bibfnamefont {A.}~\bibnamefont
  {L\"{a}uchli}}, \bibinfo {author} {\bibfnamefont {G.}~\bibnamefont {Schmid}},
  \ and\ \bibinfo {author} {\bibfnamefont {S.}~\bibnamefont {Trebst}},\ }\href
  {\doibase 10.1103/physrevb.74.144426} {\bibfield  {journal} {\bibinfo
  {journal} {Physical Review B}\ }\textbf {\bibinfo {volume} {74}} (\bibinfo
  {year} {2006}),\ 10.1103/physrevb.74.144426}\BibitemShut {NoStop}%
\end{thebibliography}%

\newpage
\onecolumngrid

\begin{center}
\textbf{{\Large Supplemental Material}}
\end{center}
\begin{center}
\textbf{\large{Anomalous spin transport in one-dimensional antiferromagnets}}
\end{center}

\tableofcontents

\section{Exact expressions for spin Drude weight and spin diffusion constant in integrable models}
Let $\hat{H}$ denote a Hamiltonian of an antiferromagnetic spin chain with the conserved total magnetization
$\hat{S}^{z} = \sum_x \hat{s}^z_x$ and local spin current $\hat{j}_x$. We shall be interested exclusively in the transport of spin (local magnetization) in
grand-canonical Gibbs equilibrium states at finite temperature $T$ and filling (external magnetic field) $h$,
specializing to low temperatures and the vicinity of \textit{half filling} $h \sim 0$.

We begin by introducing the relevant linear transport coefficients, namely
\begin{itemize}
\item the spin Drude weight,
\begin{equation}
\mathcal{D}(T,h) = \lim_{t\to \infty}\sum_{x}\expect{\hat{j}_{x}(t)j_{0}(0)}_{T,h},
\end{equation}
\item the Drude self-weight (zero-frequency noise),
\begin{equation}
\mathcal{D}^{\rm self} (T,h)= 2\int^{\infty}_{0}\dd t \expect{\hat{j}_{0}(t)\hat{j}_{0}(0)}_{T,h},
\end{equation}
\item the spin diffusion constant,
\begin{equation}
\mathfrak{D} (T,h)= \frac{1}{T\chi_{h}(T,h)}\int^{\infty}_{0}\dd t \left(\sum_{x}\expect{\hat{j}_{x}(t)\hat{j}_{0}(0)}_{T,h}-\mathcal{D}(T,h)\right).
\end{equation}
\end{itemize}
where $\chi_h$ is the spin susceptibility.
In integrable system, the above quantities can be \textit{exactly} expressed in terms of the following hydrodynamic mode resolutions
\begin{equation}
\mathcal{D}(T,h) = \sum_{s} \int d\theta\,\mathcal{D}_{s}(\theta),\qquad
\mathcal{D}^{\rm self}(T,h) = \sum_{s} \int d\theta\,\mathcal{D}^{\rm self}_{s}(\theta),\qquad
\mathfrak{D}(T,h)= \frac{1}{2} \sum_{s} \int d\theta\,\mathfrak{D}_{s}(\theta) +\mathcal{O} (h^{2}),
\end{equation}
with kernels
\begin{align}
\label{eq:spin_Drude}
\mathcal{D}_{s}(\theta) &= \chi_{s}(\theta)\Big(v^{\rm eff}_{s}(\theta)m^{\rm dr}_{s}\Big)^{2},\\
\label{eq:self_Drude}
\mathcal{D}^{\rm self}_{s}(\theta) &= \chi_{s}(\theta)|v^{\rm eff}_{s}(\theta)|(m^{\rm dr}_{s})^{2},\\
\label{eq:spin_diffusion}
\mathfrak{D}_s(\theta) &=  \chi_{s}(\theta)|v_{s}^{\rm eff}(\theta)|\left(\mathcal{W}_s(\theta)\right)^{2},
\end{align}
derived in refs. \cite{SciPostPhys.3.6.039,PhysRevB.96.081118}, \cite{SciPostPhys.3.6.039},
and \cite{DeNardis2018,1812.00767}, respectively.
In the above formulae, integer label $s$ enumerates distinct quasi-particle species in the spectrum with (bare) momenta
$k_{s}=k_{s}(\theta)$, static susceptibility $\chi_{s}(\theta)\equiv \rho_{s}(\theta)(1-n_{s}(\theta))$ , where
$\rho_{s}(\theta)$ are quasi-particle rapidity distributions in an equilibrium state characterised by
(Fermi) occupation functions $n_{s}(\theta)$. The physical meaning of $\mathcal{W}_{s}(\theta)$ will be explain in a moment.

The many-body scattering of quasi-particles is, due to integrability, fully encoded in a symmetric two-body scattering kernel
\begin{equation}
K_{s,s'}(\theta,\theta') =\frac{1}{2\pi \ii}\partial_{\theta}\log S_{s,s'}(\theta,\theta'),
\end{equation}
where $S_{s,s'}(\theta,\theta')$ are the amplitudes of the scattering matrix.
The group velocities of propagation are given by the effective dispersion relations
\begin{equation}
v^{\rm eff}_{s}(\theta)=\frac{\partial_{\theta}\varepsilon_{s}(\theta)}{\partial_{\theta}p_{s}(\theta)},
\end{equation}
where $\varepsilon_{s}(\theta)$ and $p_{s}(\theta)$ are dressed energies and momenta of interacting of quasi-particles
with respect to a finite density equilibrium state. By employing a compact vector notation (see a remark on notation in 
\footnote{Notice that in ref. \cite{DeNardis2018,1812.00767} and \cite{SciPostPhys.3.6.039} the notation used is $T = - K$ since the 
scattering matrix is taken to be the inverse of the one used here. Moreover in \cite{DeNardis2018,1812.00767}
and \cite{SciPostPhys.3.6.039} the notation $\rho_{{\rm p}, a}$, $\rho_{{\rm s},a }$ is used for $\rho_a$, $\rho^{\rm tot}_a$ here.}),
the dressed energies, momenta and spin are computed as
\begin{align}
\varepsilon' & = (1 + K n)^{-1} e' \\
p' & = (1 + K n)^{-1} k' \\
m^{\rm dr}&  = (1 + K n)^{-1} m^{\rm bare},
\end{align}
respectively, with $e_{s}(\theta)$, $k_{s}(\theta)$ and $m^{\rm bare}_{s}$ being the corresponding single particle (bare) quantities associated to the quasi-particle of type $s$. The dressed Fredholm operator,
\begin{equation}
(1 + K\,n) \equiv \delta_{s,s'}\delta(\theta  - \theta') + K_{s,s'}(\theta,\theta')n_{s'}(\theta'),
\end{equation}
is a linear integral operator which acts on both variables $\theta$ and $s$. We note that label $s$ usually pertains to the number
of constituent quasi-particle within a bound state, typically $m^{\rm bare}_{s}=s$.
The dressed momentum also specifies the total density of states,
$2 \pi \rho^{\rm tot}_{s} = p^{\prime}_{s}$, along with the hole densities, $\bar{\rho}_{s} = \rho^{\rm tot}_{s} - \rho_{s}$.
These are, unlike in non-interacting systems, non-trivial rapidity-dependent functions. 
Finally, functions
\begin{equation}\label{eq:ws}
\mathcal{W}_{s} = \lim_{s^{\prime} \to \infty}
\frac{K^{\rm dr}_{s,s^{\prime}}(\theta,\theta^{\prime})}{\rho^{\rm tot}_{s'}(\theta^{\prime})},
\end{equation}
represent renormalised (i.e. dressed) two-body scattering phase shifts, $K^{\rm dr} = (1 + K n)^{-1} K$
in the limit of large bare spin/charge. In the above formula, the large-$s$ limit indicates a correspondence between
the `giant quasi-particles' carrying bare spin $s$ and fluctuations of magnetization close to half filling,
see Sec.~\ref{sec:spinquasi}.\\

The quantity \eqref{eq:ws} is difficult to handle analytically or even to compute numerically for a generic equilibrium state.
We here first make a remarkable observation that, in the $h \to 0$ limit, the above exact expression for the spin diffusion constant
is nothing but the curvature of the \emph{Drude self-weight}
\begin{equation}\label{eq:magic1SM}
\mathfrak{D}(T,0) =  \frac{\partial^{2}\mathcal{D}^{\rm self}(T,\nu)}{\partial \nu^{2}} \Big|_{\nu=0},\qquad
\nu \equiv 4  {T}\expect{S^{z}}_{T,h}.
\end{equation} 
This is established on the basis of the following \textit{exact} identity,
\begin{equation}
\boxed{
 \mathcal{W}_{s} \Big|_{h=0} \equiv   \lim_{s^{\prime} \to \infty}
\frac{K^{\rm dr}_{s,s^{\prime}}(\theta,\theta^{\prime})}{\rho^{\rm tot}_{s'}(\theta^{\prime})} \Big|_{h=0} = \frac{1}{2 {T}\chi_h(T,0)} \lim_{h\to 0}\frac{m^{\rm dr}_{s}(T,h)}{h}, \quad \quad \chi_h(T,h) = - \partial_h^2 f(T,h),
}
\end{equation}
where we have introduced $\nu = 4T\chi_{h}(T,0) h + \mathcal{O}(h^2)$. For our convenience and future referencing, we
call it here the ``\emph{magic formula}". With aid of this identification we are able to recast the original quantity \eqref{eq:ws} 
in a simpler and more suggestive form involving the dressed magnetisation $m_s^{\rm dr}$, yielding an expression for the spin 
diffusion constant which is easier to deal with. This relation already indirectly appears in ref.~\cite{1812.02701} in
the Heisenberg XXZ chain at infinite temperature. Here we are able to prove it analytically in the high-temperature limit
for a wide class of integrable spin chain. The derivation is presented in Sec.~\ref{ref:magic}.
Additionally, by solving for $m_s^{\rm dr}$ and $T_{s,s'}^{\rm dr}$, we have verified it numerically at finite temperatures,
see Fig. \ref{Fig:CheckMagic}.

\section{Nested Bethe Ansatz for the O(3) nonlinear sigma model}

The quantum $O(3)$ nonlinear sigma model (NLSM) is a relativistic QFT for a non-abelian vector field
$\hat{\bf n}=(\hat{n}^{x},\hat{n}^{y},\hat{n}^{z})$ constrained on a unit sphere ($\hat{\bf n}\cdot \hat{\bf n}=1$),
described by the Euclidean action
\begin{equation}
\mathcal{A}_{0}[\hat{\bf n}] = \frac{1}{2g}\int \dd x\,\dd t
\left((\partial_{t}\hat{\bf n})^{2} - (\partial_{x}\hat{\bf n})^{2}\right).
\end{equation}
The action can be extended by including the topological $\Theta$-term,
\begin{equation}
\mathcal{A}_{\Theta}[\hat{\bf n}] = \mathcal{A}_{0}[\hat{\bf n}] + \ii \frac{\Theta}{4\pi}\int \dd x\,\dd t\,
\hat{\bf n}\cdot \partial_{t}\hat{\bf n}\times \partial_{x}\hat{\bf n}.
\end{equation}
\\

In the following, we are interested in describing the low-energy limit of the $SU(2)$-symmetric antiferromagnetic spin-$S$ chains.
The topological angle $\Theta = 2\pi\,S$ is an integer multiple of $\pi$ and crucially depends on whether $S$ if an integer
or half-integer. While in both cases the low-energy effective field theory is described by an
\emph{integrable} relativistic quantum sigma model, only in the former case the elementary spectrum is gapped.
The non-trivial topological term $\Theta = \pi$ prevents dynamical mass generation and yields massless excitations.\\

The quantum sigma model is the continuum low-energy theory of \emph{non-integrable} large-$S$ Heisenberg spin chains
\begin{equation}
\hat{H} = J\sum_{i}\hat{\bf s}_{i}\cdot \hat{\bf s}_{i+1},
\end{equation}
with antiferromagnetic exchange coupling $J>0$, normalization $\hat{\bf s}\cdot \hat{\bf s}=S(S+1)$, and unit lattice spacing.
In the continuum limit, the staggered and ferromagnetic fluctuations are represented by two smooth fields,
\begin{equation}
\hat{\bf s}_{i} \approx S(-1)^{i}\hat{\bf n} + \hat{\bf m},
\end{equation}
where
\begin{equation}
\hat{\bf m} = \frac{1}{g}\hat{\bf n}\times \hat{\bf p},\qquad \hat{\bf m}\cdot \hat{\bf n} = 0,
\end{equation}
generates rotations of the field $\hat{\bf n}$, and
\begin{equation}
{\bf p}=\frac{1}{g}\partial_{t}{\bf n} + \frac{\Theta}{4\pi}{\bf n}\times \partial_{x}{\bf n},
\end{equation}
is the momentum canonically-conjugate to $\hat{\bf n}$. This yields the Hamiltonian
\begin{equation}
\hat{H}_{\Sigma} = \frac{v}{2}\int \dd x \left[g\left(\hat{\bf m} + \frac{\Theta}{4\pi}\partial_{x}\hat{\bf n}\right)^{2}
+ \frac{1}{g}(\partial_{x}\hat{\bf n})^{2}\right],
\end{equation}
which matches the action $\mathcal{A}$ (in dimensionless units where the velocity $v=2JS$ is set to one).
The coupling constant $g$ is related to spin $S$ via
\begin{equation}
g=2/S.
\end{equation}

\subsection{O(3) NLSM without topological term}

The equation of motion is the conservation law for the Lorentz two-current,
\begin{equation}
\partial_{\mu}\hat{\bf j}_{\mu} = 0,
\end{equation}
with components
\begin{equation}
\hat{\bf j}_{\mu} = g^{-1}\hat{\bf n} \times \partial_{\mu}\hat{\bf n},\qquad \mu = x,t.
\end{equation}
implying the conserved Noether charges
\begin{equation}
\hat{\bf m} = \int \dd x\, \hat{\bf j}_{t}(x,t).
\end{equation}

The elementary excitations of the $O(3)$ NLSM form a spin-triplet of massive bosons with relativistic dispersion relation
\begin{equation}
e(k) = \sqrt{k^{2}+\mass^{2}},
\end{equation}
where $\mass$ denotes the non-perturbatively generated mass (spectral gap) related to the bare coupling constant $g$
via $\mass \sim J\,e^{-2\pi/g}$. We employ the usual rapidity parametrization in terms of variable $\theta$
\begin{equation}
k(\theta) = \mass \sinh{(\theta)},\qquad e(\theta) = \mass \cosh{(\theta)},
\end{equation}
in the absence of an external applied field.\\

\paragraph*{Exact $S$-matrix.}
The creation/annihilation operators for the elementary excitations in the $O(3)$ NLSM constitute the
(associative, non-commutative) Faddeev--Zamolodchikov algebra
\begin{align}
Z_{a}(\theta_{1})Z_{b}(\theta_{2}) &= \mathcal{S}^{a^{\prime}b^{\prime}}_{a b}(\theta_{1}-\theta_{2})
Z_{b^{\prime}}(\theta_{2})Z_{a^{\prime}}(\theta_{1}),\\
Z^{\dagger}_{a}(\theta_{1})Z^{\dagger}_{b}(\theta_{2}) &= \mathcal{S}^{a^{\prime}b^{\prime}}_{a b}(\theta_{1}-\theta_{2})
Z^{\dagger}_{b^{\prime}}(\theta_{2})Z^{\dagger}_{a^{\prime}}(\theta_{1}),\\
Z_{a}(\theta_{1})Z^{\dagger}_{b}(\theta_{2}) &= 2\pi \delta_{a b}\delta(\theta_{1}-\theta_{2})
+\mathcal{S}^{b^{\prime},a}_{b,a^{\prime}}(\theta_{1}-\theta_{2})Z^{\dagger}_{b^{\prime}}(\theta_{2})Z_{a^{\prime}}(\theta_{1}),
\end{align}
where $a,b\in \{x,y,z\}$ are isospin quantum numbers assigned to the $O(3)$ tripet of Bose fields.
The Fock space vacuum $\ket{0}$ is a state with the property $Z_{a}(\theta)\ket{0}=0$.
The $n$-particle scattering states are constructed in the standard manner
\begin{equation}
\ket{\theta_{n},\ldots,\theta_{2},\theta_{1}} = Z^{\dagger}_{a_{n}}(\theta_{n})\cdots
Z^{\dagger}_{a_{2}}(\theta_{2})Z^{\dagger}_{a_{1}}(\theta_{1})\ket{0}.
\end{equation}
The two-body scattering matrix of the $O(3)$ NLSM has the following structure
\begin{equation}
\mathcal{S}_{ab}^{cd}(\theta) =
\delta_{ab}\delta_{cd}\,\sigma_{1}(\theta)
+\delta_{ac}\delta_{bd}\,\sigma_{2}(\theta)
+\delta_{ad}\delta_{bc}\,\sigma_{3}(\theta),
\end{equation}
where $\theta$ designates rapidity difference of the incident particles, and
\begin{equation}
\sigma_{1}(\theta) = \frac{2\pi \ii \theta}{(\theta + \ii \pi)(\theta - 2\ii \pi)},\quad
\sigma_{2}(\theta) = \frac{\theta(\theta-\ii \pi)}{(\theta + \ii \pi)(\theta - 2\ii \pi)},\quad
\sigma_{3}(\theta) = \frac{2\pi \ii(\ii \pi -\theta)}{(\theta + \ii \pi)(\theta - 2\ii \pi)}.
\end{equation}
The non-diagonal multi-particle scattering is completely factorizable and thus fully described
by the two-particle (quantum) scattering $\mathcal{S}$-matrix obeying the celebrated Yang--Baxter relation
\begin{equation}\label{eqn:YBE}
\mathcal{S}^{b_{1}b_{2}}_{c_{1}c_{2}}(\theta-\theta^{\prime})\mathcal{S}^{c_{1}b_{3}}_{a_{1}c_{3}}(\theta)
\mathcal{S}^{c_{2}c_{3}}_{a_{2}a_{3}}(\theta^{\prime}) =
\mathcal{S}^{b_{2}b_{3}}_{c_{2}c_{3}}(\theta^{\prime})\mathcal{S}^{b_{1}c_{3}}_{c_{1}a_{3}}(\theta)
\mathcal{S}^{c_{1}c_{2}}_{a_{1}a_{2}}(\theta-\theta^{\prime}).
\end{equation}
As a consequence, the model possesses infinitely many local conservation laws.

\subsubsection*{Thermodynamic Bethe Ansatz}

Placing the field theory on space-time worldsheet of a cylinder geometry with circumference $L$ imposes
non-trivial quantization conditions for $n$ particle rapidities $\{\theta_{a}\}_{a=1}^{n}$.
The main object of the algebraic diagonalization of the many-body scattering process is the transfer matrix
$\mathcal{T}(\lambda;\{\theta_{a}\})$, acting on a $3^{n}$-dimensional Hilbert space with matrix elements
\begin{equation}
\mathcal{T}^{b}_{a}(\lambda;\{\theta_{a}\})_{i_{1}\cdots i_{n}}^{j_{1}\cdots j_{n}} =
\mathcal{S}_{a i_{1}}^{c_{1} j_{1}}(\lambda-\theta_{1})\mathcal{S}_{c_{1} i_{2}}^{c_{2} j_{2}}(\lambda-\theta_{2})\cdots 
\mathcal{S}_{c_{n-1} i_{n}}^{b j_{n}}(\lambda-\theta_{n}),
\end{equation}
where $\lambda$ is a complex spectral parameter.
The periodicity constraint for an $n$-particle wave-function amplitudes $\Psi(\{\theta_{a}\})$ yields the
celebrated Bethe equations
\begin{equation}
\left(\mathcal{T}(\theta_{j};\{\theta_{a}\})+e^{\ii p(\theta_{j})L}\right)\Psi(\{\theta_{a}\}) = 0.
\end{equation}
By virtue of Eq.~\eqref{eqn:YBE}, which implies
\begin{equation}
\mathcal{T}^{a^{\prime \prime}}_{a}(\theta)\mathcal{T}^{b^{\prime \prime}}_{b}(\theta^{\prime})
\mathcal{S}^{a^{\prime} b^{\prime}}_{a^{\prime \prime} b^{\prime \prime}}(\theta-\theta^{\prime})
= \mathcal{S}^{a^{\prime \prime}b^{\prime \prime}}_{ab}(\theta-\theta^{\prime})
\mathcal{T}^{b^{\prime}}_{b^{\prime \prime}}(\theta^{\prime})\mathcal{T}^{a^{\prime}}_{a^{\prime \prime}}(\theta),
\end{equation}
the traces $\mathcal{T(\theta)}=\sum_{a}\mathcal{T}^{a}_{a}(\theta)$ are in involution for all values of the spectral parameters,
\begin{equation}
\left[\mathcal{T}(\theta),\mathcal{T}(\theta^{\prime})\right]=0.
\end{equation}

Presently we deal with a \emph{non-diagonal} scattering theory, referring to non-trivial mixing
of internal (spin) degrees of freedom upon elastic quasi-particle collisions.
The scattering can nonetheless be transformed to a diagonal one at expense of introducing auxiliary magnonic particles,
in effect resulting in the so-called nested Bethe equations. Instead of describing the entire procedure here we refer
the reader to e.g. \cite{ZZ1992}.

In the case of the $O(3)$ NLSM, the nested Bethe equations have been originally obtained in \cite{Wiegmann1985}.
In the sector with $M_{\theta}$ physical excitations and $M_{\lambda}$ auxiliary magnon rapidities, they take the form
\begin{align}
e^{\ii k(\theta_{a})L}\prod_{b=1}^{M_{\theta}}S(\theta_{a},\theta_{b})\prod_{c=1}^{M_{\lambda}}S^{-1}(\theta_{a},\lambda_{c}) &= 1,\\
\prod_{b=1}^{M_{\theta}}S^{-1}(\lambda_{a},\theta_{a})\prod_{c=1}^{M_{\lambda}}S(\lambda_{a},\lambda_{c})&=-1,
\label{eqn:sigma_Bethe_equations}
\end{align}
where
\begin{equation}
S(\theta) = \frac{\theta - \ii \pi/2}{\theta + \ii \pi /2},
\end{equation}
is the elementary scattering amplitude which depends on the difference of the incident quasi-particles' rapidities $\theta$.\\

\paragraph*{Auxiliary magnons.}
We wish to stress that the so-called auxiliary rapidities $\lambda_{a}$ do not describe physical (i.e. momentum-carrying) degrees
of freedom. Rather, they are the internal spin degrees of freedom in the form of spin waves (defined with respect to the fully 
polarised reference state). These can be most easily pictured as fictitious quasi-particles which propagate in the static
reference frame of physical particles. Recall that the latter have been introduced in order to transform the original non-diagonal 
scattering theory to a diagonal one.

The magnonic degrees of freedom play a pivotal role and are crucial, in particular, for understanding and explaining
anomalous properties of spin transport. Most importantly, auxiliary rapidities can take complex values which signals
formation of bound states. Fusion properties of the scattering amplitudes imply that in the large-$L$ limit
the only allowed complex rapidities $\lambda_{a}$ are those in the form the `$k$-string compounds',
\begin{equation}
\lambda^{(k)i}_{b} = \left\{\lambda_{b} + \frac{\ii \pi}{2}(k+1-2i)\right\},
\end{equation}
with index $i=1,2,\ldots,k$ running over the constituent complex $\lambda$-rapidities.
Therefore, as far as the spin dynamics is concerned, the auxiliary bound states (labelled by a real-valued center $\lambda_{b}$)
play a similar role to physical multi-magnon bound states in the integrable Heisenberg spin chains, except that in the
quantum sigma model these propagate in an inhomogeneous background of physical excitations.
To this end, by appropriately shifting the poles of the elementary amplitude $S$, we introduce the elementary fused amplitudes
\begin{equation}
S_{n}(\theta) = \frac{\theta - n\,\ii \pi/2}{\theta + n\,\ii \pi/2},
\end{equation}
with the associated scattering phases
\begin{equation}
\Theta_{n}(\theta) = - \ii \log{\left(-S_{n}(\theta)\right)} = 2\,{\rm arctan}\left(\frac{2\theta}{n\pi}\right).
\end{equation}\\

\paragraph*{Bethe--Yang equations.}

We are interested in the finite-density limit of Eqs.~\eqref{eqn:sigma_Bethe_equations}. This amounts to take the
$L\to \infty$ limit while keeping ratios $M_{\theta}/L$ and $M_{\lambda}/L$ finite.
Thermodynamic states are understood as ensembles of locally indistinguishable microstates. These are referred to as macrostates
and are characterised by finite densities of physical and auxiliary excitations, denoted here
by $\rho_{0}(\theta)$ and $\rho_{s\geq 1}(\theta)$, respectively. Following the standard procedure,
namely taking the logarithmic derivative with respect to rapidity $\theta$ and converting the discrete summations over rapidities
to convolution-type integrals, we arrive at the Bethe--Yang equations of the form (suppressing rapidity dependence for clarity)
\begin{align}
\rho^{\rm tot}_{0} &= \frac{k^{\prime}}{2\pi} + \mathcal{K}\star \rho_{0} - K_{s}\star \rho_{s},\\
\rho^{\rm tot}_{s} &= K_{s}\star \rho_{0} - K_{s,s^{\prime}}\star \rho_{s^{\prime}}.
\label{eqn:sigma_Bethe-Yang}
\end{align}
with functions $\rho^{\rm tot}_{s\geq 0}$ denoting the total densities of available states for both the physical and auxiliary
quasi-particles. Here and subsequently we use a compact notation for summations over repeated indices,
\begin{equation}
K_{s}\star g_{s} = \sum_{s=1}^{\infty}\int_{-\infty}^{\infty}\dd \theta^{\prime}K_{s}(\theta-\theta^{\prime})
g_{s}(\theta^{\prime}),\qquad
K_{s,s^{\prime}}\star g_{s^{\prime}}=\sum_{s^{\prime}=1}^{\infty}\int_{-\infty}^{\infty} \dd \theta^{\prime}K_{s,s^{\prime}}
(\theta-\theta^{\prime})g_{s^{\prime}}(\theta^{\prime}).
\end{equation}
The convolution kernels
\begin{align}\label{eq:sigma_convolution_kernels}
K_{s}(\theta) &= \kappa_{s-1}(\theta) + \kappa_{s+1}(\theta),\\
K_{s,s^{\prime}}(\theta) &= \sum_{\ell = |s-s^{\prime}|}^{s+s^{\prime}+2}\kappa_{\ell}(\theta) + \kappa_{\ell+2}(\theta),
\end{align}
with $K_{s\geq 0}\equiv 0$, are given in terms of the differential scattering phases
\begin{equation}
\kappa_{s}(\theta) = \frac{1}{2\pi \ii}\partial_{\theta}\Theta_{s}(\theta) = \frac{2s}{s^{2}\pi^{2}+4\theta^{2}}.
\end{equation}
Using convention $\hat{f}(k)=\int_{-\infty}^{\infty}\dd \theta e^{-\ii k \theta}f(\theta)$, we have
the following Fourier-space representation
\begin{equation}
\hat{\kappa}_{s}(k) = e^{-(\pi/2)s|k|}.
\end{equation}\\

\paragraph*{Quasi-local form.}
The canonical Bethe--Yang integral equations \eqref{eqn:sigma_Bethe-Yang} can be further simplified with aid of the
fusion identities. The second equation in \eqref{eqn:sigma_Bethe-Yang} can be presented as a Fredholm-type integral
equation
\begin{align}
(1 + K)_{s,s^{\prime}}\star \rho_{s} = K_{s}\star \rho_{0} - \ol{\rho}_{s}.
\label{eqn:BY_canonical_spin}
\end{align}
Remarkably, the corresponding resolvent $R$, defined via matrix equation,
\begin{equation}
(1-R)\cdot(1+K) = 1,
\end{equation}
admits the following compact representation
\begin{equation}
R_{s,s^{\prime}}\star g_{s^{\prime}} \equiv I^{A_{\infty}}_{s,s^{\prime}}\fs\star g_{s^{\prime}}.
\end{equation}
Here we have introduced an infinite-dimensional incidence (adjacency) matrix of the $A_{\infty}$ Dynkin diagram,
\begin{equation}
I^{A_{\infty}}_{s,s^{\prime}} = \delta_{s,s^{\prime}-1} + \delta_{s,s^{\prime}+1},
\end{equation}
where the $\fs$-kernel, defined as the solution to $\kappa_{1}-\fs\star \kappa_{2}=\fs$, reads explicitly
\begin{equation}
\fs(\theta) = \frac{1}{2\pi \cosh{(\theta)}}.
\end{equation}
and has a simple Fourier representation $\hat{\fs}(k)=(2\cosh{(k\pi/2)})^{-1}$.
Moreover the convolution with the inverse of the Fredholm operator $(1+K)$ has the following important properties
\begin{align}
\left(1- I^{A_{\infty}}\,\fs\right)_{s,s^{\prime}}\star \kappa_{s^{\prime}} &= \kappa_{s}-\fs\star (\kappa_{s-1}+\kappa_{s+1})
= \delta_{s,1}\fs,\\
\left(1- I^{A_{\infty}}\,\fs\right)_{s,s^{\prime}}\star K_{s^{\prime}} &= K_{s} - \fs\star (K_{s-1}+K_{s+1}) = \delta_{s,2}\fs,\\
\left(1- I^{A_{\infty}}\,\fs\right)_{s,s^{\prime \prime}}\star K_{s^{\prime \prime},s^{\prime}} &= I^{A_{\infty}}_{s,s^{\prime}}\fs,
\end{align}
which are straightforward to prove. Inverting Eqs.~\eqref{eqn:BY_canonical_spin}, we find immediately
$\rho^{\rm tot}_{s}=\fs \star I^{A_{\infty}}_{s,s^{\prime}}\ol{\rho}_{s^{\prime}}$ for $s\geq 1$.
The remaining equation for the momentum-carrying particle can be simplified with aid of
\begin{equation}
K_{s}\star \rho_{s}=K_{2}\star \fs \star \rho_{0} - \fs \star \ol{\rho}_{2}.
\end{equation}
This way, we obtain the following quasi-local form of the Bethe--Yang equations
\begin{align}
\rho^{\rm tot}_{0} &= \frac{k^{\prime}}{2\pi} + \fs \star \ol{\rho}_{2},\\
\rho^{\rm tot}_{s} &= \delta_{s,2}\fs \star \rho_{0} + \fs \star I^{A_{\infty}}_{s,s^{\prime}}\ol{\rho}_{s^{\prime}}.
\end{align}
Notice that the self-coupling term in the canonical equation for $\rho^{\rm tot}_{0}$ disappears thanks to
$\kappa_{2}-\fs \star (\kappa_{1}+\kappa_{3})=0$.\\

\paragraph*{The universal dressing transformation.}

By identifying the total state densities with the dressed momentum derivatives,
\begin{equation}
2\pi \rho^{\rm tot}_{s} = p^{\prime}_{s},\qquad s=0,1,2,\ldots,
\end{equation}
the Bethe--Yang equation be recast in a more suggestive form
\begin{align}
p^{\prime}_{0} - \fs \star \ol{n}_{2}p^{\prime}_{2} &= k^{\prime}_{0},\\
p^{\prime}_{s} - \fs \star I^{A_{\infty}}_{s,s^{\prime}} - \delta_{s,2} \fs \star n_{0}p^{\prime}_{0} &= 0.
\label{eqn:sigma_dressing_momentum}
\end{align}
Physically speaking, these linear integral equations describe renormalization of particle's bare momenta $k_{0}$
with respect to an equilibrium macrostate, $k^{\prime}_{s}\mapsto p^{\prime}_{s}=\mathcal{F}^{\rm dr}_{\{n_{s}\}}(k^{\prime}_{s})$,
where the dressing transformation $\mathcal{F}^{\rm dr}$ is a linear functional which depends on the mode occupation functions
\begin{equation}
n_{s}(\theta) = \frac{\rho_{s}(\theta)}{\rho^{\rm tot}_{s}(\theta)},\qquad s\geq 0.
\end{equation}
Here $\ol{n}_{s}(\theta) \equiv 1 - n_{s}(\theta)$ are interpreted as the occupations functions of hole excitations.\\

\paragraph*{Thermodynamic Bethe Ansatz.}

In the formalism of the Thermodynamic Bethe Ansatz (TBA), the equilibrium partition sum is expressed as a functional integral
over particle rapidity distributions $\rho_{s\geq 0}(\theta)$. Specifically, the equilibrium free energy density
\begin{equation}
f[\{\rho_{s}\}] = e[\{\varrho_{s}\}] - s[{\varrho_{s}}],
\end{equation}
which is a functional of the energy and entropy densities,
\begin{equation}
e = \int_{-\infty}^{\infty}\dd \theta\,(\mathfrak{m} \cosh{(\theta)}-h)\rho_{0}(\theta),\qquad
s = \sum_{s\geq 0}\int_{-\infty}^{\infty}\dd \theta\left(\rho^{\rm tot}_{s}\log \rho^{\rm tot}_{s}-
\rho_{s}\log \rho_{s}-\ol{\rho}_{s}\log \ol{\rho}_{s}\right),
\end{equation}
respectively, is minimised by demanding a vanishing variational derivative, $\delta f = 0$.

\medskip

There is no need of performing an explicit derivation. One can simply resort to universality of the dressing equation, which
is a neat way to formulate the TBA equations once the dressing equations for the bare momenta (that is the Bethe--Yang equations)
are known. The free-energy minimization is essentially nothing but the energy counterpart of the momentum dressing, namely
$\varepsilon^{\prime}_{s}\mapsto \mathcal{F}^{\rm dr}_{\{n_{s}\}}(e^{\prime}_{s})$, reading
\begin{align}
\varepsilon^{\prime}_{0} - \fs \star \ol{n}_{2}\varepsilon^{\prime}_{2} &= e^{\prime}_{0},\\
\varepsilon^{\prime}_{s} - \fs \star I^{A_{\infty}}_{s,s^{\prime}}\ol{n}_{s^{\prime}}\varepsilon^{\prime}_{s^{\prime}} &= 0.
\end{align}
By introducing the the TBA $Y$-functions,
\begin{equation}
Y_{s}(\theta) = \frac{\ol{\rho}_{s}(\theta)}{\rho_{s}(\theta)},\qquad s\geq 0,
\end{equation}
and identifying them with the dressed energies $\varepsilon_{s}$ via $\log Y_{s}=\beta \varepsilon_{s}$,
we readily obtain the quasi-local form of the TBA equations
\begin{align}\label{eq:sigma_model_TBA}
\log Y_{0} &= \beta\,e - \fs \star \log(1+Y_{2}),\\
\log Y_{s} &= \delta_{s,2}\,\fs \star \log(1+1/Y_{0}) + \fs \star I_{s,s^{\prime}}\log(1+Y_{s^{\prime}}),
\end{align}
where we have used
\begin{equation}
\partial_{\theta}\log(1+Y_{s})=\ol{n}_{s}\varepsilon_{s},\qquad
\partial_{\theta}\log(1+1/Y_{s})=-n_{s}\varepsilon_{s}.
\end{equation}
With the additional particle-hole transformation on the massive node, the latter can be brought into the group-theoretic form
\begin{equation}
\log Y_{s} = -\delta_{s,0}\beta\,e^{\prime} + \fs \star I^{D_{\infty}}_{s,s^{\prime}}\log(1+Y_{b}),\qquad s\geq 0,
\end{equation}
compatible with the so-called $Y$-system hierarchy associated to the $D_{\infty}$ Dynkin diagram.

The free energy density is only a functional of energy-carrying $Y$-function
\begin{equation}
f = -T\int_{-\infty}^{\infty}\dd \theta \frac{k^{\prime}(\theta)}{2\pi}\log(1+1/Y_{0}(\theta)).
\end{equation}

\subsection{O(3) NLSM with topological term}

Now we consider the addition of the topological $\Theta$-term in the $O(3)$ NLSM action.
This now gives a $SU(2)$-symmetric massless relativistic quantum field theory of with completely factorizable
non-diagonal scattering \cite{ZZ1992}. We use subscripts $\pm$ to denote the internal quantum label of the $SU(2)$ doublet, with $+$ designating the `right movers' ($k>0$) and $-$ the `left movers' ($k<0$). Their bare dispersion relations are
\begin{equation}
e_{\pm}(\theta) = \pm k(\theta) = \frac{\Lambda}{2}e^{\pm \theta},\qquad -\infty < \theta < \infty.
\end{equation}
Here $\Lambda \sim e^{-2\pi/g}$ sets the cut-off scale at which the asymptotically free UV behavior changes into the
scale-invariant IR regime.
The scattering relations are provided by the following Faddeev--Zamolodchikov algebra~\cite{ZZ1992}
\begin{align}
R_{a}(\theta_{1})R_{b}(\theta_{2}) &= \mathcal{S}^{a^{\prime}b^{\prime}}_{ab}(\theta_{1}-\theta_{2})
R_{b}(\theta_{2})R_{a^{\prime}}(\theta_{1}),\\
L_{a}(\theta_{1})L_{b}(\theta_{2}) &= \mathcal{S}^{a^{\prime}b^{\prime}}_{ab}(\theta_{1}-\theta_{2})
L_{b}(\theta_{2})L_{a^{\prime}}(\theta_{1}),\\
R_{a}(\theta_{1})L_{\ol{a}}(\theta_{2}) &= \mathcal{U}^{b\ol{b}}_{a\ol{a}}(\theta_{1}-\theta_{2})L_{\ol{b}}(\theta_{2})R_{b}(\theta_{1}),
\end{align}
with scattering amplitudes of the form
\begin{align}
\mathcal{S}^{a^{\prime}b^{\prime}}_{a b}(\theta) &= \frac{\mathcal{S}^{(\pi)}(\theta)}{\theta-\ii \pi}
\left(\theta \delta_{a,a^{\prime}}\delta_{b,b^{\prime}}-\ii \pi \delta_{a,b^{\prime}}\delta_{b,a^{\prime}}\right),\\
\mathcal{U}^{b\ol{b}}_{a\ol{a}}(\theta) &= \frac{\ii\,\mathcal{S}^{(\pi)}(\theta)}{\theta - \ii \pi}
\left(\theta \delta_{a,b}\delta_{\ol{a},\ol{b}} -\ii \pi \delta_{a,\ol{b}}\delta_{\ol{a},b}\right).
\end{align}
The scattering phase shift of the model at $\Theta = \pi$, denoted by $\mathcal{S}^{(\pi)}(\theta)$,
has the following useful integral representation
\begin{equation}
\log\left(-\mathcal{S}^{(\pi)}(\theta)\right) = \int^{\infty}_{0}\dd k\,\frac{e^{-\pi k/2}}{2\cosh{(\pi k/2)}}
\frac{\sin{(\theta k)}}{k}.
\end{equation}

Note that in the UV regime, $\theta_{2}-\theta_{1}\to \infty$, the left-right scattering trivializes, and
the scattering process becomes indistinguishable from that of the trivial topologically angle $\Theta = 0$.
\\

\paragraph*{Bethe, Bethe--Yang and TBA equations.}

In the periodic box of size $L$, the quasi-particle rapidities $\{\theta_{\alpha}\}_{\alpha=1}^{N}$ are subjected to
the Bethe quantization constraints
\begin{equation}
e^{\ii p(\theta_{\alpha})L}\prod_{\beta=1}^{M_{\lambda}}\frac{\theta_{\alpha}-\lambda_{\beta}+\ii \pi/2}{\theta_{\alpha}-\lambda_{\beta}-\ii \pi/2}
\prod_{\gamma=1}^{M_{\theta}}\mathcal{S}^{(\pi)}(\theta_{\alpha}-\theta_{\gamma}) = 1.
\end{equation}
As usual, here $\{\theta_{\alpha}\}$ denote a set of physical rapidities which parametrise momenta of the right and left movers,
while $\lambda_{\beta}$ pertain to auxiliary magnons which diagonalise the $SU(2)$-invariant scattering.
\\

In the thermodynamic limit $L\to \infty$ (keeping ratios $M_{\theta}/L$ and $M_{\lambda}/L$ finite), one arrives at the following
equations for the physical and auxiliary quasi-particle densities $\rho_{\pm}$ and $\rho_{s\geq 1}$,
\begin{align}
\rho^{\rm tot}_{\pm} &= L\frac{k^{\prime}_{\pm}}{2\pi} + \mathcal{K}\star \rho_{\pm} - K_{s}\star \rho_{s},\\
\rho^{\rm tot}_{s} &= K_{s}\star (\rho_{+}+\rho_{-}) + K_{s,s^{\prime}}\star \rho_{s^{\prime}}.
\end{align}
Here $k^{\prime}_{\pm}(\theta)=(\Lambda/2)e^{\theta}$ are rapidity derivatives of bare momenta of physical excitations,
the convolution kernels $K_{s}$ are given by \eqref{eq:sigma_convolution_kernels}, and
\begin{equation}
\mathcal{K}(\theta) = \frac{1}{2\pi \ii}\partial_{\theta}\log \mathcal{S}^{(\pi)}(\theta)
= \frac{1}{\pi}(\fs\star K_{1})(\theta).
\end{equation}
The equivalent quasi-local form yields
\begin{align}
\rho^{\rm tot}_{\pm} &= \frac{L}{2\pi}\frac{\Lambda}{2}e^{\pm \theta} +  \fs \star \ol{\rho}_{1},\\
\rho^{\rm tot}_{s} &= \delta_{s,1}\fs \star(\rho_{+}+\rho_{-}) + \fs \star I^{A_{\infty}}_{s,s^{\prime}}\ol{\rho}_{s^{\prime}}.
\end{align}
The TBA equations for the thermodynamic free energy in a finite volume $L=1/T$ take the form
\begin{align}
\log Y_{\pm} &= L\frac{\Lambda}{2}e^{\pm \theta}+ \fs \star \log(1+Y_{{1}}),\\
\log Y_{s} &= \delta_{s,1}\fs \star \log(1+{Y_{+}})(1+{Y_{-}})
+ \fs\star I^{A_{\infty}}_{s,s^{\prime}} \log(1+Y_{s^{\prime}}).
\end{align}

\section{Low-temperature expansion of the spin diffusion constant of the spin-$1/2$ XXZ chain and $O(3)$ sigma model}

\subsection{Heisenberg XXZ chain}

We consider the Heisenberg spin-$1/2$ XXZ chain,
\begin{equation}
\hat{H} = \sum_{x}\hat{s}^{x}_{x}\hat{s}^{x}_{x+1} + \hat{s}^{y}_{x}\hat{s}^{y}_{x+1} + \Delta\,\hat{s}^{z}_{x}\hat{s}^{z}_{x+1},
\end{equation}
in the gapped phase $|\Delta|=\cosh \eta >1$.
The TBA equations for the grand-canonical Gibbs equilibrium are of the form
\begin{equation}
\log Y_{s} = -T^{-1} \fs^{(\eta)} \delta_{s,1}  +  \fs \star I^{A_{\infty}}_{s,s^{\prime}}\log(1 + Y_{s^{\prime}}),
\qquad  \lim_{s \to \infty} s^{-1}\log Y_{s} =  h/T,
\end{equation}
where
\begin{equation}
\mathfrak{s}(\theta) = \frac{1}{2 \pi} \sum_{k \in \mathbb{Z}} \frac{e^{2 \ii k \theta}}{\cosh{(k \eta)}},\qquad
\fs^{(\eta)}(\theta) \equiv \pi\sinh{(\eta)}\,\fs(\theta),
\end{equation}
are the usual and the deformed $\fs$-kernels, respectively.

Below we carry out the low-temperature expansion of the TBA quantities,
as previously done in e.g. \cite{Johnson1972,Bertini2018}. For the subsequent analysis it is important
to assume $h/T \gg 1$. In this regime, in the $T\to 0$ limit we have the following behaviour
\begin{align}
Y_{1}(\theta) &= e^{-\fs^{(\eta)}(\theta)/T} \frac{\sinh (h/T)}{\sinh (h/2T)}\times \big(1 + \mathcal{O}(e^{-h/T})\big),\\
Y_{s>1}(\theta) &=  \left(\frac{\sinh^{2}(s\,h/2T)}{\sinh^{2} (h/2T)} - 1 \right) \times \big(1 + \mathcal{O}(e^{-h/T})\big),
\end{align}
from where we obtain
\begin{equation}
\rho^{\rm tot}_{1} = \fs,\qquad \rho^{\rm tot}_{s>1}(\theta) = K_{s+1} \star (\fs\,Y_{1}),
\end{equation}
for the state densities and
\begin{equation}
v^{\rm eff}_{1} =  - \frac{\sinh{(\eta)}}{2}\,\frac{\mathfrak{s}^{\prime}}{\fs},\qquad 
v^{\rm eff}_{s>1} =  -\frac{\sinh{(\eta)}}{2} \frac{K_{s+1} \star (\fs^{\prime} Y_{1})}{K_{s+1} \star (\fs\,Y_{1}) },
\end{equation}
for the dressed velocities, up to subleading corrections which are of the order $\mathcal{O}(e^{-h/T})$.
The low-temperature limit of the Gibbs thermodynamic free energy,
$f=- T \int_{-\pi/2}^{\pi/2} \dd\theta\,\fs(\theta) \log(1 + Y_{1}(\theta))$, yields
\begin{equation}\label{eq:ffintegral}
f = -T\int_{-\pi/2}^{\pi/2} \dd\theta\, \fs(\theta)  Y_{1} (\theta) e^{- \fs^{(\eta)}(\theta)/T}
\frac{\sinh(h/T)}{\sinh (h/2T)} \times \big(1 + \mathcal{O}(e^{-h/T})\big).
\end{equation}
In the limit $T\to 0$, the latter can evaluated with the saddle point technique.
The spectral gap $\mathfrak{m}$ is given by
\begin{equation}
\mathfrak{m} = \fs^{(\eta)}(\pm \pi/2).
\end{equation}
The dressed dispersion relation for the unbound magnons ($1$-strings), expanded around points $\theta = \pm \pi/2$, reads
\begin{equation}
\varepsilon_1(\theta) = \mathfrak{m} + \frac{(\theta \pm \pi/2)^2}{2}
\partial^2_\theta \varepsilon_1(\theta)\Big|_{\theta=\pm \pi/2} + \ldots,
\end{equation}
with curvature
\begin{equation}
\partial^2_\theta \varepsilon_1(\theta)\Big|_{\theta=\pm \pi/2}
= \partial^{2}_{\theta}\fs^{(\eta)}(\theta)\Big|_{\theta=\pm \pi/2}
=  \pi \sinh{(\eta)}\,\fs^{\prime \prime}(\pi/2).
\end{equation}
By approximating the dispersion relation in the vicinity of the spectral gap, we find the following
low-temperature behavior of the free energy,
\begin{equation}
f(T,h) = -T \fs(\pi/2) e^{-\mathfrak{m}/T} \frac{\sqrt{2\pi\,T}}{\sqrt{\partial^{2}_{\theta} \varepsilon_{1}(\theta)
\big|_{\theta=\pm \pi/2}  }} \frac{\sinh(h/T)}{\sinh(h/2T)} \times \big(1 + \mathcal{O}(e^{-\mathfrak{m}/T})\big).
\end{equation}
In a similar manner, we obtain the following low-$T$ limit of the static spin susceptibility
\begin{equation}
 {\chi_h (T,h)} =-\frac{\partial^{2}f(T,h)}{\partial h^{2}} = \frac{\fs(\pi/2)}{2} \sqrt{\frac{2\pi}{T \partial^2_\theta \varepsilon_1(\theta)\big|_{\theta=\pm \pi/2}}}
e^{- (\mathfrak{m} - h/2)/T}\big(1 + \mathcal{O}(e^{-\mathfrak{m}/T})\big).
\end{equation}

\medskip

We proceed by expressing the spin diffusion $\mathfrak{D}(T,h)$ in the vicinity of the half filled state $h \sim 0 $,
expressed the curvature of the spin Drude self-weight. To this end, we employ the exact hydrodynamic mode decomposition
\begin{equation}\label{eq:diffusionXXZ1}
\mathfrak{D}=\frac{1}{8  {(T\chi_h(T,h))}^2} \sum_{s \geq 1 } \int_{-\pi/2}^{\pi/2} \dd\theta\,\rho_{s}^{\rm tot}(\theta) n_{s}(\theta) (1-n_{s}(\theta))
|v_{s}^{\rm eff}(\theta)| \left( \lim_{h \to 0 } \lim_{T \to 0} \frac{m_{s}^{\rm dr}(T,h)}{h} +\mathcal{O}(h^2) \right)^{2}.
\end{equation}
In the limit $T\to 0$ limit, we found the following dressed value of magnetization
\begin{equation} 
\lim_{T \to 0} \lim_{h \to 0 } \frac{m_{1}^{\rm dr}(T,h)}{h} = \frac{1}{2},\qquad
\lim_{T \to 0} \lim_{h \to 0 } \frac{m_{s>1}^{\rm dr}(T,h)}{h} = \frac{s^{2}}{3}.    
\end{equation}
Moreover, at low temperatures we have the following total state densities,
\begin{equation}
\rho^{\rm tot}_{1}(\theta) \sim \mathcal{O}(1),\qquad  \rho^{\rm tot}_{s>1}(\theta) \sim \mathcal{O}(\sqrt{T}e^{-\mathfrak{m}/T}),
 \end{equation}
and mode occupation functions
\begin{equation}
n_{1}(\theta)(1- n_{1}(\theta)) \sim e^{- \mathfrak{m}/T},\qquad
n_{s>1}(\theta)(1- n_{s>1}(\theta)) \sim e^{- (s-1)h/T}.
\end{equation}
The above relations imply that the bound-state contributions pertaining to quasi-particles with $s>1$
get exponentially suppressed as $\sim e^{-s\,h/T}$.
The leading contribution thus comes from the unbound magnons ($s=1$) and reads
\begin{equation}
\mathfrak{D}_{\rm XXZ}  =  \frac{1}{8 (T\chi_h(T,h))^2} \frac{1}{4}\int_{-\pi/2}^{\pi/2} \dd\theta \ \mathfrak{s}(\theta)
Y_{1}(\theta)  |v^{\rm eff}_1(\theta)| \times  \big(1 + \mathcal{O}(e^{-h/T})\big).
\end{equation}
All the higher contributions coming from the magnonic bound states ($s>1$) are contained in the correction term.

It is important to stress at this point that in the gapped in the XXZ chain, the summation over $s>1$ converges for
any value of $h$. This is a corollary of an exponential suppression of the dressed velocities
$\int d\theta |v^{\rm eff}_{s}| \sim e^{-s\,\eta}$ for large $s$.
In stark contrast, in the the isotropic (XXX), where $\int \dd\theta |v^{\rm eff}_{s}| \sim 1/s $,
the sum only converges strictly away from half filling, whereas exactly at $h=0$ the higher-order contributions due to the spectrum
of bound states can no longer be discarded.

\medskip

Let us further analyse the dominant contribution which comes from $s=1$.
The saddle points of $Y_{1}(\theta)$ are located at $\theta = \pm \pi/2$ where the effective velocity vanishes.
From the saddle-point analysis we deduce
\begin{equation}\label{eq:integralu}
  \int_{-\pi/2}^{\pi/2} d\theta\  \mathfrak{s}(\theta)|v^{\rm eff}_1(\theta)|e^{-\mathfrak{s}^{(\eta)}(\theta) /T }  \frac{\sinh (h/T)}{\sinh (h/2T)}
  =  \frac{\sinh (h/T)}{\sinh (h/2T)} \frac{2T \mathfrak{s}(\pi/2)   e^{- \mathfrak{m}/T} }{\partial^{2}_{\theta} \varepsilon_{1}(\theta)
  \big|_{\theta = \pm \pi/2}
   } |\partial_\theta v_{1}^{\rm eff}(\theta)|_{\theta=\pi/2} \times \big(1 + \mathcal{O}(e^{- \mathfrak{m}/T})\big),
\end{equation}
and the spin diffusion constant can be expressed as
\begin{equation}
\mathfrak{D}_{\rm XXZ} =   \frac{1}{4} \frac{\sinh (h/T)}{\sinh (h/2T)} \frac{|\partial_{\theta} v_{1}^{\rm eff}(\theta)|_{\theta=\pi/2}}{2\pi \mathfrak{s}(\pi/2)} 
e^{( \mathfrak{m} - h )/T} \times \big( 1 + \mathcal{O}(e^{-\mathfrak{m}/T}) \big).
\end{equation}
The corrections of the order $\mathcal{O}(e^{-h/T})$ are here due to the bound states ($s>1$), while
the corrections of the order $\mathcal{O}(e^{-\mathfrak{m}/T})$ are a consequence of the saddle-point approximation of
the rapidity integration in Eqs.~\eqref{eq:ffintegral} and \eqref{eq:integralu}. Notice moreover that
\begin{equation}
{\partial_{k}^{2} \varepsilon_{1}(k)}\Big|_{k=0}  = \partial_{k} v_{1}^{\rm eff}(k)\Big|_{k=0},
\end{equation}
where
\begin{equation}
\frac{\partial_{\theta}  v_{1}^{\rm eff}}{2\pi \rho^{\rm tot}_{1}} = \partial_{k} v^{\rm eff}(k) .
\end{equation}
Finally, in the gapped phase of the XXZ spin-$1/2$ chain, the leading low-$T$ and low-$h$ behaviour of the spin diffusion
constant, neglecting terms $\mathcal{O}(e^{- h/T})$, is given by
\begin{equation}
 \mathfrak{D}_{\rm XXZ} = \frac{1}{2}\partial_{k}^{2} \varepsilon_{1}(k)\Big|_{k=0}
e^{\mathfrak{m}/T } \left( 1 + \mathcal{O}(e^{- \mass/T}) \right) \equiv \mathfrak{D}_{\rm cl},
\end{equation}
with
\begin{equation}
{\partial_k^{2} \varepsilon_1(k)}\Big|_{k=0} =\Big| \frac{\sinh{(\eta)}\fs^{\prime \prime}(\theta)}{4 \pi \fs(\theta)^2}\Big|_{\theta=\pi/2}.
\end{equation}

\subsection{$O(3)$ NLSM}
Now we come to the non-topological ($\Theta = 0$) $O(3)$ NLSM, cf. Eqs.~\eqref{eq:sigma_model_TBA}, and
analyze the low-temperature regime. Assuming first $h/T\gg 1$ and $T\ll \mass$, and expanding the magnonic $Y$-functions (see also \cite{Konik2003}),
\begin{equation}
Y_{s\geq 1} =  \left( \frac{\sinh^{2} (h(s+1)/2T)}{\sinh^{2}(h/2T)} - 1 \right) \times \big(1 + \mathcal{O}(e^{-h/T})\big),
\end{equation}
we obtain
\begin{equation}
Y_{0}(\theta)^{-1}= e^{-\,e(\theta)/T} \left(\frac{\sinh (3h/2T)}{\sinh(h/2T)}  \right) \times \big(1 + \mathcal{O}(e^{-h/T})\big).
\end{equation}
The dressed magnetization behaves as
\begin{equation}
\lim_{T \to 0} \lim_{h \to 0} \frac{m_0^{\rm dr}(h)}{h} =  \frac{4}{3},  \qquad
\lim_{T \to 0} \lim_{h \to 0} \frac{m_{s\geq 1}^{\rm dr}(h)}{h} =  \frac{1}{3}(s+1)^{2}.
\end{equation}
and implies vanishing spin Drude weight at half filling.
The static spin susceptibility,
\begin{equation}
\chi_h(T,h) =-\partial_{h}^{2} f(h) =  \sqrt{\frac{2\mathfrak{m}}{T\pi}} e^{-(\mathfrak{m}-h)/T} \times
\big(1+ \mathcal{O}(e^{-\mathfrak{m}/T}) + \mathcal{O}(e^{-h/T})\big),
\end{equation}
follows from the free-energy density
$f(T,h) =-T\int_{-\infty}^{\infty}\dd \theta\,\frac{k^{\prime}(\theta)}{2\pi}\log(1+1/Y_{0}(\theta))$.

In precise analogy with the above calculation in the gapped Heisenberg spin chain, by neglecting the contribution of order $\mathcal{O}(e^{- h/T})$, the spin diffusion constant comes solely from the physical excitations ($s=0$),
\begin{equation}
 \mathfrak{D}_0 = \frac{1}{\mathfrak{m}}  \left(\frac{\sinh (h/2T)}{\sinh(3h/2T)}  \right)  e^{\mathfrak{m}  /T}\left( 1 + \mathcal{O}(e^{- \mass/T}) \right) \equiv \mathfrak{D}_{\rm cl},
\end{equation}
where we used that curvature of the dispersion relation is now given by
\begin{equation}
{\partial_k^{2} \varepsilon_0(k)} \Big|_{k=0}
= \Big| \frac{e^{\prime \prime}(\theta)}{(k^{\prime}(\theta)^{2})}\Big|_{\theta=0}
= \frac{1}{\mathfrak{m}}.
\end{equation}
We wish to stress once again that the corrections $ \mathcal{O}(e^{- h/T} )$ due to magnonic degrees of freedom ($s>0$) are only negligible provided $h/T \gg 1$. Analogously to the XXX chain, in the half filling limit $h\to 0$ they yield
a $\sim 1/h$ type of divergence of $\mathfrak{D}$. 

\medskip

The spin Drude weight is given by 
\begin{equation}
\mathcal{D}_{\Sigma} = \sum_{s} \int \dd\theta \ \rho(\theta)(1- n(\theta)) ( v^{\rm eff}_s(\theta) m_s^{\rm dr})^2,
\end{equation}
and, analogously to the XXX spin 1/2 chain, it goes to zero at small $h$ as $h^2\log h$, due to $ m_s^{\rm dr} \sim h$.
By following the previous reasoning and including only the contribution from the $s=0$ quasi-particle, we obtain
\begin{equation}
\mathcal{D}_{\Sigma}/T =\frac{8 \mathfrak{m}}{3}     \left(h^2 + O(h^4)\right)    \sqrt{\frac{2 T }{\pi} }      e^{- \mathfrak{m} /T}\times \big(1  +  \mathcal{O}(e^{-h/T})+ \mathcal{O}(e^{-\mathfrak{m}/T}) \big).
\end{equation}
\section{Divergence of spin diffusion constant at the isotropic point}

We now specialise to the case of isotropic interactions, namely the spin-$1/2$ XXX chain with $\Delta = 1$.
In approaching the half filling $h \to 0$, we show that the spin diffusion constant $\mathfrak{D}$ diverges when as $h^{-1}$.
In order to show this is sufficient to analyse the asymptotic behaviour at large $s$ of the summand in \eqref{eq:diffusionXXZ1}.
We have
\begin{equation}\label{eq:occfun}
 \mathfrak{D}  \simeq  \frac{1}{8 (T\chi_{h}(T,h))^2} \sum_{s\geq 1} \int \dd\theta\,\rho^{\rm tot}_{s}(\theta) n_s(\theta) (1- n_{s}(\theta)) |v_{s}^{\rm eff}(\theta)|
\left(  s^2 + \mathcal{O}(s)   \right)^{2}.
\end{equation}
Since $n_{s} \sim e^{- s h/T }$, the sum is convergent for non-zero $h$. However, the $h \to 0$ limit is rather subtle and eventually 
yields a divergent $\mathfrak{D}$. This type of anomaly can be attributed to the large-$s$ behaviour of the integrand which reveals
that contributions in the limit of infinitely long strings saturates with increasing the bare spin $s$. The sum \eqref{eq:siummm}
is a clear signature of non-perturbative physics in the vicinity of half filling: the sum over quasi-particle spices $s$ must be 
evaluated \textit{before} taking the limit $h \to 0$. Using the asymptotic form of the occupation functions\textit{ for large $s$ and $h>0$},
\begin{equation}
n_{s}(\theta) \simeq \left(\frac{\sinh{(h/(2T))}}{\sinh{(s\,h/(2T))}} \right)^{2},
\end{equation}
one can readily extract the type of divergence as $h \sim 0$,
\begin{align}\label{eq:siummm}
 \mathfrak{D} \simeq & \frac{1}{(T\chi_{h}(T,0))^2}  \sinh^{2}\left( \frac{h}{2T}\right) \sum_{s\geq 1} s^{4}  e^{-s h/T }
\int \dd\theta\, |\varepsilon_s^{\rm dr}(\theta)| \nonumber  \\&  \sim \frac{1}{(T\chi_{h}(T,0))^2}  \sinh^{2}\left( \frac{h}{2T}\right)\tanh\left(\frac{h}{2 T}\right) \sum_{s\geq 1}s^{3} e^{-s h/T }
\sim \frac{1}{(T\chi_{h}(T,0))^2} \frac{T}{h}.
\end{align}
We have used
\begin{equation}\label{eq:vv}
\int \dd\theta\, |\varepsilon_s^{\rm dr}(\theta)| = \tanh\left(\frac{h}{2 T}\right) \frac{1}{s} + \mathcal{O}(s^{-2}).
\end{equation}
which is valid only at the isotropic point $\Delta=1$.
Since the spin susceptibility at the isotropic point and at $h=0$ is $\chi_{h}(T,0) \sim {T^{-1/2}}$, we immediately
have that at $h \sim 0$
\begin{equation}
\mathfrak{D}(T,h) \chi_{h} (T,h) \sim \frac{\kappa(T)}{h} + \mathcal{O}(h^0).
\end{equation}
with $\kappa(T) \sim  (T\chi_h)^{-1} \simeq T^{-1/2}$ at low $T$ but in general $\kappa(T) >0$ for any $T$, implying super-diffusive spin transport at any temperature. An analogous result can be found with a similar calculation in the $O(3)$ non-linear sigma model since the structure of the dressing equations for its thermodynamic quantities are analogous. 

\section{Spin fluctuations from giant magnons}\label{sec:spinquasi}

Here we explain why fluctuations of the Fermi functions $\delta n_{s}$ pertaining to quasi-particle in the limit of
infinitely large bare spin $s$ are directly connected to fluctuations of the local magnetization
$\delta \langle s^{z}_{x} \rangle = \langle s^{z}_{x} \rangle -  \langle s^z \rangle_{T,h=0}$ with respect to a half-filled thermal state. First, recall that the occupation functions $n_{s}(\theta)$ are linked to the TBA  $Y_s(\theta)$ functions via
\begin{equation}
n_{s}(\theta) = \frac{1}{1 + Y_{s}(\theta)}.
\end{equation}
Information about the filling is contained in the large-$s$ asymptotics, namely
\begin{equation}
h/T =  \lim_{s \to \infty} \frac{\log Y_{s}}{s}.
\end{equation}
Combining the two, we find close to half filling
\begin{equation}
\langle s^{z} \rangle = T\,\chi_{h}(T,h) \frac{h}{T} + \mathcal{O}(h^{2}) =  T\,\chi_h(T,h)  \lim_{s \to \infty} \frac{\log(Y_{s})}{s} + \mathcal{O}(h^{2}).
\end{equation}
Considering small fluctuations of local magnetization in the vicinity of a half-filled state,
$\delta \langle  s^{z} \rangle = T\,\chi_{h} \delta (h/T)$, we deduce
\begin{equation}
 \delta  \langle  s^{z} \rangle = T\,\chi_h(T,h)    \lim_{s \to \infty} \delta(\log Y_{s}/s)
 = T\,\chi_h(T,h)  \delta n_{\infty}.
\end{equation}
where
\begin{equation}
\delta n_{\infty} \equiv \lim_{s \to \infty} \frac{1}{s} \frac{\delta n_{s}}{n_{s}(n_{s}-1)},
\end{equation}
pertain to fluctuations of `giant magnons'. As a concrete example of this principle we mention the spin diffusion in the XXZ 
spin-$1/2$ chain, where $\delta n_{\infty}(x,t)$ satisfies the equation of motion
\begin{equation}
\partial_t \delta n_{\infty}(x,t)  =  \widetilde{w}_\infty \partial_x^2 \delta n_{\infty}(x,t).
\end{equation}
Here the coefficient $\widetilde{w}_{\infty}$ corresponds to the variance of the fluctuations of the giant magnons and it
is precisely the spin diffusion constant at half filling, $\tilde{w}_\infty  = \mathfrak{D}$, see Eq. (6.38) in \cite{1812.00767}.

\section{Spin diffusion constant and ``the magic formula''}\label{ref:magic}

\begin{figure}[t!]
\center
\includegraphics[width=0.49\textwidth]{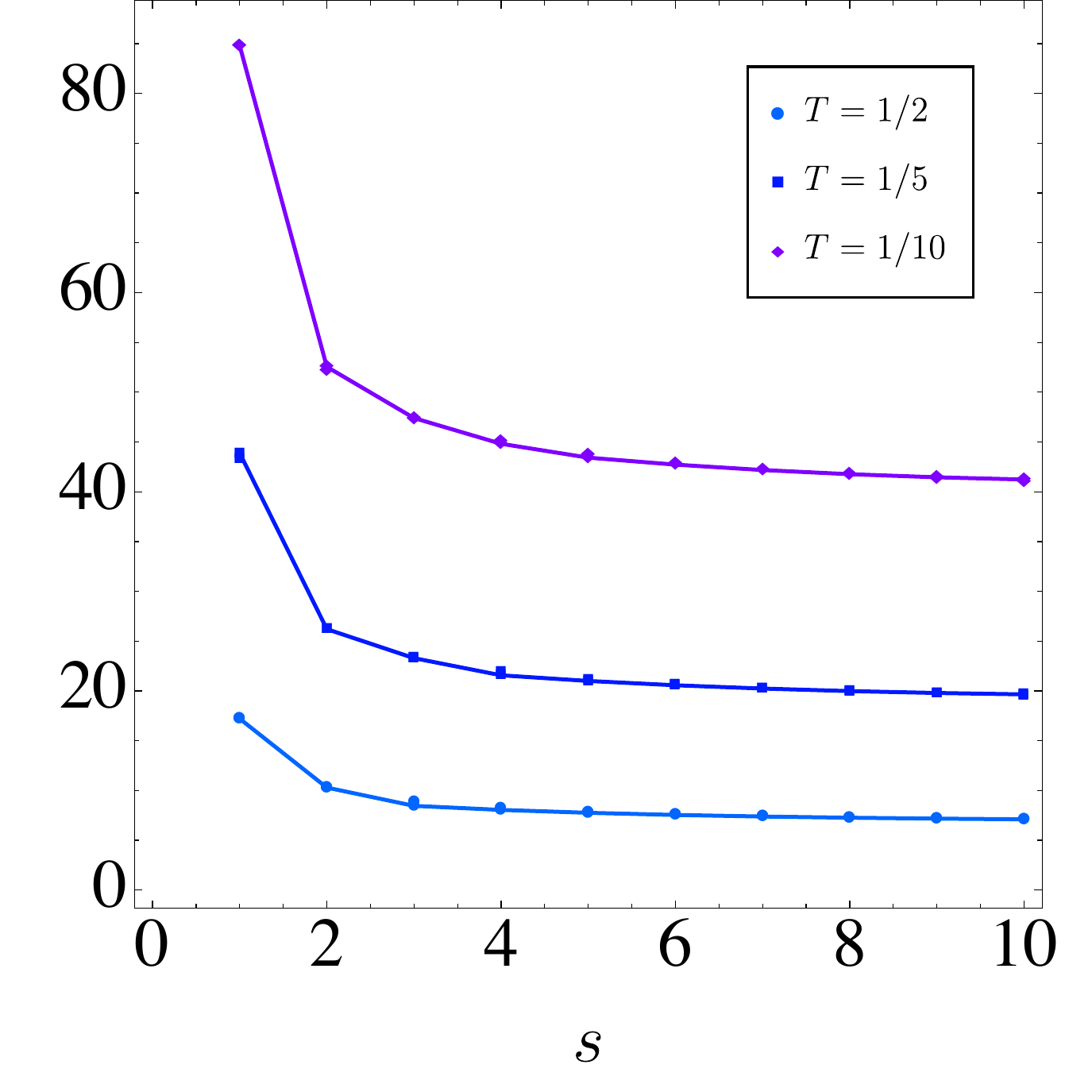}
\caption{For an XXX spin-$1/2$ chain, we plot $\mathcal{W}_s/(s+1)^2$ obtained from the numerical solution of the dressing equations 
for different temperatures and values of $s$ (continuous lines) versus $\frac{1}{2 {T} \chi_{h}(T,0)} \times \left[ \lim_{h \to 0} \frac{m^{\rm dr}_{s}(T,h)}{h}\right]/(s+1)^2$, also obtained by numerical solving for $m^{\rm dr }_s$ and numerically
evaluating $\chi_h$ (points). The perfect agreement between the two (up to the precision of the numerical solutions)
confirming the validity of the `magic formula' at general temperature. The same comparison can be done for the dressed functions
in the $O(3)$ non-linear sigma model.}
\label{Fig:CheckMagic}
\end{figure}

We begin by the exact hydrodynamic formula for the spin diffusion constant
valid in the vicinity of the half-filled thermal state,
\begin{equation}\label{eq:resultfinallXXZ}
\mathfrak{D}(T,h)  = \frac{1}{2} \sum_{s} \int \dd \theta\,
\rho_{s} (\theta) (1-n_{s}(\theta))| v_{s}^{\rm eff}(\theta)| [\mathcal{W}_s]^{2} + \mathcal{O}(h^2),
\end{equation}
obtained in \cite{1812.00767} via the thermodynamic form-factor expansion. Here the rapidity-independent weights
$\mathcal{W}_{s}$ are related to the suitably normalised dressed differential scattering phases,
\begin{equation}
\mathcal{W}_{s} = \lim_{b \to \infty}  \frac{K^{\rm dr}_{bs}(\alpha,\theta)}{\rho^{\rm tot}_{b}(\alpha)},
\end{equation}
where in the limit of large $s$ it becomes a constant function of $\theta$ and $\alpha$.
We subsequently demonstrate that formula \eqref{eq:resultfinallXXZ} can be rewritten as follows
\begin{equation}\label{eq:curvature_bound}
\mathfrak{D}(T,h)  = \frac{1}{8  {(T\chi_{h}(T,0))^2}} \sum_s \int \dd\theta\, \rho_{s}(\theta) (1-n_{s}(\theta)) |v_s^{\rm dr}(\theta)|
\left(\lim_{h \to 0} \frac{m^{\rm dr}_{s}}{h} \right)^{2} +\mathcal{O}(h^2),
\end{equation}
where the prefactor,
\begin{equation}
\chi_{h}(T,0) = \frac{\partial \expect{s^{z}}}{\partial h}\big|_{h=0} = -\partial_{h}^{2} f(T,h)\Big|_{h=0},
\end{equation}
is the static spin susceptibility at half filling, see Fig. \ref{Fig:CheckMagic}.
The formulae \eqref{eq:resultfinallXXZ} and \eqref{eq:curvature_bound} can be identified provided
\begin{equation}
\mathcal{W}_{s}(T) =  \frac{1}{2 {T} \chi_{h}(T,0)} \times \left[ \lim_{h \to 0} \frac{m^{\rm dr}_{s}(T,h)}{h}\right],
\end{equation}
holds true at half filling.

In the next section, we establish the above identity in the limit of infinite temperature \textit{where the dressing equations
take an algebraic form}. We to this for two representative models, (i) the Heisenberg spin-$1/2$ XXZ chain and (ii) the integrable 
$SU(3)$-symmetric Lai--Sutherland spin chain. Notice that in the $T\to \infty$ limit the prefactor simplifies,
\begin{equation}
\chi_{\mu}(T,0) = T\chi_{h}(T,0),\qquad
\lim_{T\to \infty}\chi_{\mu}(T,\mu=0) =  \partial^{2}_{\mu}\log \chi_{\square}(\mu)\Big|_{\mu=0} =  \frac{d^{2}-1}{12},
\end{equation}
where $\chi_{\square}(h)$ is the fundamental character and $d$ is the dimension of the local Hilbert space.
Therefore, writing $\mu \equiv h/T$, the relation which we shall prove below reads
\begin{equation}
\lim_{T \to \infty} \mathcal{W}_{s}(T)  \Big|_{h=0} = \frac{6}{d^{2}-1} \times \left[\lim_{\mu \to 0 } \frac{m^{\rm dr}_{s}(\mu)}{\mu}\right] + \mathcal{O}(\mu^2).
\end{equation}

\subsection{Proof of the magic formula}

\subsubsection{Isotropic Heisenberg chain}

The core part of the proof is based on the explicit calculation of the dressed differential scattering phase shifts
$K^{\rm dr}_{s,s^{\prime}}(\theta,\theta^{\prime})$. We consider the large temperature $T \to \infty$
the dressing transformation becomes a coupled system algebraic equations which can be solved in a closed analytic form.\\

\paragraph*{Identities for the scattering amplitudes.}
Using the fusion identities amongst the scattering amplitudes, the calculation boils down to computing the dressed momenta
of quasi-particle excitations for the entire familyof integrable $SU(2)$ spin chains with higher-spin local Hilbert spaces.
To this end, let the representation label $s^{\prime}\in \mathbb{N}$ denote the physical spin $S=s^{\prime}/2$ degrees of freedom
of the spin chain. The bare momenta of physical excitations (i.e. unbound magnons) are then given by
\begin{equation}
k^{(s^{\prime})}_{1}(\theta) = \ii \log S_{s^{\prime}}(\theta),
\end{equation}
where we have introduced the single-index `magnon-string' scattering amplitudes
\begin{equation}
S_{s}(\theta) = \frac{\theta - s\,\ii/2}{\theta + s\,\ii/2}.
\end{equation}
The inter-particle interactions allow for formation of bound states. These correspond to the so-called $s$-stings compounds,
consisting of $s$ magnons each carrying bare spin $s^{\prime}$, with bare momenta
\begin{equation}
k^{(s^{\prime})}_{s}(\theta) = \ii \log S_{s,s^{\prime}}(\theta).
\end{equation}
Here the two-particle scattering amplitudes are obtained from fusion,
\begin{equation}
S_{s,s^{\prime}}(\theta) = S_{|s-s^{\prime}|}(\theta)S_{s+s^{\prime}}(\theta)
\prod_{\ell = 1}^{{\rm min}(s,s^{\prime})-1}S^{2}_{|s-s^{\prime}|+2\ell}(\theta),
\end{equation}
and as usual depend only on the difference of the incident rapidities. Accordingly, we introduce the elementary scattering kernels,
\begin{equation}
K_{s}(\theta) = \frac{1}{2\pi \ii}\partial_{\theta}\log S_{s}(\theta),
\end{equation}
whose Fourier representation, defined through $\hat{f}(k)=\int_{\mathbb{R}} \dd \theta f(\theta)e^{-\ii k \theta}$, reads
\begin{equation}
\hat{K}_{s}(k) = e^{-|k|s/2}.
\end{equation}
Similarly, the kernels for the two-body differential scattering phases are given by
\begin{equation}
K_{s,s^{\prime}}(\theta) = \frac{1}{2\pi \ii}\partial_{\theta}\log S_{s,s^{\prime}}(\theta).
\end{equation}
It is worthwhile noticing the following two important kernel identities
\begin{align}
(1+K)_{s,s^{\prime}}\star K_{s^{\prime}} = \delta_{s,1}\fs,\qquad
(1+K)_{s,s^{\prime \prime}}\star K_{s^{\prime \prime},s^{\prime}} = I^{A_{\infty}}_{s,s^{\prime}}\fs.
\end{align}

Moreover, for later purpose it is convenient to define the `bare momentum tensor',
\begin{equation}
G_{s,s^{\prime}}(\theta) = \sum_{\ell = 1}^{{\rm min}(s,s^{\prime})}K_{|s-s^{\prime}|-1+2\ell}(\theta),\qquad
G_{s,s^{\prime}}(\theta,\theta^{\prime}) = G_{s^{\prime},s}(\theta^{\prime},\theta),
\end{equation}
given by
\begin{equation}
G_{s,s^{\prime}}(\theta) = \frac{1}{2\pi}|\partial_{\theta}k^{(s^{\prime})}_{s}(\theta)|.
\end{equation}
Notice that the two-particle scattering phase decompose as
\begin{equation}
K_{s,s^{\prime}}(\theta) = G_{s-1,s^{\prime}}(\theta) + G_{s+1,s^{\prime}}(\theta)
= G_{s,s^{\prime}-1}(\theta) + G_{s,s^{\prime}+1}(\theta).
\end{equation}

\paragraph*{Dressed magnetization.}

In the infinite temperature limit $T \to \infty$ with the $U(1)$ chemical potential $\mu \equiv h/T$,
the Fermi occupation function become rapidity independent and only depend on $\mu$.
The solution can be compactly expressed in terms of classical $SU(2)$ characters $\chi_{s}=\chi_{s}(h)$
(to not be confused with spin susceptibility $\chi_{h}(T,h)$),
\begin{equation}
n^{(0)}_{s} = \frac{1}{\chi^{2}_{s}(h)},\qquad \ol{n}^{(0)}_{s}(h) = 1-n^{(0)}_{s}(h).
\end{equation}
Introducing the variable $z\equiv e^{\mu}$, the characters of irreducible $(s+1)$-dimensional representation read
\begin{equation}
\chi_{s}(\mu) = \frac{z^{-(s+1)}-z^{s+1}}{z^{-1}-z}.
\end{equation}
The dressed magnetization can most easily extracted from the log-derivative of the infinite-temperature $Y$-functions
\begin{equation}
Y^{(0)}_{s}(\mu) = \chi^{2}_{s}(\mu) - 1,
\end{equation}
as
\begin{equation}
m^{\rm dr}_{s}(\mu) = \partial_{\mu}\log Y^{(0)}_{s}(\theta;\mu),
\end{equation}
At half filling, the lattice behave as
\begin{equation}
m^{\rm dr}_{s}(\mu) = \frac{1}{3}(s+1)^{2}\mu + \mathcal{O}(\mu^{3}).
\end{equation}

\paragraph*{Dressed momenta.}
By splitting the dressed scattering kernel into two parts,
\begin{equation}
K^{\rm dr }_{s,s^{\prime}}(\theta) 
= G^{\rm dr }_{s,s^{\prime}-1}(\theta) + G^{\rm dr }_{s,s^{\prime}+1}(\theta),
\end{equation}
we proceed by calculating the the dressed values of the bare energy tensor $G^{\rm dr}_{s,s^{\prime}}$.
Recall that the latter provides the dressed rapidity derivatives, $\partial_{\theta}p^{(s^{\prime})}_{s}(\theta)$.

By proceed by analytically solving the dressing equations. To this end, we represent them in the quasi-local form.
In this respect, it is crucial to determined the position of the source node depending on the spin label $s^{\prime}$.
This can be inferred by convolving tensor $G$ with the (pseudo)inverse of the Fredholm kernel, that is
\begin{equation}
\left(1+K\right)^{-1}_{s,s^{\prime} \prime} \star G_{s^{\prime \prime},s^{\prime}}
= \left(1-I^{A_{\infty}}\fs \right)_{s,s^{\prime \prime}}\star G_{s^{\prime \prime},s^{\prime}} = \delta_{s,s^{\prime}}\fs.
\end{equation}
The source term thus resides at the $s^{\prime}$-th node. Therefore, to find the dressed momentum tensor $G^{\rm dr}$,
one has to solve the following system
\begin{equation}
(1 - I^{A_{\infty}}\,\ol{n}\,\fs)_{s,s^{\prime \prime}}\star G^{\rm dr}_{s^{\prime \prime},s^{\prime}} = \delta_{s,s^{\prime}}\fs.
\end{equation}
By introducing variables $F^{(s^{\prime})}_{s} \equiv G^{\rm dr}_{s,s^{\prime}}$ and transferring to Fourier space,
$F^{(s^{\prime})}_{s}(\theta;z) \mapsto \hat{F}^{(s^{\prime})}_{s}(k;z)$,
we arrive at the following three-point inhomogeneous recurrence relation
\begin{equation}
\fs^{-1}\cdot \hat{F}^{(s^{\prime})}_{s}
- I^{A_{\infty}}_{s,s^{\prime \prime}}\ol{n}^{(0)}_{s^{\prime \prime}}\hat{F}^{(s^{\prime})}_{s^{\prime \prime}}
= \delta_{s,s^{\prime}},
\end{equation}
where $n^{(0)}_{s}$ denote the infinite-temperature mode occupation functions and $\fs^{-1}(k)=2\cosh{(k/2)}$.
We first obtain the homogeneous solution to the above recurrence, which is given by
\begin{equation}
\hat{\Phi}^{(s^{\prime})}_{s}(k;z|C_{-},C_{+}) =
\sum_{\alpha=\pm}\frac{\chi_{s}(z)}{\chi_{1}(z)}
\left[\frac{e^{\alpha(s+1)k/2}}{\chi_{s-1}(z)}-\frac{e^{\alpha(s+1)k/2}}{\chi_{s+1}(z)}\right]C_{\alpha}(k;z),
\end{equation}
for two unknown functions $C_{\pm}(k;z)$.
The particular solution is singled out by imposing appropriate initial and boundary conditions.
To satisfy the large-$s$ asymptotics, $\lim_{|\theta|\to \infty}F^{(s^{\prime})}_{s}(\theta)=0$, we put
\begin{equation}
\hat{F}^{(s^{\prime})}_{s\geq s^{\prime}} \leftarrow \hat{\Phi}^{(s^{\prime})}_{s}(k;z|\mathcal{C},0),\qquad
\hat{F}^{(s^{\prime})}_{s< s^{\prime}} \leftarrow \hat{\Phi}^{(s^{\prime})}_{s}(k;z|\mathcal{A},\mathcal{B}),
\end{equation}
and write a closed system of equations at the initial node and the two gluing conditions at nodes $s^{\prime}-1$ and $s^{\prime}$.
The solution for the fundamental particles is simply given by
\begin{equation}
s^{\prime}=1:\qquad\qquad \mathcal{C}(k;z) = e^{-k/2}.
\end{equation}
The general solution for higher representations, namely for $s^{\prime}\geq 3$, is more unwieldy and reads
\begin{align}
\mathcal{A}(k;z) &= \frac{(1+z^{2})\left(e^{k}(z^{4-2s^{\prime}}-1)-z^{2}(z^{2s^{\prime}-1})\right)}
{(e^{k}-1)(z^{2(s^{\prime}+1)}-1)\left(z^{2}(1+e^{2k})-e^{k}(1+z^{4})\right)}e^{-s^{\prime}k/2},\\
\mathcal{B}(k;z) &= -e^{k}\mathcal{A}(k;z),\\
\mathcal{C}(k;z) &= \frac{(1+z^{2})\left(z^{2}(z^{2s^{\prime}}-1)-z^{2}(z^{2s^{\prime}}-1)e^{(s^{\prime}+2)k}+(z^{4+2s^{\prime}}-1)(e^{(s^{\prime}+1)k}-e^{k})\right)}{(e^{k}-1)(z^{2(s^{\prime}+1)}-1)(z^{2}(1+e^{2k})-e^{k}(1+z^{4}))}.
\end{align}
We will need $s\gg s^{\prime}$, hence only $\mathcal{C}(k;z)$ will be of our interest. The full $k$-dependent
solution $\hat{F}^{(s^{\prime})}_{s}(k;z)$ is quite lengthy and we thus suppress it here. Importantly however, since the
final solution, after taking the $s\to \infty$ limit contains no rapidity dependence,
it suffices to consider only the $k\to 0$ limit. In particular, one can explicitly verify that
\begin{equation}
\lim_{s\to \infty}\frac{\hat{K}^{\rm dr}_{s,s^{\prime}}(k)}{\eta_{s,s^{\prime}}} = \delta(k),\qquad 
\eta_{s,s^{\prime}} \equiv \frac{2}{3}(s^{\prime}+1)^{2}\frac{s+1}{s(s+2)},
\end{equation}
which implies that in Fourier space the rescaled dressed scattering kernels converge towards a delta function.
In the $k\to 0$ limit, we find a simpler expression
\begin{equation}
\lim_{k\to 0}\hat{F}^{(s^{\prime})}_{s>s^{\prime}}(k;z) =
\frac{(z^{2(s+1)}-1)(z^{2(s+1)}+1)\left(s^{\prime}(z^{2}-1)(z^{2(s^{\prime}+1)}+1)-2z^{2}(z^{2s^{\prime}}-1)\right)}
{(z^{2}-1)(z^{2(s^{\prime}+1)}-1)(z^{2s}-1)(z^{4+2s}-1)},
\end{equation}
and, using the relation
\begin{equation}
K^{\rm dr}_{s,s^{\prime}}(\theta;z) = G^{\rm dr}_{s,s^{\prime}-1}(\theta;z) + G^{\rm dr}_{s,s^{\prime}-1}(\theta;z)
= F^{(s^{\prime}-1)}_{s}(\theta;z) + F^{(s^{\prime}+1)}_{s}(\theta;z),
\end{equation}
we obtain
\begin{equation}
\lim_{k\to 0}\hat{K}^{\rm dr}_{s,s^{\prime}}(k;z) = 
\frac{2(z^{2(b+1)}-1)(z^{4(s+1)}-1)\left(b(z^{2}-1)(z^{2(b+1)}+1)-2z^{2}(z^{2b}-1))\right)}
{(z^{2}-1)(z^{2b}-1)(z^{4+2b}-1)(z^{2s}-1)(z^{4+2s}-1)}.
\end{equation}
Taking furthermore the limit of half filling, $\mu\to 0$ ($z\to 1$), the above result reduces to
\begin{equation}
\lim_{z\to 1}\lim_{k\to 0}\hat{K}^{\rm dr}_{s,s^{\prime}}(k;z) = \eta_{s,s^{\prime}}.
\end{equation}
In particular, $\hat{K}^{\rm dr}_{s,s^{\prime}}(k=0,z=e^{\mu})$ decays to zero in both the large-$s$ and small-$\mu$ limit:
\begin{align}
\lim_{\mu\to 0}\lim_{k\to 0}\hat{K}^{\rm dr}_{sb}(\mu) &= \frac{2}{3}(s^{\prime}+1)^{2}\frac{1}{s} + \mathcal{O}(s^{-2}),\\
\lim_{s\to \infty}\lim_{k\to 0}\hat{K}^{\rm dr}_{s,s^{\prime}}(\mu) &= \frac{2}{3}(s^{\prime}+1)^{2}\mu + \mathcal{O}(\mu^{3}).
\end{align}
Likewise, for the total state densities $\rho^{\rm tot}_{s}(\theta)=\tfrac{1}{2\pi}|\partial_{\theta}p^{(s^{\prime}=1)}_{s}(\theta)|$,
we find
\begin{equation}
\lim_{\mu\to 0}\lim_{k\to 0}\hat{\rho}^{\rm tot}_{s}(k;\mu) = \frac{1}{s} + \mathcal{O}(s^{-2}),\qquad
\lim_{s\to \infty}\lim_{k\to 0}\hat{\rho}^{\rm tot}_{s}(k;\mu) = \mu + \mathcal{O}(\mu^{3}).
\end{equation}

\subsubsection{Integrable SU(N) spin chains}

In this section we extend the above computation to a class of model solvable with the nested Bethe Ansatz.
We consider integrable the $SU(N)$-symmetric spin chains made of fundamental particles.
The quasi-particle spectrum now arranges on vertices of an infinite lattice known as `the T-strip'.
Since the nodes are in one-to-one correspondence with rectangular irreducible unitary representations of $\mathfrak{su}(N)$
Lie algebra we will label them by $(a,s)$, with integers $1\leq a \leq N$ and $s\in \mathbb{N}$.
In particular, the row label $a$ runs over different species (flavors) of particles, while the column label $s$
belongs to bound states with $s$ constituent particles.
By convention, the momentum-carrying particles belong are assigned to the bottom row $a=1$.

In the \emph{fundamental} $SU(N)$ spin chains, the elementary magnon excitations have momenta
$k_{1,s}(\theta) = -\ii \log S_{1}(\theta)$. All other magnons ($a=2,\ldots,N$) can be though of as auxiliary particles which carry
no momenta and energy, that is $p_{a>1,s}=e_{a>1,s}=0$. Each particle species participate in the formation of bound
state (Bethe strings). The mechanism is analogous to that of the $SU(2)$ chain. The momenta of $s$-strings read
$k_{a,s}(\theta) = -\ii\,\delta_{a,1}\log S^{-1}_{s}(\theta)$.
Therefore, we have $|\partial_{\theta}k_{a,s}(\theta)/2\pi|=\delta_{a,1}G_{1,s}(\theta)=\delta_{a,1}K_{s}(\theta)$.
\\

\paragraph*{Bethe equations.}

The Bethe equations for the fundamental $SU(N)$ chain of length $L$ have the nested form
\begin{equation}
e^{\ii \delta_{\ell,1}k(\theta^{(\ell)}_{\alpha})L}
\prod_{\beta\neq \alpha}^{M_{\ell}}S_{2}(\theta^{(\ell)}_{\alpha},\theta^{(\ell)}_{\beta})
\prod_{r=1;r = \ell \pm 1}^{N-1}\prod_{\beta = 1}^{M_{r}}S^{-1}_{1}(\theta^{(\ell)}_{\alpha},\theta^{(r)}_{\beta}) = 1,\qquad
\ell=1,\ldots,N-1,
\label{eqn:Bethe_equations_SU(N)}
\end{equation}
where $\{\theta^{(\ell)}_{\alpha}\}_{\alpha=1}^{M_{\ell}}$ denote (complex) rapidity variables for different quasi-particle species
$\ell = 1,2,\ldots N-1$.
\\

\paragraph*{Bethe--Yang equations.}
In the thermodynamic limit, obtained by taking $L\to \infty$ while keeping all filling fractions $M_{\ell}/L \sim \mathcal{O}(1)$ 
finite, Bethe equations \eqref{eqn:Bethe_equations_SU(N)} can be reformulated as the Bethe--Yang equations for
analytic rapidity densities $\rho_{a,s}$,
\begin{equation}
\rho_{a,s} + \ol{\rho}_{a,s} = \left|\frac{k^{\prime}_{a,s}}{2\pi}\right|
- K_{(a,s),(a^{\prime},s^{\prime})}\rho_{a^{\prime}s^{\prime}}.
\label{eqn:Bethe-Yang_SUN(N)}
\end{equation}
This follows from \eqref{eqn:Bethe_equations_SU(N)} by
(i) taking the logarithmic rapidity derivative, (ii) reducing the product of scattering amplitudes by using string configurations,
and (iii) passing to continuum description by approximating large sum with convolution-type integrals.
Another (equivalent) form of Eqs.~\eqref{eqn:Bethe-Yang_SUN(N)} is
\begin{equation}
(1+K)_{(a,s),(a^{\prime},s^{\prime})}\star \rho_{a^{\prime},s^{\prime}}
= \left|\frac{k^{\prime}_{a,s}}{2\pi}\right| - \ol{\rho}_{a,s}.
\end{equation}
Kernels $K_{(a,s),(a^{\prime},s^{\prime})}(\theta)$ encode differential scattering phases associated
to the scattering even between $(a,s)$ and $(a^{\prime},s^{\prime})$ string excitations with rapidity difference $\theta$,
where $a$ is the flavour label and $s$ the number of constituent magnons.
As can be seen from the structure of the (nested) Bethe equations, the Fredholm kernel $(1+K)$ is such that only the neighbouring 
species interact among each other, namely $K_{(a,s),(a^{\prime},s^{\prime})}$ is non-zero only if $a^{\prime}=a\pm 1$.
\\

\paragraph*{Higher representations.}
Next, we consider a family of spin chains whose local Hilbert spaces belong to the one-row tableaux with $s^{\prime}$ boxes.
In analogy to the $N=2$ case, we introduce the bare momentum tensor
$G_{(a,s),(a^{\prime},s^{\prime})}(\theta)\equiv G^{(a^{\prime},s^{\prime})}_{a,s}(\theta)$, which carries all
information about the bare momenta $k^{(a^{\prime},s^{\prime})}_{a,s}(\theta)\equiv k^{(s^{\prime})}_{s}$ of
elementary magnon excitations in a spin-$s^{\prime}/2$ chain (including their $s$-magnon bound states, namely
\begin{equation}
G_{s,s^{\prime}}(\theta) = \left|\frac{\partial_{\theta}k^{(s^{\prime})}_{s}(\theta)}{2\pi}\right|.
\end{equation}

\paragraph*{Dressed magnetization.}

The rank of $\mathfrak{su}(N)$ Lie algebra is $N-1$, which is the number of globally conserved Noether charges.
This means that the whole Cartan sector is parametrised by $N-1$ distinct chemical potentials.
Here we focus only to a single charge, namely the total magnetization $S^{z}_{\rm tot} = \sum_{i}S^{z}_{i}$, with
local density $S^{z} = {\rm diag}(S,S-1,\ldots,-S)$. The conjugate chemical potential will be denoted by $\mu$.

The classical characters of rectangular Young tableau are functions of the Cartan elements
\begin{equation}
G={\rm diag}(x_{1},\ldots,x_{N}) = \exp{(-\mu S^{z})},
\end{equation}
and can be compactly expressed with help of the Weyl formula
\begin{equation}
\chi^{(N)}_{a,s}(\mu) = \frac{{\rm Det}\left(x^{N-j+s+\Theta_{s,j}}_{k}\right)_{1\leq j,k,\leq N}}
{{\rm Det}\left(x^{N-j}_{k}\right)_{1\leq j,k\leq N}},
\end{equation}
where $\Theta_{i,j}=1$ if $i\geq j$ and zero otherwise. In fact, all $\chi^{(N)}_{a\geq 2,s}(\mu)$ are uniquely determined by
the symmetric functions $\chi^{(N)}_{1,s}(\mu)$ by virtue of the Giambelli--Jacobi--Trudi formula
\begin{equation}
\chi^{(N)}_{a,s}(\mu) = {\rm Det}\left(\chi^{(N)}_{1,s+j-k}(\mu)\right)_{1\leq j,k,\leq a}.
\end{equation}

The occupation functions of the grand-canonical Gibbs ensembles at infinite-temperature are
encoded in the classical $Y$-functions $Y^{(0)}_{a,s}(\mu)$. The latter are the following non-linear combinations of the 
$\mathfrak{su}(N)$ characters,
\begin{equation}
Y^{(0)}_{a,s}(\mu) = \frac{\chi_{a,s-1}(\mu)\chi_{a,s+1}(\mu)}{\chi_{a-1,s}(\mu)\chi_{a+1,s}(\mu)},
\qquad a \in \{1,2\},\quad s\in \mathbb{N},
\end{equation}
with boundary conditions $\chi_{a,s}\equiv 0$ for $a\in \{0,3\}$.
For example, to extract the dressed magnetization for $N=3$, we set $G={\rm diag}(e^{-\mu},1,e^{\mu})$ and compute
\begin{equation}
m^{\rm dr}_{a,s}(\mu) = \partial_{\mu}\log Y^{(0)}_{a,s}(\mu).
\end{equation}
In the vicinity of a half-filled state, we find
\begin{equation}
m^{\rm dr}_{a,s}(\mu) = \frac{1}{6}(s+1)(s+2)\mu + \mathcal{O}(h^{3}),\qquad a\in \{1,2\}.
\end{equation}\\

\paragraph*{Dressed momenta.}

In the $SU(3)$ case, the dressing equation for the rapidity derivatives of the bare momenta become (written in Fourier space)
a coupled system of recurrence relations for functions $\{F_{1,s},F_{2,s}\}_{s\geq 1}$,
\begin{align}
\hat{\fs}^{-1}\cdot \hat{F}_{1,s} - \ol{n}^{(0)}_{s-1}\hat{F}_{1,s-1} - \ol{n}^{(0)}_{s+1}\hat{F}_{1,s+1}
- n^{(0)}_{s-1}\hat{F}_{2,s-1} &= \delta_{s,s^{\prime}},\\
\hat{\fs}^{-1}\cdot \hat{F}_{2,s} - \ol{n}^{(0)}_{s-1}\hat{F}_{2,s-1} - \ol{n}^{(0)}_{s+1}\hat{F}_{2,s+1}
- n^{(0)}_{s-1}\hat{F}_{1,s-1} &= 0.
\label{eqn:SU3_recurrence}
\end{align}
In the half filled case $\mu=0$ we consider here, the infinite-temperature occupation functions read
\begin{equation}
n^{(0)}_{a,s} = \frac{2}{(s+1)(s+2)}.
\end{equation}
In Eqs.~\eqref{eqn:SU3_recurrence} ,the position of the source node, located at $(a,s)=(1,S)$, is prescribed by
the one-row tableau associated to local physical degrees of freedom in the spin chain.

By introducing two independent linear combinations
\begin{equation}
\hat{F}^{\pm}_{s} = \hat{F}_{1,s} \pm \hat{F}_{2,s},
\end{equation}
the system of equation \eqref{eqn:SU3_recurrence} reduces to a one-dimensional recurrence,
\begin{equation}
\hat{s}^{-1}\cdot \hat{F}^{\pm}_{s} - \ol{n}^{(0)}_{s-1}\hat{F}^{\pm}_{s-1} - \ol{n}^{(0)}_{s+1}\hat{F}^{\pm}_{s+1}
\mp n^{(0)}_{s}\hat{F}^{\pm}_{s} = \delta_{s,1},
\label{eqn:SU3_recurrence_merged}
\end{equation}
which can in turn be solved using a similar strategy as previously in the $N=2$ case.
Specifically, for the fundamental representation $S=1$, we find
\begin{equation}
\hat{F}^{\pm}_{1}(k) = \hat{K}_{1}(k) \pm \frac{1}{3}\hat{K}_{1}(k) - \frac{1}{3}\hat{K}_{3}(k) \mp \frac{1}{6}\hat{K}_{4}(k),\qquad
\hat{C}^{\pm}(k) = 2\hat{K}_{1}(k),
\end{equation}
implying
\begin{align}
\hat{F}_{1,s}(k) &= \frac{1}{3s}\left((s+2)\hat{K}_{s}(k)-s\,\hat{K}_{s+2}(k)\right),\\
\hat{F}_{2,s}(k) &= \frac{1}{3(s+3)}\left((s+3)\hat{K}_{s+1}(k)-(s+1)\hat{K}_{s+3}(k)\right).
\end{align}

Next, we obtain solutions for generic one-row tableaux with $S$ boxes. These can be found with a similar strategy, except that
the source term in Eq.~\eqref{eqn:SU3_recurrence_merged} now jumps to the $s^{\prime}$-th node. Although it is not difficult
to obtain closed-form expressions, e.g. with assistance with symbolic algebra routines, we unfortunately could not
display them in a sufficiently economic way. Their general structure, valid for $S\geq 3$, is however of the form
\begin{equation}
\hat{F}^{(s^{\prime})}_{1,s} = \sum_{k=0}^{s^{\prime}}c^{(1)}_{s,k}\hat{K}_{s-s^{\prime}+1+2k},\qquad
\hat{F}^{(s^{\prime})}_{2,s} = \sum_{k=0}^{s^{\prime}}c^{(2)}_{s,k}\hat{K}_{s-s^{\prime}+2+2k}.
\end{equation}
The $k\to 0$ limits are nonetheless rather simple,
\begin{align}
\lim_{k\to 0}\hat{F}^{(s^{\prime})}_{1,s}(k) &= \frac{s^{\prime}(s^{\prime}+3)(5s+3s^{\prime}+12)}{30s(s+3)},\\
\lim_{k\to 0}\hat{F}^{(s^{\prime})}_{2,s}(k) &= \frac{s^{\prime}(s^{\prime}+3)(5s-3s^{\prime}+3)}{30s(s+3)},
\end{align}
Moreover, we have the following large-$s$ limits (with $s>s^{\prime}$)
\begin{align}
\hat{F}^{(s^{\prime})}_{a,s}(0) &= \frac{1}{6}s^{\prime}(s^{\prime}+3)\frac{1}{s} + \mathcal{O}\left(s^{-2}\right),\\
\hat{F}^{(s^{\prime}-1)}_{a,s}(0)+\hat{F}^{(s^{\prime}+1)}_{a,s}(0) &\simeq
\frac{1}{3}\left((s^{\prime})^{2}+3s^{\prime}+1\right)\frac{1}{s} + \mathcal{O}\left(s^{-2}\right),
\end{align}
which do not depend on label $a$. To complete the proof of the magic formula,
the dressed scattering kernels $K^{\rm dr}_{(a,s),(a^{\prime},s^{\prime})}$ are finally expressed in terms of the
dressed bare momentum tensor $G^{\rm dr}_{s,s^{\prime}}$. Specifically,
\begin{align}
\lim_{s\to \infty}\lim_{k\to 0}\hat{K}^{\rm dr}_{(a,s),(a^{\prime},s^{\prime})}(k)
&=\lim_{s\to \infty}\lim_{k\to 0}
\left[\delta_{a,a^{\prime}}(\hat{G}^{\rm dr}_{s,s^{\prime}-1}(k)+\hat{G}^{\rm dr}_{s,s^{\prime}+1}(k))
-I^{A_{2}}_{a,a^{\prime}}\hat{G}^{\rm dr}(k)_{s,s^{\prime}}\right],\\
&= \frac{1}{6}(s^{\prime}+1)(s^{\prime}+2)\frac{1}{s} + \mathcal{O}(s^{-2}).
\end{align}

\newpage 
\section{Additional numerical data}
\begin{figure}[t!]
\center
\includegraphics[width=0.49\textwidth]{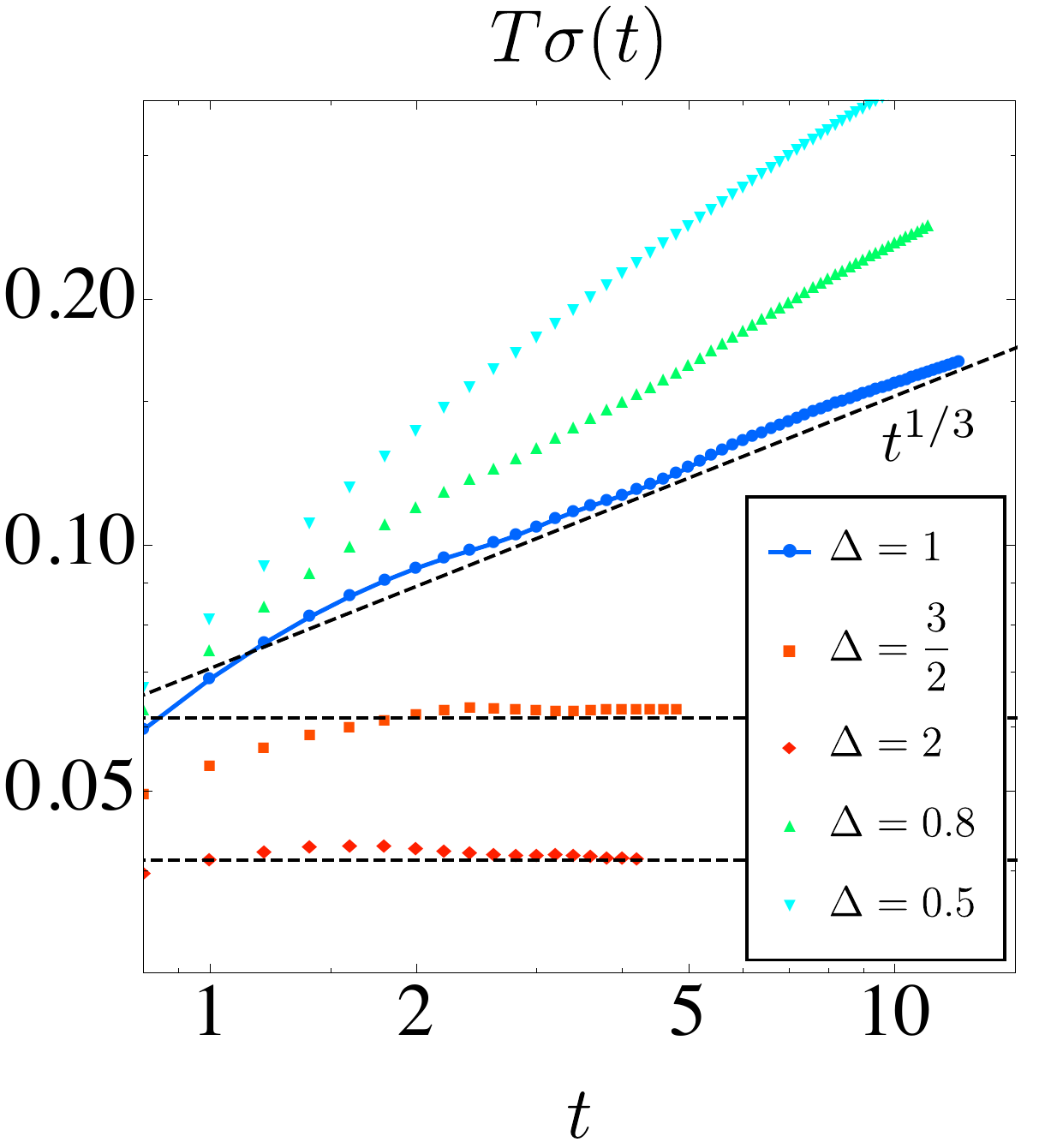}
\caption{Log-Log plot of the spin conductivity  for the (non-integrable) spin-1 XXZ chain $\hat{H}_{\Delta}$
at $\Delta=\{1,1.5,2,0.5,0.8\}$ and $T=10$ and $h=0$.}
\label{Fig:spin1Delta}
\end{figure}

\begin{figure}[t!]
\center
\includegraphics[width=0.49\textwidth]{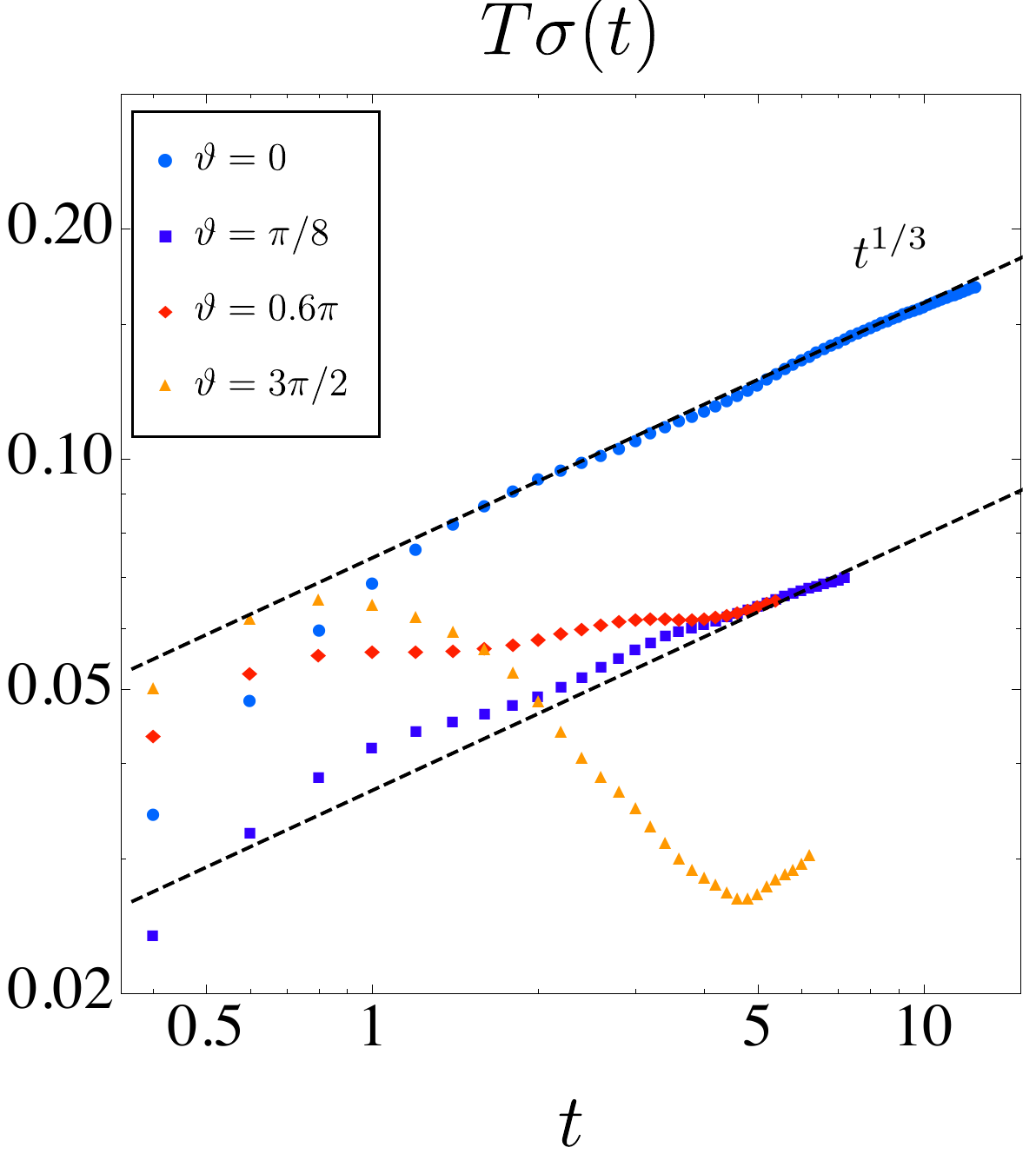}
\caption{Log-Log plot of the spin conductivity at $T=10$ and $h=0$ for the $SU(2)$ spin-1 chains $\hat{H}_{\vartheta}$.   }
\label{Fig:spin1theta}
\end{figure}

\begin{figure}[t!]
\center
\includegraphics[width=0.49\textwidth]{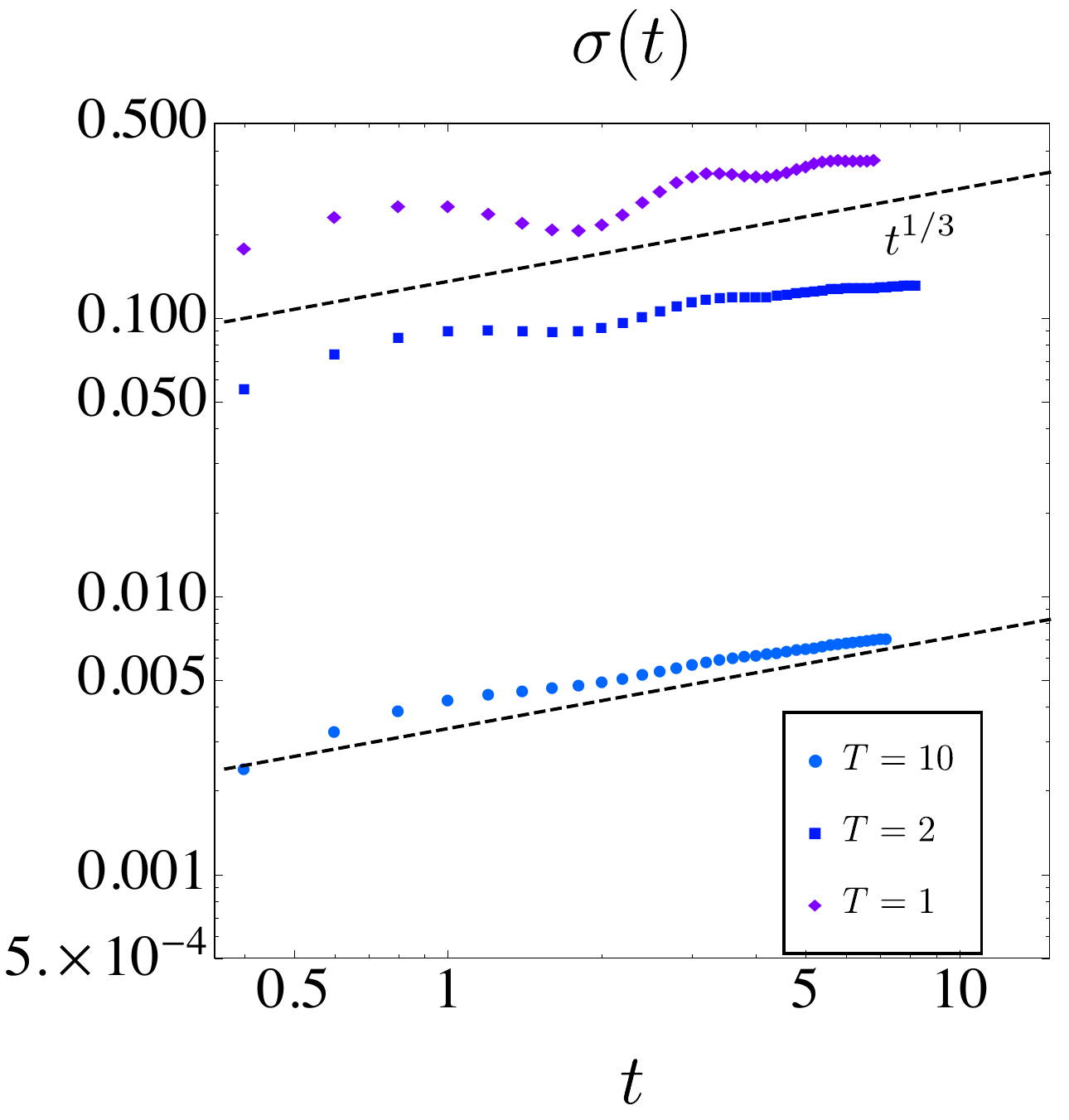}
\caption{Log-Log plot of the spin conductivity at $h=0$ for the $SU(2)$ spin-1 chain $\hat{H}_{\vartheta=\pi/8}$
for three different values of temperature.}
\label{Fig:spin1125}
\end{figure}

We here report additional numerical data on the time dependent conductivity of the spin current $\hat{j}$
\begin{equation}\label{eq:condu2}
\sigma(t) =\frac{1}{T}\int_{0}^{t}\!\!\dd t'    \big\langle \hat{J}(t')\hat{j}_{0}(0) \big\rangle_{T,h=0}  \quad  \quad \hat{J} = \sum_i \hat{j}_i.
\end{equation}
 We first examine the restoration of normal diffusion upon explicitly breaking
the interaction isotropy. We consider the uniaxially anisotropic version of the Haldane \textit{spin-$1$ chain},
see Fig. \ref{Fig:spin1Delta}, namely 
\begin{equation}
\hat{H}_{\Delta} =  \sum_x \hat{s}^x_x \hat{s}^x_{x+1} + \hat{s}^y_x \hat{s}^y_{x+1} + \Delta \hat{s}^z_x \hat{s}^z_{x+1}.
\end{equation}
The $SU(2)$-symmetric point $\Delta=1$ displays super-diffusion with $\sigma(t) \sim t^{1/3}$ as predicted by the low-energy theory. For $\Delta>1$ we instead find
normal diffusion with $\sigma(t) \sim \mathfrak{D} \chi_h + b/t^{1/2}$ with $ \mathfrak{D}$ finite. For $\Delta<1$ instead the situation is less clear as super-diffusion seems also to be present, although with some larger exponent. While this phenomenon could be related to the presence of a deformation of the $O(3)$ sigma model which also displays super-diffusion, we postpone these questions to further studies.  In Fig. \ref{Fig:spin1theta} we study the growth of the time-dependent conductivity 
for a one-parametric family
of $SU(2)$-invariant spin-1 Hamiltonians
\begin{equation}
\hat{H}_{\vartheta} = \sum_{i} \left( \cos(\vartheta) \hat{\bf s}_{i}\cdot \hat{\bf s}_{i+1} +  \sin(\vartheta)( \hat{\bf s}_{i}\cdot \hat{\bf s}_{i+1})^2 \right),
\end{equation}
for several different values of $\vartheta$, including the Haldane gapped phase, $\vartheta=0,\pi/8$, the ferromagnetic phase, $\vartheta = 0.6 \pi$, and the dimerised phase, $\vartheta =3\pi/2$, see for example \cite{Luchli2006}. While withing
the Haldane-gapped phase we find clear evidence of super-diffusion (as expected from the low-lying $O(3)$ non-linear sigma model theory), the results for the other two phases are less conclusive.
Finally, in Fig. \ref{Fig:spin1125}, we display the spin conductivity in the $SU(2)$ spin-1 chains
$\hat{H} = \sum_{i} \cos(\vartheta) \hat{\bf s}_{i}\cdot \hat{\bf s}_{i+1} +  \sin(\vartheta)( \hat{\bf s}_{i}\cdot \hat{\bf s}_{i+1})^2 $ at $\vartheta=\pi/8$ (inside the Haldane phase) for two different values of temperature, both compatible
with super-diffusion.

\newpage 

\end{document}